\pdfoutput=1
\documentclass[11pt,final,twoside]{memoir}
\usepackage{amsmath,amssymb,bm,amsthm,bbold}
\usepackage{ifdraft}
\usepackage[final]{graphicx}
\usepackage{enumitem}
\usepackage[final]{listings}
\usepackage{xmpincl}
\usepackage{latexsym}
\usepackage[utf8]{inputenc}
\usepackage[T1]{fontenc}
\usepackage{mathrsfs}
\usepackage{tikz}

\usepackage{color}
\definecolor{Black}{rgb}{0,0,0}
\definecolor{MidnightBlue}{rgb}{0.023,0.375,0.559}
\definecolor{Gray}{rgb}{0.4,0.4,0.4}
\definecolor{ForestGreen}{rgb}{0.3,0.6,0.3}
\definecolor{CadetBlue}{rgb}{0.3,0.3,0.6}
\definecolor{Blue}{rgb}{0,0,1}
\definecolor{White}{rgb}{1,1,1}

\usepackage[LowerCaseTrace,SquareTraceBrackets]{quantum}

\newcommand{\zero}{\mathnormal{0}}
\renewcommand{\one}{\mathnormal{1}}
\newcommand{\vac}{\ket{\varnothing}}
\newcommand{\modsq}[1]{\Abs{#1}^2}
\newcommand{\lb}[1]{\log_2\left( #1 \right)}

\newcommand{\qr}{q_\text{R}}
\newcommand{\AR}{\overline{R}}
\newcommand{\ro}{r}
\newcommand{\ud}{\mathrm{d}}
\DeclareMathOperator{\Li}{Li}
\DeclareMathOperator{\artanh}{artanh}
\DeclareMathOperator{\csch}{csch}
\DeclareMathOperator{\sech}{sech}
\usetikzlibrary{shapes,arrows}
\tikzstyle{decision} = [diamond, draw, text width=8em, text badly centered, node distance=4cm]
\tikzstyle{block} = [rectangle, draw, text width=10em, text centered, rounded corners, minimum height=4em]
\tikzstyle{line} = [draw, -latex']
\tikzstyle{endpoint} = [draw, rectangle, minimum height=2em, text width=6em, text centered]
\tikzstyle{decision answer}=[near start,color=black]

\usepackage[backend=biber,
            sortcase=false,
            sorting=nyt,
            sortcites=true,
            hyperref=true,
            backref=true,
            backrefstyle=none,
            abbreviate=false,
            alldates=iso8601,
            arxiv=abs,
            isbn=false,
            url=false
                      ]{biblatex}
\AtEveryBibitem{%
  \ifentrytype{online}
    {\clearfield{url}}
    {}%
}

\usepackage[colorlinks=true,final,
            unicode=true,
            pdftex,
            pdfauthor={Dominic Hosler},
            pdftitle={Relativistic Quantum Communication},
            pdfsubject={Relativistic Quantum Information Theory},
            pdfkeywords={},
            pdfproducer={LaTeX with hyperref and Memoir},
            pdfcreator={pdflatex}]{hyperref}
\usepackage[all]{hypcap}

\hypersetup{linkcolor=MidnightBlue,citecolor=MidnightBlue,urlcolor=MidnightBlue}

\usepackage[xindy,nomain,section=chapter,toc,nonumberlist,style=super]{glossaries}
\renewcommand{\glsnamefont}[1]{\mdseries #1}
\renewcommand{\glsgroupskip}{} 
\newglossary[slg]{symbolslist}{syi}{syg}{List of Symbols}
\makeglossaries

\newglossaryentry{sym:zero}{name={\ensuremath{\zero}},description={A logical zero},type=symbolslist,sort={0}}

\newglossaryentry{sym:one}{name={\ensuremath{\one}},description={A logical one},type=symbolslist,sort={1}}

\newglossaryentry{sym:alice}{name={\ensuremath{A}},description={The inertial observer Alice},type=symbolslist,sort={a}}

\newglossaryentry{sym:a}{name={\ensuremath{a}},description={Rob's proper acceleration},type=symbolslist,sort={aa}}

\newglossaryentry{sym:aoperators}{name={\ensuremath{\hat{a},\hat{a}^\dagger}},description={Creation and annihilation operators, subscripts denote to which mode it applies},type=symbolslist,sort={ab}}

\newglossaryentry{sym:alphabeta}{name={\ensuremath{\alpha,\beta}},description={Coefficients of $\ket\zero$ and $\ket\one$ respectively in a state},type=symbolslist,sort={alpha}}

\newglossaryentry{sym:alphajk}{name={\ensuremath{\alpha_{jk}}},description={Bogoliubov transformation coefficient between the modes of $j$ and $k$},type=symbolslist,sort={alphajk}}

\newglossaryentry{sym:boperators}{name={\ensuremath{\hat{b},\hat{b}^\dagger}},description={Creation and annihilation operators, for the transformed modes \gls{sym:fmodes}, subscripts denote to which mode it applies},type=symbolslist,sort={ba}}

\newglossaryentry{sym:betajk}{name={\ensuremath{\beta_{jk}}},description={Bogoliubov transformation coefficient between the modes of $j$ and $k$},type=symbolslist,sort={betajk}}

\newglossaryentry{sym:capacity}{name={\ensuremath{C}},description={Restricted channel capacity, this is the optimised communication rate when the channel is restricted to a particular encoding method, for a full discussion see \cref{subsec:inf:quantumchannelcapacity}},type=symbolslist,sort={capacity}}

\newglossaryentry{sym:probmetric}{name={\ensuremath{d^{\mu \nu}}},description={The metric defined on the space of probability distributions, or the space of quantum states},type=symbolslist,sort={d}}

\newglossaryentry{sym:eta}{name={\ensuremath{\eta}},description={Timelike coordinate used in the Rindler coordinate system},type=symbolslist,sort={eta}}

\newglossaryentry{sym:fmodes}{name={\ensuremath{f_l(x)}},description={Transformed mode functions with discrete index $l$ written in the coordinate system $x$},type=symbolslist,sort={fmodefunctions}}

\newglossaryentry{sym:fidelity}{name={\ensuremath{F}},description={Fidelity between two quantum states},type=symbolslist,sort={fidelity}}

\newglossaryentry{sym:fisher}{name={\ensuremath{\mathscr{F}(\theta)}},description={Fisher information of a state with respect to a parameter $\theta$},type=symbolslist,sort={fisher}}

\newglossaryentry{sym:information}{name={\ensuremath{i(x)}},description={The information contained in the outcome $x$ of a random variable $X$. Defined as $i(x) = \lb{p(x)}$, where $p(x)$ is the probability that the symbol $x$ occurs},type=symbolslist,sort={information}}

\newglossaryentry{sym:metric}{name={\ensuremath{g^{\mu \nu}}},description={The general spacetime metric. The metric signature convention used in this thesis is $(-,+,+,+)$. It is also used as the metric for the distances defined on quantum states and probability distributions},type=symbolslist,sort={g}}

\newglossaryentry{sym:Icoherentinfo}{name={\ensuremath{I\left(A\rangle R\right)_\rho}},description={Coherent information for Alice to Rob, subscript denotes that it depends on the state of the field},type=symbolslist,sort={Ia}}

\newglossaryentry{sym:Imutualinfo}{name={\ensuremath{I\left(A;R\right)_\sigma}},description={Holevo information between Alice and Rob, subscript denotes that it depends on the state of the field. This is the same notation and definition as mutual information, for a full discussion see \cref{subsec:inf:quantuminfomeasures}},type=symbolslist,sort={Ib}}

\newglossaryentry{sym:ketA0}{name={\ensuremath{\Ket[A]{\zero}}},description={Logical zero state of Alice's field as a ket},type=symbolslist,sort={keta0}}

\newglossaryentry{sym:ketA1}{name={\ensuremath{\Ket[A]{\one}}},description={Logical one state of Alice's field as a ket},type=symbolslist,sort={keta1}}

\newglossaryentry{sym:ketbraR}{name={\ensuremath{\Ket[R]{n}\!\Bra{m}}},description={Rob's field mode expressed in the Fock basis, where the density matrix element is not necessarily on the diagonal},type=symbolslist,sort={ketbraR}}

\newglossaryentry{sym:lambda}{name={\ensuremath{(\lambda_A)_k}},description={The $k$'th eigenvalue of the density matrix $\rho_A$},type=symbolslist,sort={lambda}}

\newglossaryentry{sym:LOCC}{name={\ensuremath{\mbox{LOCC}}},description={Local Operations and Classical Communication, a common restriction on the allowed set of operations one can do when performing quantum communication protocols. Any local operation is allowed, and any amount of classical communication, either one-way or two-way as specified in the protocol. These operations cannot affect the intrinsic type or amount of entanglement, which is why they are important for studying quantum communication},type=symbolslist,sort={locc}}

\newglossaryentry{sym:loweringoperator}{name={\ensuremath{\mathscr{L}_\rho(B)}},description={The Lowering superoperator as applied to the density matrix $B$ given the state or metric $\rho$. For a full discussion see \cref{subsec:meas:distancebetweendensityoperators}},type=symbolslist,sort={loweringoperator}}

\newglossaryentry{sym:phi}{name={\ensuremath{\hat{\phi}(x)}},description={Quantum field operator, expressed in coordinate system $x$},type=symbolslist,sort={phi}}

\newglossaryentry{sym:polylog}{name={\ensuremath{\Li_n(z)}},description={Polylog function, defined as $\Li_n(z)=\sum_{k=1}^\infty \frac{z^k}{k^n}$},type=symbolslist,sort={li}}

\newglossaryentry{sym:conditionalprobability}{name={\ensuremath{p(x|\theta)}},description={The conditional probability distribution of measurement results $x$ given the parameter $\theta$},type=symbolslist,sort={px}}

\newglossaryentry{sym:omega}{name={\ensuremath{\omega}},description={Positive parameter used in the definition of a Rindler mode. It can be interpreted as the Rindler frequency of the field mode in Rob's frame of reference},type=symbolslist,sort={omegaa}}

\newglossaryentry{sym:Omega}{name={\ensuremath{\Omega}},description={Used in $\ud \Omega^2$ as the line element of a 2-dimensional spherical shell},type=symbolslist,sort={omegab}}

\newglossaryentry{sym:ql}{name={\ensuremath{q_\text{L}}},description={Coefficient of the left Rindler wedge creation operator, when transforming from the Minkowski creation operator},type=symbolslist,sort={qla}}

\newglossaryentry{sym:qr}{name={\ensuremath{\qr}},description={Coefficient of the right Rindler wedge creation operator, when transforming from the Minkowski creation operator},type=symbolslist,sort={qlb}}

\newglossaryentry{sym:r}{name={\ensuremath{r}},description={Squeezing parameter used in the transformation between Alice's frame and Rob's accelerated frame},type=symbolslist,sort={r}}

\newglossaryentry{sym:rho}{name={\ensuremath{\rho}},description={Density matrix for a general field state representing quantum information},type=symbolslist,sort={rhoa}}

\newglossaryentry{sym:rhoAR}{name={\ensuremath{\rho_{AR}}},description={Combined density matrix of Alice and Rob},type=symbolslist,sort={rhob}}

\newglossaryentry{sym:rhoARbar}{name={\ensuremath{\rho_{A\overline{R}}}},description={Combined densiy matrix of Alice and AntiRob},type=symbolslist,sort={rhoba}}

\newglossaryentry{sym:rhoS}{name={\ensuremath{\rho^{(s)}}},description={Field state using the single rail encoding method},type=symbolslist,sort={rhoc}}

\newglossaryentry{sym:rhotheta}{name={\ensuremath{\rho^{(s)}_R(\theta)}},description={Parametrised field state on Rob's subsystem, encoded using the single rail, where the parameter is $\theta$},type=symbolslist,sort={rhotheta}}

\newglossaryentry{sym:rhoone}{name={\ensuremath{\rho(\one)}},description={Field state representing a logical one},type=symbolslist,sort={rhod}}

\newglossaryentry{sym:rhoexample}{name={\ensuremath{\rho_{R}^{(d)}(\zero)}},description={Example field state, represented by a density matrix in Rob's basis, where the encoding is dual rail and it represents a logical zero},type=symbolslist,sort={rhoe}}

\newglossaryentry{sym:rob}{name={\ensuremath{R}},description={The non-inertial observer Rob, either accelerating or hovering near a black hole},type=symbolslist,sort={rob}}

\newglossaryentry{sym:antirob}{name={\ensuremath{\overline{R}}},description={The non-inertial observer AntiRob, accelerating equal and opposite to Rob such that they are causally disconnected and exist in opposite regions of the Rindler space},type=symbolslist,sort={roba}}

\newglossaryentry{sym:Svonneumann}{name={\ensuremath{S\left(\rho\right)}},description={Von Neumann entropy of the state $\rho$},type=symbolslist,sort={sa}}

\newglossaryentry{sym:Scondentropy}{name={\ensuremath{S\left(A|R\right)_\rho}},description={Conditional entropy of Alice given Rob's state, subscript denotes that it depends on the state of the field},type=symbolslist,sort={sc}}

\newglossaryentry{sym:sigma}{name={\ensuremath{\sigma}},description={Density matrix for a general field state representing classical information. Extra information follows layout of \gls{sym:rho}},type=symbolslist,sort={sigma}}

\newglossaryentry{sym:SWM}{name={\ensuremath{\mbox{SWM}}},description={Single Wedge Mapping, a choice of $\qr=1$ in the transformation to Rindler modes so that the excitation Alice creates is mapped directly to the Rindler region I. For a full discussion, see \appref{ch:SMA}},type=symbolslist,sort={swm}}

\newglossaryentry{sym:theta}{name={\ensuremath{\theta}},description={A continuous parameter used in certain general states},type=symbolslist,sort={theta}}

\newglossaryentry{sym:uriomega}{name={\ensuremath{u_{k,\text{R}_\text{I}}^{(\omega)}(x)}},description={Mode function for the Rindler mode with single Rindler frequency $\omega$ in region I, written in terms of the $x$ coordinate system with wavenumber $k$},type=symbolslist,sort={uri}}

\newglossaryentry{sym:uur}{name={\ensuremath{u_{k,\text{U}_\text{L}}^{(\omega)}(x)}},description={Mode function for the Unruh mode in the left wedge which maps to a Rindler mode with single frequency $\omega$, written in terms of the $x$ coordinate system},type=symbolslist,sort={uur}}

\newglossaryentry{sym:v}{name={\ensuremath{v_{k,n}}},description={Coefficient of a transformed vacuum state, for the $n$th state in the Fock basis of the mode $k$},type=symbolslist,sort={v}}

\newglossaryentry{sym:xi}{name={\ensuremath{\xi}},description={Spacelike coordinate in the direction of acceleration used in the Rindler coordinate system},type=symbolslist,sort={xi}}

\glsaddall

\setstocksize{297mm}{210mm}
\settrimmedsize{\stockheight}{\stockwidth}{*}
\setbinding{20mm}
\settypeblocksize{*}{145mm}{1.35}
\setlrmargins{*}{*}{1.9}
\setulmargins{*}{*}{1.7}

\checkandfixthelayout[nearest]
\checkthelayout[nearest]
\fixthelayout
\baselineskip \topskip

\nouppercaseheads
\headstyles{dowding}

\setsecheadstyle{\large\bfseries\centering}
\setsubsecheadstyle{\bfseries\centering}

\makeevenhead{headings}{\textit{\thepage}}{}{\textit{\leftmark}}
\makeoddhead{headings}{\textit{\rightmark}}{}{\textit{\thepage}}

\makeoddfoot{plain}{}{\textit{\thepage}}{}
\makeevenfoot{plain}{}{\textit{\thepage}}{}

\ifdraftdoc
\makeoddfoot{headings}{}{}{\textit{Draft: \today}}
\makeevenfoot{headings}{\textit{Draft: \today}}{}{}
\fi

\usepackage[noabbrev]{cleveref}

\newcommand{\chapref}[1]{\cref{#1}}
\newcommand{\appref}[1]{\cref{#1}}
\newcommand{\secref}[1]{\cref{#1}}
\newcommand{\myeqref}[1]{\cref{#1}}
\newcommand{\figref}[1]{\cref{#1}}

\crefname{section}{Section}{Sections}
\crefname{chapter}{Chapter}{Chapters}
\crefname{appendix}{Appendix}{Appendices}
\crefname{align}{equation}{equations}
\Crefname{align}{Equation}{Equations}
\crefname{pluralequations}{equations}{equations}
\Crefname{pluralequations}{Equations}{Equations}
\crefname{figure}{figure}{figures}
\Crefname{figure}{Figure}{Figures}

\creflabelformat{pluralequations}{(#2#1#3)}

\newcommand{\degree}[1]{%
\gdef\stC{#1}}
\newcommand{\stC}{}
\newcommand{\degreehook}{%
\begin{flushright} \textsc{Thesis prepared for the degree of}\\ \Large \stC \end{flushright}}

\newcommand{\department}[1]{%
\gdef\stD{#1}}
\newcommand{\stD}{}
\newcommand{\departmenthook}{%
\begin{flushright} \large \stD \end{flushright}}

\renewcommand{\maketitlehookc}{\degreehook \vskip 2pc \departmenthook}

\newcommand{\university}[1]{%
\gdef\stA{#1}}
\newcommand{\stA}{}
\renewcommand{\maketitlehooka}{%
\begin{flushright} \Large\textsc \stA \end{flushright}}

\pretitle{\begin{flushright} \Huge \vskip 4pc}
\posttitle{\par\end{flushright}}
\preauthor{\begin{flushright} \huge \vskip 4pc}
\postauthor{\par  \vskip 4pc \end{flushright}}
\predate{\begin{flushright} \vskip 4pc \textsc}
\postdate{\par\end{flushright}}

\university{The University Of Sheffield}
\title{Relativistic Quantum Communication}
\degree{Doctor of Philosophy}
\department{Department of Physics and Astronomy}
\author{Dominic Jason Hosler}
\date{June 2013}

\includexmp{metadata}

\begin{document}

\frontmatter

\begin{titlingpage}
\maketitle
\end{titlingpage}

\chapter{Declaration of Authorship}
I, \textsc{Dominic Hosler}, declare that the work presented in this thesis, except where otherwise stated, is based on my own research and has not been submitted previously for a degree in this or any other university.
Parts of the work reported in this thesis have been published as follows:

\section{Publications}
\begin{enumerate}
 \item Dominic Hosler, Carsten van de Bruck and Pieter Kok. ``Information gap for classical and quantum communication in a Schwarzschild spacetime.'' In: \emph{Physical Review A} 85.4 (2012-04-24), page 042312.
 \item Eduardo Mart\'in-Mart\'inez, Dominic Hosler and Miguel Montero. ``Fundamental limitations to information transfer in accelerated frames.'' In: \emph{Physical Review A} 86.6 (2012-12-07), page 062307.
 \item Dominic Hosler and Pieter Kok. ``Parameter estimation over a relativistic quantum channel.'' (2013-06-13) arXiv: 1306.3144. Submitted to: \emph{Physical Review A}.
\end{enumerate}

\bigskip\noindent
Signed:
\rule{7cm}{1pt}

\bigskip\noindent
Dated:
\rule{7cm}{1pt}

\chapter{Abstract}
In this Ph.D. thesis, I investigate the communication abilities of non-inertial observers and the precision to which they can measure parametrized states.
I introduce relativistic quantum field theory with field quantisation, and the definition and transformations of mode functions in Minkowski, Schwarzschild and Rindler spaces.
I introduce information theory by discussing the nature of information, defining the entropic information measures, and highlighting the differences between classical and quantum information.
I review the field of relativistic quantum information.
We investigate the communication abilities of an inertial observer to a relativistic observer hovering above a Schwarzschild black hole, using the Rindler approximation.
We compare both classical communication and quantum entanglement generation of the state merging protocol, for both the single and dual rail encodings.
We find that while classical communication remains finite right up to the horizon, the quantum entanglement generation tends to zero.
We investigate the observers' abilities to precisely measure the parameter of a state that is communicated between Alice and Rob.
This parameter was encoded to either the amplitudes of a single excitation state or the phase of a NOON state.
With NOON states the dual rail encoding provided greater precision, which is different to the results for the other situations.
The precision was maximum for a particular number of excitations in the NOON state.
We calculated the bipartite communication for Alice-Rob and Alice-AntiRob beyond the single mode approximation.
Rob and AntiRob are causally disconnected counter-accelerating observers.
We found that Alice must choose in advance with whom, Rob or AntiRob she wants to create entanglement using a particular setup.
She could communicate classically to both.

\chapter{Copyright notice}
\begin{center}
\copyright Dominic Hosler (2013)
\end{center}

This work is licensed under the Creative Commons Attribution 3.0 Unported License. To view a copy of this license, visit http://creativecommons.org/licenses/by/3.0/ or send a letter to Creative Commons, 444 Castro Street, Suite 900, Mountain View, California, 94041, USA.

\begin{center}
 \includegraphics{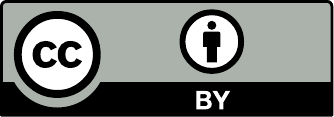}
\end{center}

The author, Dominic Hosler, reserves his moral right to be identified as the author of this work.
This is reflected in the attribution clause of the applied license.

\chapter{Acknowledgements}
Firstly I would like to thank my supervisor, Pieter Kok for all the support and direction he has given me throughout my studies.
I have really enjoyed working with someone who has such an enthusiasm for physics.
We have shared many interesting discussions.
It is safe to say I would not have achieved all that I have without him.

I have been fortunate enough during my studies towards my Ph.D. to work with some really talented physicists.
Carsten van de Bruck, with whom I collaborated when working on communication near a black hole.
During my trip to Madrid, I was hosted by Juan Le\'o{}n with whom I worked on interesting photon localization problems.
The same trip started a collaboration with Eduardo Mart\'i{}n-Mart\'i{}nez and Miguel Montero, I am grateful to be able to work with both of them.

My thanks also goes to my friends and co-workers in the office, who have all supported me.
These were, Carlos P\'e{}rez-Delgado, Marcin Zwierz, Samantha Walker, Mark Pearce, Sam Coveney, Michael Woodhouse and Matt Hewitt.
We have shared many in-depth discussions about related and non-related subjects.
They also provided plenty of welcome relief from study.

I would like to thank my family.
My wife Sascha for her continual support while I was studying, her understanding for the nature of my chosen career path and her encouragement when I hit stumbling blocks.
She is always there for me, to listen and help with anything she can.
My daughter Caoimhe for being a source of inspiration and motivation.

I would like to thank my uncle, Akin Kumoluyi for making the possibility of getting a Ph.D. seem more achievable.
I remember his Ph.D. graduation ceremony when I was about 7, it was then I decided that I too could get a Ph.D. if I chose.
I would like to extend gratitude to my grandparents, Mary and Charles Hosler for loaning me their spare room to use as an office whilst writing up.

Finally, I wish to express sincere gratitude to my parents Adam and Sandra Hosler, who dedicated a lot of their time to home educating me and making me believe I could achieve anything.
Without them I would not be where I am today.

For the publication ``Information gap for classical and quantum communication in a Schwarzschild spacetime'', Pieter Kok, Carsten van de Bruck and I would like to thank M. Wilde for valuable comments on the manuscript, and J. Dunningham and V. Palge for stimulating discussions.

For the publication ``Parameter estimation using NOON states over a relativistic quantum channel'', Pieter Kok, and I would like to thank Michael Skotiniotis for helpful discussions.

For the publication ``Fundamental limitations to information transfer in accelerated frames'', Eduardo Mart\'i{}n-Mart\'i{}nez, Miguel Montero and I would like to thank Pieter Kok for useful discussions and Mark Wilde for helpful suggestions.
During the research and preparation of this manuscript, I visited Madrid and I would like to thank, IFF, CSIC, Madrid for their hospitality.
I also acknowledge funding support for the trip from a Santander Research Mobility Award.

The stick people used in a number of my figures are inspired by characters from the web-comic xkcd\footnote{http://xkcd.com}.

\printglossary[type=symbolslist]

\cleardoublepage
\setcounter{tocdepth}{2}
\tableofcontents
\listoffigures

\mainmatter

\setcounter{secnumdepth}{2}

\chapter{Relativistic quantum field theory}
\label{ch:fields}

The universe we live in appears flat to the naked eye, with time disconnected as a separate parameter.
However, this is not fundamentally true.
The universe consists of a spacetime that bends and stretches according to the arrangement and motion of energy inside it.
General relativity is the study of the relationship between the energy and the spacetime \cite{Carroll2004}.
It is also the study of how this spacetime affects anything inside it.

All matter is made up of twelve fundamental particles and their corresponding antiparticles that interact using three fundamental forces on a background of curved spacetime.
Often it will be quoted that there are four forces, with gravity being the fourth but here we will treat gravity as purely a spacetime curvature.
These forces can be described by fields.
The source of the force creates a field that is felt by the test particle.
It does not take too much of a leap to describe the particles themselves by fields.
Every particle is an excitation of its field and every interaction is mediated by the exchange of (virtual) excitations of the force field.
This is what is described in great detail using the Standard Model of particle physics \cite{Mann2009a}.

Combining all of this results in quantum fields on a background of curved spacetime.
This is the field of relativistic quantum field theory, a discussion of which is given by Parker and Toms \cite{Parker2009}.
We introduce some key concepts in this chapter, which will be required in the rest of the thesis.
\secref{sec:fields:fieldtheoryintro} introduces field theory, the mode functions, inner product and creation and annihilation operators.
We introduce the coordinate systems and spacetimes required in \cref{sec:fields:rindler,sec:fields:schwarzschild}.
We describe the mode functions we will require in \cref{subsec:fields:rindlermodes,sec:fields:unruhmodes} and the transformation between them in \secref{sec:fields:transformation}.
Finally, we discuss some general transformations in \cref{sec:fields:bogoliubov}.
Throughout this thesis we will use natural units where $c=\hbar=G=1$.

\section{Quantum field theory}
\label{sec:fields:fieldtheoryintro}

A field is a mathematical object which associates a value with every point in spacetime.
Depending on the nature of the field this value could be a scalar, vector, tensor or operator.
These fields are affected by the curved nature of the spacetime that they exist in.
However, in this thesis we use the common assumption that the spacetime acts as a static background that is unaffected by the fields that exist inside it.
This is a very good approximation for the relative masses of the things we study here.

It is very difficult to do anything with completely arbitrary spacetimes.
Almost everything that we do must assume that some symmetries and structures exist in the spacetime.
For example we always require that the spacetime has a causal structure.
We only use linear fields as we require the principle of superposition for the quantization process.
In this thesis, we consider only massless scalar fields corresponding to bosons.
These fields are always solutions of a wave equation.

\subsection{Mode functions}
\label{subsec:fields:modefunctions}
We start with a field $\phi$ and its canonical momentum $\pi$.
These fields live in a spacetime with coordinates $x^\mu$.
This particular classical field is a solution to the massless Klein Gordon equation,
\begin{equation}
  \label{eq:fields:kleingordonminkowski}
  \nabla^\mu \nabla_\mu \phi(x^\mu) = 0.
\end{equation}
This is a linear wave equation so the superposition principle holds and the field can be written as a superposition of functions, $u_k(x^\mu)$,
\begin{equation}
  \label{eq:fields:phimodefunctions}
  \phi(x^\mu)=\sum_k a_k u_k(x^\mu) + a^*_k u^*_k(x^\mu),
\end{equation}
where the $u_k(x^\mu)$ are known as the mode functions and generally taken to be a simple family of solutions to the Klein Gordon equation.
The plane wave solutions are a good example of a set of mode functions,
\begin{equation}
  \label{eq:fields:planewaves}
  u_k(x^\mu)= A e^{i k_\mu x^\mu},
\end{equation}
where $A$ is some appropriate normalization factor.
In the field expansion of \myeqref{eq:fields:phimodefunctions}, there is a sum over $k$, but depending on the nature of the mode function set, the parameter may be continuous.
If the mode function parameter is continuous, we replace the sum over $k$ with an integral over the parameter.

\subsection{Field quantization}
\label{subsec:fields:quantization}
Canonical quantization of fields is where we replace the fields with field operators.
We require the following equal-time commutation relations between the field operator and its canonical momentum operator \cite{Kok2010},
\begin{subequations}
\begin{align}
  \label{eq:fields:fieldcommutation}
  \left[\hat{\phi}(\mathbf{x},t),\hat{\phi}(\mathbf{x}^{'},t)\right]& = 0 ,\\
  \left[\hat{\pi}(\mathbf{x},t),\hat{\pi}(\mathbf{x}^{'},t) \right]& = 0 ,\\
  \left[\hat{\phi}(\mathbf{x},t),\hat{\pi}(\mathbf{x}^{'},t) \right]& = i \delta(\mathbf{x}-\mathbf{x}^{'}).
\end{align}
\end{subequations}

In this procedure, we quantize the amplitudes of the mode functions, which gives us the following,
\begin{equation}
  \label{eq:fields:phimodefunctionsquantized}
  \hat{\phi}(x^\mu)=\sum_k \hat{a}_k u_k(x^\mu) + \hat{a}^\dagger_k u^*_k(x^\mu).
\end{equation}
The operators $\hat{a}^\dagger$ and $\hat{a}$ are known as the creation and annihilation operators respectively and will be explained in \secref{subsec:fields:creationoperators}.

\subsection{Inner product}
\label{subsec:fields:innerproduct}
We can refine our definition of mode functions once we have defined an inner product.
The inner product is dependent on the spacetime, so we define it as, \cite{Parker2009},
\begin{equation}
  \label{eq:fields:innerproductgeneral}
  \left(\phi,\psi \right)\equiv i \int_\sigma \! \ud \sigma\, \sqrt{-g} n^\nu \phi^* \overleftrightarrow{\delta_\nu} \psi,
\end{equation}
where $g$ is the determinant of the spacetime metric $g^{\mu\nu}$, $\overleftrightarrow{\delta_\nu}$ is the antisymmetric derivative such that $a\overleftrightarrow{\delta_\nu}b=a \delta_\nu b - (\delta_\nu a) b$, and $\sigma$ is a spacelike hypersurface with $n^\nu$ its future directed normal unit vector.
It is trivial to see that this reduces to the standard inner product in Minkowski space, as it also does in the metrics we use in this thesis.
This is given here,
\begin{equation}
  \label{eq:fields:innerproduct}
  \left(\phi,\psi \right)\equiv i \int \! \ud \mathbf{x} \, \phi^* \overleftrightarrow{\partial_t} \psi,
\end{equation}
with $\overleftrightarrow{\partial_t}$ the antisymmetric time derivative and the integral performed over the spatial variables.

\subsection{Bases and orthonormality of mode functions}
\label{subsec:fields:bases}
With a definition of an inner product, we can add structure to the mode functions.
A set of mode functions can form a basis if they span the entire space.
If they do, the set is called complete.
This means if any possible state can be written as a superposition of mode functions from the set, that set is a complete basis.
We will restrict ourselves here to only looking at complete orthonormal sets of mode functions.
First, we need to explain orthonormality and mathematically define both the completeness and orthonormality conditions.

Orthonormality is a combination of orthogonality and normalization.
Orthogonality is the condition that the inner product between any two different mode functions is zero.
Normality is where every mode function has unit size, they do not change the size of other states when taking inner products with them.
Mathematically orthonormality is achieved when the following condition is satisfied,
\begin{equation}
  \label{eq:fields:orthonormalitycondition}
  \left(u_j(x^\mu),u_k(x^\mu) \right) = \delta_{jk}.
\end{equation}
The mathematical condition for achieving completeness is,
\begin{equation}
  \label{eq:fields:completenesscondition}
  \sum_k u_k(x^\mu) u^*_k(x^{\prime\mu})=\delta(x^\mu-x^{\prime\mu}),
\end{equation}
where we have used the Kronecker delta $\delta_{jk}$ and the Dirac delta function $\delta(x^\mu-x^{\prime\mu})$ in both these conditions.

One can construct a basis for a field, which could be any set of functions obeying the conditions above.
In our case here the basis is the set of mode functions described in \myeqref{eq:fields:planewaves}.
These mode functions satisfy the completeness and orthogonality conditions, but we must modify the normalization to use periodic boundary conditions.

It is also possible and useful here, to construct a basis for the quantum states of the field.
Useful bases for quantum states are often eigenvectors of observable operators.
This means the basis automatically satisfies the required conditions for a basis and allows easier calculation of the expectation value of that operator.

In our case we are interested in the quantum state of one particular mode of the field.
We choose to work in the Fock basis, which is defined as the set of eigenvectors of the number operator.
This is a particularly useful basis to choose because we are dealing with noise that creates excitations in the mode we are studying.
We could take the tensor product of this Fock basis with the other Fock bases (one for each field mode) to construct a basis for the entire field.
However, the full field's basis construction is beyond the scope of this thesis.

\subsection{Creation and annihilation operators}
\label{subsec:fields:creationoperators}
Using the definition of the field from \myeqref{eq:fields:phimodefunctionsquantized} and the commutation relations in \myeqref{eq:fields:fieldcommutation} it can easily be shown that the commutation relations of the creation and annihilation operators are,
\begin{subequations}
  \label{eq:fields:creationcommutation}
\begin{align}
  \left[\hat{a}_k,\hat{a}_{k^\prime}\right]& = 0 ,\\
  \left[\hat{a}^\dagger_k,\hat{a}^\dagger_{k^\prime}\right]& = 0 ,\\
  \left[\hat{a}_k,\hat{a}^\dagger_{k^\prime} \right]& = \delta_{kk^\prime}.
\end{align}
\end{subequations}

The creation and annihilation operators can be found by taking inner products of the field with the mode functions,
\begin{align}
  \label{eq:fields:creationfrominnerproduct}
  \hat{a}_k = \left(u_k(x^\mu),\hat{\phi}(x^\mu)\right), \quad
  \hat{a}^\dagger_k = \left(u^*_k(x^\mu),\hat{\phi}(x^\mu)\right).
\end{align}

\subsection{The vacuum and particles}
\label{subsec:fields:vacuumandparticles}
We can think of all particles as excitations of a field.
These particles can be excitations that are spatially and temporally localized, with possibly some motion and interactions with other fields.
Or these particles can be mathematical idealisations where we define them as living entirely in one mode.
We can define the vacuum as being the absence of any particles in all modes.
The vacuum state $\ket{\varnothing}$ is mathematically defined by the equation,
\begin{equation}
  \label{eq:fields:vacuumdefinition}
  \hat{a}_k \ket{\varnothing} = 0 \,\,\,\forall k.
\end{equation}
We can then mathematically define particles as single excitations.
This is written as creation operators acting on the vacuum,
\begin{subequations}
\begin{align}
  \label{eq:fields:particledefinition}
  &\ket[k]{1} = \hat{a}^\dagger_k \ket{\varnothing} \\
  \label{eq:fields:particleabitraryform}
  &\ket{1} = \sum_k c_k \hat{a}^\dagger_k \ket{\varnothing}
\end{align}
\end{subequations}
where \myeqref{eq:fields:particleabitraryform} describes a single particle with an arbitrary wavefunction, with the condition $\sum_k \modsq{c_k} = 1$.
This arbitrary wavefunction particle exists in just the same way as any other quantum superposition.
There may exist a particular basis where $c_k = \delta_{jk}$ for some particular $j$, i.e., the particle exists entirely in a single mode.
This would then be written with an equation similar to \myeqref{eq:fields:particledefinition}.

In quantum physics we are generally only concerned with observables.
The same is true for quantum field theory.
We can construct informative observables from the creation and annihilation operators.
The number operator is given by,
\begin{equation}
  \label{eq:fields:numberoperator}
  \hat{N}_k = \hat{a}^\dagger_k \hat{a}_k.
\end{equation}
It measures the number of excitations in field mode $u_k$.
As mentioned previously we can use its eigenvectors as a basis for that field mode.
We also find that its expectation value and variance are useful quantities.

\section{Constantly accelerated Rindler observer}
\label{sec:fields:rindler}
One of the simplest examples of field theory in a general relativistic setting is where we quantise the field from the point of view of a constantly accelerated observer in flat space.
These observers are known as Rindler observers.
They measure their own proper acceleration as constant and travel in flat Minkowski space with a hyperbolic spacetime path.
We first define their coordinate system, then treat the quantization of a field from their perspective.
We finally compare the different views of the same physical field.

\begin{figure}[htpb]
  \begin{center}
      \includegraphics[width=272pt]{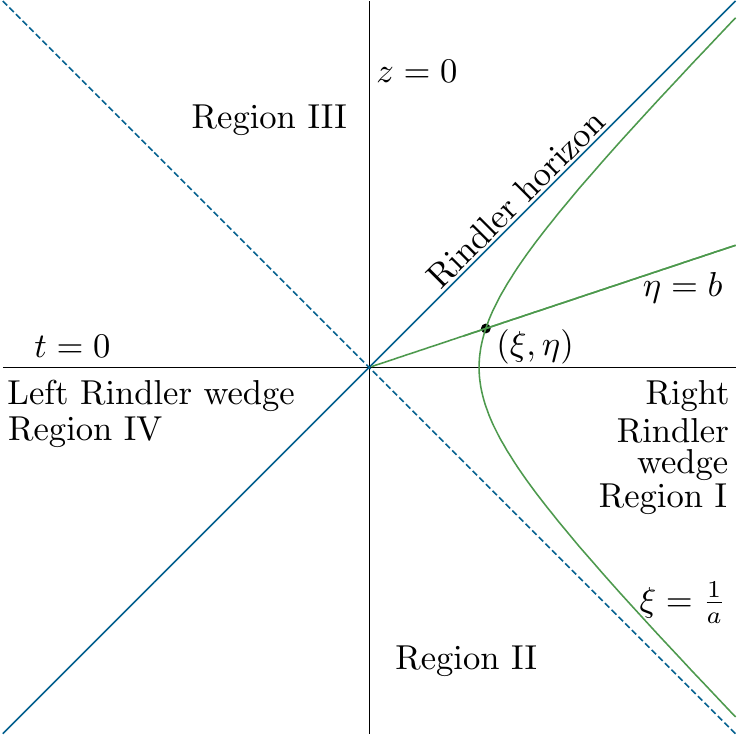}
  \caption[Rindler coordinates]{This represents the Rindler coordinates of an event on a background of Minkowski coordinates. The Rindler horizon is included, which is an event horizon for observers travelling on paths of constant $\xi$ in the right Rindler wedge. These observers will measure their proper acceleration as $a$. The four regions of Rindler spacetime are labelled, which correspond to the four Rindler coordinate systems required. In this thesis we only are concerned with region I and region IV.}  \label{fig:fields:rindlercoords}
  \end{center}
\end{figure}

\subsection{Rindler coordinates}
\label{subsec:fields:rindlercoords}
Everything is happening inside Minkowski spacetime so we define the Rindler coordinates using their transformation from Minkowski coordinates \cite{Rindler1966}.
We take $z$ as the direction in which the observer is accelerating.
The transformation only affects the $z$ and $t$ coordinates,
\begin{subequations}
\begin{align}
  \label{eq:fields:rindlercoords}
  \eta=& \frac1a \artanh\left(\frac{t}{z} \right),\\
  \xi =& \sqrt{z^2-t^2}.
\end{align}
\end{subequations}
The new time coordinate is $\eta$, and $\xi$ is the new space coordinate in the $z$ direction.
The coordinates $x$ and $y$ remain unchanged as there is no acceleration happening in their plane.

Rindler coordinates can be represented graphically on a background of Minkowski coordinates by the graph in \figref{fig:fields:rindlercoords}.
An observer travelling along a path of constant $\xi$ will measure a proper acceleration $a$.
The line $t=z$ is a future Rindler horizon for any observer following a path of constant $\xi$ in the right Rindler wedge.
We can also see that the coordinates are only valid in the right Rindler wedge.
There is an equivalent coordinate system construction for the left Rindler wedge, and the future and past regions of the spacetime.
With these four coordinate sets we can fully map the whole Minkowski spacetime.

An observer Rob, travelling along the path of constant $\xi_0=a^{-1}$ will see a metric given by the line element,
\begin{equation}
  \label{eq:fields:rindlermetric}
  \ud s^2 = - \frac{\xi^2}{a^2} \ud \eta^2 + \ud \xi^2 + \ud x^2 + \ud y^2.
\end{equation}
Rob can only observe the field in his own Rindler wedge, region I.
To fully specify a field transformation from Minkowski space we must introduce another equivalent observer in the other Rindler wedge, region IV.
This observer we call AntiRob, with a trajectory mirroring Rob's across the $z=0$ axis.

\subsection{Rindler mode functions}
\label{subsec:fields:rindlermodes}
If we write the massless Klein-Gordon equation in the Rindler observer's coordinates we can quantize from the perspective of the accelerated observer.
The first stage is again to find a set of simple single frequency solutions to the wave equation.
Just as we constructed the coordinates by using four separate regions with different coordinate sets, we solve the massless Klein-Gordon equation in each of these related coordinate systems.
This ensures that the Rindler modes can represent the whole field in any part of the Minkowski spacetime.
Each region of spacetime contains its own complete set of Rindler modes.
There will be one mode for a particular frequency in both region I and region IV.
The form of these modes is given in full in \cref{ch:rindler-modes}.

We can now expand the field in terms of this new basis, just like we did for flat space only we now have four coefficients, two for each region \cite{Takagi1986},
\begin{equation}
  \label{eq:fields:fieldinrindlerbasis}
  \phi=\int_0^\infty \ud \omega \int \ud^2 k \left( a_{k,\text{R}_\text{I}}^{(\omega)} u_{k,\text{R}_\text{I}}^{(\omega)} + a_{k,\text{R}_\text{I}}^{(\omega)*} u_{k,\text{R}_\text{I}}^{(\omega)*} + a_{k,\text{R}_\text{IV}}^{(\omega)} u_{k,\text{R}_\text{IV}}^{(\omega)} + a_{k,\text{R}_\text{IV}}^{(\omega)*} u_{k,\text{R}_\text{IV}}^{(\omega)*}\right).
\end{equation}

We use the same canonical quantization procedure as we did for flat space.
We write the field as an operator, and make the coefficients of each mode function operators too.
It is easy to confirm that these operators are also creation and annihilation operators satisfying the conditions in \myeqref{eq:fields:creationcommutation}.

\section{Schwarzschild metric}
\label{sec:fields:schwarzschild}
We introduce here the spherically symmetric solution to Einstein's equations resulting from a point mass at the origin.
Named the Schwarzschild metric, it represents any spherically symmetric mass including black holes \cite{Schwarzschild1916}.
If the spherically symmetric object extends beyond the event horizon of the metric, then the metric is only valid for the spacetime outside of the object.
In spherical polar coordinates the metric is given by its line element,
\begin{equation}
 \label{eq:fields:schwarzschildmetric}
 \ud s^2 = - \left(1-\frac{r_s}{r} \right) \ud t^2 + \frac{1}{1-\frac{r_s}{r}}\ud r^2 + \ud \Omega^2,
\end{equation}
where $\ud \Omega$ is the line element on a 2-dimensional spherical shell and $r_s=2M$ is the Schwarzschild radius, only dependent on the mass of the object.

\subsection{Approximation using Rindler metric}
\label{subsec:fields:rindlerapprox}
It is easy to show that the Schwarzschild metric of \cref{eq:fields:schwarzschildmetric} involves a coordinate singularity at $r=r_s$, which is also called the event horizon of the black hole.
However, in this thesis we are only interested in effects outside the horizon.
It is very difficult to study field theory in the Schwarzschild metric directly.
For a sufficiently large black hole with an observer sufficiently close to the horizon, we can approximate the observer and his local spacetime with a Rindler observer accelerating in flat space.
This Rindler observer has a proper acceleration equal to the proper acceleration of the hovering observer.

For this derivation we restrict ourselves to the radial and temporal coordinates only because the other two are trivial.
First, we need to write the Schwarzschild metric in terms of the proper time $\tau$ of the observer hovering at $r_0$.
This transforms the metric into,
\begin{equation}
  \label{eq:fields:schwarzschildmetricpropertime}
  \ud s^2 = - \frac{\left(1-\frac{r_s}{r} \right)}{\left(1-\frac{r_s}{r_0} \right)} \ud \tau^2 + \frac{1}{1-\frac{r_s}{r}}\ud r^2.
\end{equation}
Then we introduce a replacement coordinate $z_*$ to use in the radial direction,
\begin{equation}
  \label{eq:fields:zstardefinition}
  r - r_s = \frac{z_*^2}{4 r_s}.
\end{equation}
Using both of these replacements we can write the Schwarzschild metric again,
\begin{equation}
  \label{eq:fields:schwarzschildmetricintauz}
  \ud s^2 = - \frac{z_*^2}{(4 r_s^2 + z_*^2)\left(1-\frac{r_s}{r_0} \right)} \ud \tau^2 + \left(1 + \frac{z_*^2}{4r_s^2} \right)\ud r^2.
\end{equation}
The approximation comes about by expanding this metric to the first non trivial order in $z_*$ around $z_*=0$ (the event horizon),
\begin{equation}
  \label{eq:fields:schwarzschildmetricapprox}
  \ud s^2 = - \frac{z_*^2}{4 r_s^2 \left(1-\frac{r_s}{r_0}\right)} \ud \tau^2 + \ud z_*^2.
\end{equation}
We can now see by comparing with \myeqref{eq:fields:rindlermetric}, that this is a Rindler metric with acceleration $a=2 r_s \sqrt{1-\frac{r_s}{r_0}}$.
This is the proper acceleration of an observer hovering at $r=r_0$, to second order approximation.

We conclude by stating that the Schwarzschild metric near to the event horizon can be approximated by a Rindler metric.
We must match the proper accelerations of the observers hovering in the Schwarzschild metric and accelerating in the Rindler metric.
This approximation is only valid for $\left( \frac{z_{*0}}{2 r_s}\right) \ll 1$.
The region of validity corresponds to observers hovering at a radius $r_0 - r_S \ll 2 r_S$ in standard Schwarzschild coordinates.
The region of Schwarzschild spacetime inside the black hole horizon is not covered by this approximation, neither are observers hovering further away from the horizon.
We do not study, in this thesis, communication involving observers that are inside the black hole horizon.

This approximation is very useful for studying informational properties of fields because all of the interesting effects happen in this region.
It has been shown previously that the entanglement effects are localized near the horizon \cite{Martin-Martinez2010a}.
We show in \cref{ch:blackholes,ch:measure} that the effects introduced by the curvature and the horizon are also localized near the horizon.
As such, we find it very convenient to use this approximation because it is valid in exactly the region of interest.

\section{Unruh modes}
\label{sec:fields:unruhmodes}
The Unruh modes are constructed with the requirements that they are superpositions of purely positive frequency modes in flat spacetime.
This means that they share a vacuum with the Minkowski modes.
It is also required that they map $1:1$ with Rindler modes.
We must therefore separate the Unruh modes into equivalent regions to the Rindler regions I and IV.
We refer to these as the right (R) Unruh modes and the left (L) Unruh modes.
Each single frequency Rindler mode has a mode in region I and region IV, the equivalent Unruh mode has one in region R and one in region L.
The Rindler modes are constructed using the acceleration of a given observer.
The same applies to the Rindler modes when they are constructed as an approximation for an observer hovering near a Schwarzschild horizon.
It is easy to see why this is the case.
Each set of Rindler modes defines a particular vacuum, different observers with different accelerations do not share a vacuum and hence cannot share a set of mode functions.

The fact that the Rindler modes, and therefore the Unruh modes as well, can only be defined in relation to a particular observer with a particular acceleration does present some problems.
First, to be able to use these modes, we must assume that Rob either knows or can accurately measure his acceleration.
We also require Rob and Alice to share this knowledge before any experiment or communication starts, so Alice knows what mode function to use when creating the states.

Due to the delocalized nature of Unruh modes they are arguably not completely measurable, and in the best case scenario they can be determined only approximately by means of localized measurements.
Also, the Unruh modes, as seen from any inertial observer, behave in a highly oscillatory way near the acceleration horizon.
This makes them bad candidates for physically feasible states.
Finally, we are choosing a different Unruh mode for each acceleration, so as to keep the Unruh to Rindler change of basis always simple.

We use Unruh modes as the states built from them allow direct Bogoliubov transformations to Rindler modes.
This maximises the information transfer of the channels we study providing a limit to the entropic functions, and makes the diagonalisation of the density matrices possible in \chapref{ch:measure}.
Also, these modes in this setting have been used widely throughout the literature in recent years, so this allows us to compare and discuss with regard to previous works \cite{Bruschi2010}.

\section{Transformation between Unruh and Rindler modes}
\label{sec:fields:transformation}
Alice can directly create excitations in Unruh modes because these share a vacuum with the Minkowski modes.
Therefore, we do not look at any transformation between the Minkowski and Unruh modes.
The transformation where all the interesting effects happen is between the Unruh modes and the Rindler modes.
This transformation is given by a squeezing operator between two regions of Rindler space \cite{Unruh1976}.

To perform the transformation, in the Single Wedge Mapping, Alice creates an excitation in her Unruh R modes which map to region I.
Then we perform squeezing between the modes in region I and the equivalent modes in region IV.
This calculation of two-mode squeezing follows the notation of Kok and Lovett \cite{Kok2010}, original derivation is found in \cite{Truax1985}.
The squeezing parameter is $r$ and is given by Rob's acceleration and the Rindler frequency of the modes being used,
\begin{equation}
  \label{eq:fields:squeezingparameter}
  \tanh r = \exp \left(\frac{- \omega \pi}{a} \right).
\end{equation}
The two mode squeezing transformation is given by,
\begin{equation}
 \label{eq:fields:twomodesqueezing}
\widehat{U} = \exp\left[i r \hat{a}_I \hat{a}_{IV} - i r \hat{a}_I^\dagger \hat{a}_{IV}^\dagger \right].
\end{equation}

The transformation mixes creation and annihilation operators, and so does not preserve photon number.
It is however, a unitary operator so will fully preserve information, and keep pure states pure.
No information is lost in this part of the process.

We use a special case of the Baker Campbell Hausdorff formula, which allows us to separate the operators into different exponents,
\begin{equation}
  \label{eq:fields:BCHrelation}
 e^{\kappa K_+ - \kappa^* K_-}=e^{\tau K_+}e^{-2\nu K_0}e^{-\tau^* K_-}.
\end{equation}
This special case is only valid when the $K$ operators have the properties,
\begin{subequations}
\begin{align}
  \label{eq:fields:BCHKpKmrelation}
  K_+^\dagger &= K_-, \\
  K_0^\dagger &= K_0,
\end{align}
\end{subequations}
and their commutation relations are of the forms,
\begin{subequations}
\begin{align}
  \label{eq:fields:BCHcommutationofK}
  \left[K_-,K_+\right]&=2 K_0, \\
  \left[K_0,K_\pm \right] &=\pm K_\pm.
\end{align}
\end{subequations}

We apply this operator ordering to the two mode squeezing operator to simplify it into the following set of three unitary operators,
\begin{subequations}
  \label[pluralequations]{eq:fields:orderedtransformation}
\begin{align}
\widehat{U} = & \widehat{U}_1 \widehat{U}_2 \widehat{U}_3, \\
\widehat{U}_1 = &\exp \left[ i \frac{r}{\left| r \right|} \tanh(r) \hat{a}_I^\dagger \hat{a}_{IV}^\dagger \right], \\
\widehat{U}_2 = &\exp \left[ - \ln\left(\cosh (r)\right) \left( \hat{a}_I^\dagger \hat{a}_I + \hat{a}_{IV}^\dagger \hat{a}_{IV} +1 \right)\right], \\
\widehat{U}_3 = &\exp \left[ i \frac{r}{\left| r \right|} \tanh(r) \hat{a}_I \hat{a}_{IV} \right].
\end{align}
\end{subequations}
This vastly simplifies the transformation as in this thesis we will either act on a vacuum in at least one mode, either in Region I or Region IV.
For communication protocols, we always assume that Alice is able to prepare a vacuum everywhere in advance of the protocol.
$\widehat{U}_3$ always acts as the identity on these states because it is only the exponential of annihilation operators acting on a vacuum.
$\widehat{U}_2$ includes just number operators which turn into a simple factor when they only act on states with a fixed number of particles.
This factor can be evaluated for these simple states as,
\begin{subequations}
\begin{align}
  \label{eq:fields:coshfactorvacuum}
\exp \left[ - \ln \left( \cosh(r)\right) \right] \Ket{\varnothing_I,\varnothing_{IV}} &= \frac{1}{\cosh(r)} \Ket{\varnothing_I,\varnothing_{IV}},\\
  \label{eq:fields:coshfactorparticle}
\exp \left[ - \ln \left( \cosh(r)\right) (1+1)\right] \Ket{1_I,\varnothing_{IV}} &= \frac{1}{\cosh^2(r)} \Ket{1_I,\varnothing_{IV}}.
\end{align}
\end{subequations}

We can expand $\widehat{U}_1$ as a series because all the operators in the exponential commute.
This then combines with the factor from the second operator, and provides us with a state described in the Fock bases of both mode I and mode IV,
\begin{align}
  \label{eq:fields:transformvacuum}
\widehat{U} \vac &= \frac{1}{\cosh(r)} \sum_{n=0}^\infty \frac{(-i r)^n}{\left| r \right|^n} \tanh^n |r| \ket{n_I,n_{IV}} ,\\
  \label{eq:fields:transformparticle}
\widehat{U} \hat{a}^\dagger_I \vac &= \frac{1}{\cosh^2(r)} \sum_{n=0}^\infty \frac{(-i r)^n}{\left| r \right|^n} \sqrt{n+1}  \tanh^n |r|\ket{(n+1)_I,n_{IV}} .
\end{align}
The mixing is between two of the Rindler modes, one in spacetime region I (Rob) and the other in region IV (AntiRob).
Rob is causally disconnected from region IV and has no access to information there so we must trace out all region IV modes.
This results in an effective non-unitary transformation and local information loss.
When dealing with density matrices after the partial trace, we can also ignore the phase factor because it is always multiplied by its complex conjugate.

Mathematically we can treat the state as a sum of coefficients multiplied by matrix elements.
The transformation then acts independently on each element.
After tracing out the inaccessible modes, we are left with an entangled, partially mixed state of Alice and Rob's fields.
There are only four types of element, so we present the full transformations here, written in the Fock basis,
\begin{subequations}
\label[pluralequations]{eq:fields:transformmatrixelements}
\begin{align}
 \label{eq:fields:transformelement00}
 \ket{0}\bra{0} \to& \sum_{n=0}^\infty \frac{\tanh^{2n}\left|r\right|}{\cosh^2\left|r\right|}\ket{n}\bra{n},\\
 \label{eq:fields:transformelement01}
 \ket{0}\bra{1} \to& \sum_{n=0}^\infty \frac{\tanh^{2n}\left|r\right|}{\cosh^3\left|r\right|}\sqrt{n+1}\ket{n}\bra{n+1},\\
 \label{eq:fields:transformelement10}
 \ket{1}\bra{0} \to& \sum_{n=0}^\infty \frac{\tanh^{2n}\left|r\right|}{\cosh^3\left|r\right|}\sqrt{n+1}\ket{n+1}\bra{n},\\
 \label{eq:fields:transformelement11}
 \ket{1}\bra{1} \to& \sum_{n=0}^\infty \frac{\tanh^{2n}\left|r\right|}{\cosh^4\left|r\right|}\left(n+1\right)\ket{n+1}\bra{n+1}.
\end{align}
\end{subequations}
These are for one mode, the single mode system requires all of these transformations.
In the dual rail system, there are two modes, each mode has its own set of transformations.

Applying the transformation operators in \cref{eq:fields:orderedtransformation} to the annihilation operator we find that the annihilation operator becomes,
\begin{equation}
  \label{eq:fields:transformannihilation}
  \hat{a}_\text{U} = \cosh r \, \hat{a}_{\text{R}_\text{I}} - \sinh r \, \hat{a}^\dagger_{\text{R}_\text{IV}}
\end{equation}
We have made an implicit assumption that Alice wants to map her Unruh mode to Rob's region I directly, setting $\hat{a}_\text{U}=\hat{a}_{\text{U}_\text{R}}$.
We refer to this choice as the Single Wedge Mapping and it is not the most general way of transforming the modes.

Alice spans the entire spacetime so she is able to choose to construct Unruh modes that map to both regions I and IV.
Alice has the freedom to construct an Unruh mode that map to any combination of the two Rindler wedges.
This has been known in the literature as `the Single Mode Approximation'.
When it has not been used, it has been referred to as going `beyond the Single Mode Approximation' \cite{Bruschi2010}.
We are always talking about single frequency Rindler modes, so this nomenclature is misleading.
I propose to rename this choice as the Single Wedge Mapping to describe the way Alice creates excitations that map only to Rindler region I.
This is discussed in detail in \cref{ch:SMA}.

To go beyond the Single Wedge Mapping, Alice creates excitations that are composed of superpositions of excitations from each wedge.
This is represented by her general Unruh mode annihilation operator being formed from the annihilation operators of each region,
\begin{equation}
  \label{eq:fields:unruh-general-annihilation}
  \hat{a}_{\text{U}}=q_\text{L}\hat{a}_{\text{U}_\text{L}}+\qr \hat{a}_{\text{U}_\text{R}},
\end{equation}
where the parameters $q_\text{L}$ and $\qr$ are complex numbers such that $\modsq{q_\text{L}}+\modsq{\qr}=1$.
The Unruh left and right annihilation operators each map directly to the respective Rindler annihilation operators for regions IV and I respectively.
Similar to \cref{eq:fields:transformannihilation}, we now define the annihilation operators explicitly,
\begin{subequations}
  \label[pluralequations]{eq:fields:UrandUl}
\begin{align}
  \hat{a}_{\text{U}_\text{R}}&=\cosh r\, \hat{a}_{\text{R}_\text{I}} - \sinh r\, \hat{a}^\dagger_{\text{R}_\text{II}},\\
  \hat{a}_{\text{U}_\text{L}}&=\cosh r\, \hat{a}_{\text{R}_\text{II}} - \sinh r\, \hat{a}^\dagger_{\text{R}_\text{I}},
\end{align}
\end{subequations}
This is discussed in detail in \cref{ch:SMA}.

\section{General Bogoliubov transformations}
\label{sec:fields:bogoliubov}
We study the general Bogoliubov transformations of a field $\hat{\phi}(x)$.
We start with an arbitrary set of complete orthonormal mode functions $u_k(x)$ to describe the field.
The field is then transformed to a different arbitrary mode function and we study how the annihilation operator transforms.
Finally we define the vacuum for the transformed observer, and see what it looks like using the original mode functions.
We find that an arbitrary Bogoliubov transformation can only ever create an even number of field excitations.
We let the original states be labelled $A$ and the transformed states be labelled $B$ with annihilation operators $\hat{a}_k$ and $\hat{b}_l$ respectively.

\subsection{Bogoliubov Transformation of Annihilation Operator}
\label{subsec:fields:bogoannihilationtransform}
We look at the general Bogoliubov transformation of mode functions or coordinates.
The transformed mode function $f_l(x)$ can be written in terms of the original mode functions as,
\begin{equation}
\label{eq:generaltransform}
f_l(x) =  \sum_k \left[ \alpha_{lk} u_k(x) + \beta_{lk} u^*_k(x)\right],
\end{equation}
where $u_k(x)$ can be any complete orthonormal set of mode functions.

To extract the annihilation operator $\hat{b}_l$ for the mode function $f_l(x)$, we use the inner product,
\begin{equation}
\label{eq:bogoinnerprod}
\hat{b}_l = \left(f_l(x),\hat{\phi}(x)\right).
\end{equation}
If $f_l(x)$ and $\hat{\phi}(x)$ are both expanded in terms of the original mode functions, we can use the orthonormality to get this general representation of the annihilation operator,
\begin{equation}
\label{eq:generalbfroma}
\hat{b}_l = \sum_k \left[ \alpha_{lk}^* \hat{a}_k + \beta_{lk}^* \hat{a}_k^\dagger\right].
\end{equation}
This is the most general transformation of the annihilation operator, given the transformation coefficients $\alpha_{lk}$ and $\beta_{lk}$.

\subsection{Bogoliubov Transformation of Vacuum State}
\label{subsec:fields:bogovacuumtransform}
To find out how the vacuum state transforms, we need to look at the definition of the vacuum state, as defined in the frame with mode functions $f_l(x)$.
\begin{equation}
\label{eq:vacuumdefinition}
\hat{b}_l\ket[B]{0} = 0.
\end{equation}

This vacuum state can be written as an arbitrary state in the original basis of mode functions,
\begin{equation}
\label{eq:vacuuminbstateina}
\ket[B]{0} = \ket[A]{\psi}=\sum_{jn} v_{j,n}\ket[A]{n_{\otimes j}},
\end{equation}
where $v_{jn}$ are the coefficients of the vacuum state.

If we substitute in for both the annihilation operator and the general state, we can see what form the state takes,
\begin{equation}
\label{eq:definitionofAstate}
\sum_k \left[ \alpha_{lk}^* \hat{a}_k + \beta_{lk}^* \hat{a}_k^\dagger\right] \sum_{jn} v_{j,n}\ket[A]{n_j}=0.
\end{equation}

The creation and annihilation only act on their own mode, so delta functions $\delta_{jk}$ will remove one of the sums.
If we substitute this description in, and allow the creation and annihilation operators to act on the Fock basis states, we can re-label the sum to $m$ and collect the Fock basis terms,
\begin{equation}
\label{eq:Astateequation}
\sum_{km}\left[\alpha_{lk}^* v_{k,m+1}\sqrt{m+1} + \beta_{lk}^* v_{k,m-1}\sqrt{m}\, \right]\ket[A]{m_k}=0.
\end{equation}
For this to be equal to $0$ each coefficient must be equal to zero because the states $\ket[A]{m_k}$ are independent.
We first take the special case of $m=0$
\begin{equation}
\label{eq:n0partequal0}
\alpha_{lk}^* v_{k,1}\equiv0, \forall k,
\end{equation}
which is an identity because it must be true for all $k$.
We require $\alpha_{lk}^*\ne 0$ in general, so we conclude that,
\begin{equation}
\label{eq:v1is0}
v_{k,1}\equiv0, \forall k.
\end{equation}

Taking the rest of the terms in general, and relabelling the index again to $p$, leads to the recursion relation,
\begin{equation}
\label{eq:vrecursion}
v_{k,p+2}\equiv -\frac{\beta_{lk}^*}{\alpha_{lk}^*}\frac{\sqrt{p+1}}{\sqrt{p+2}}v_{k,p}.
\end{equation}
The initial condition $v_{k,0}$ is calculated from normalization considerations.
Combining with \cref{eq:v1is0}, the other required initial condition, we find
\begin{align}
\label{eq:vzerooddp}
  v_{k,p}\equiv 0, \,\, \forall \,\mbox{odd}\, p.
\end{align}
A vacuum state in one frame of reference can only ever be represented in another frame of reference by a superposition of even numbered states.

As can be seen from the factor in this recurrence relation, if $\beta_{lk}^*\equiv0$ there are no higher number states present, and the transformed state is the vacuum state in the other frame.
This point is worth emphasising, if $\beta_{lk}^*\equiv0$ the two sets of mode functions share a vacuum.
The vacuum state is therefore constant when there is no mixing between creation and annihilation operators.

\chapter{Information theory}
\label{ch:information}
Information theory is the study of the use, manipulation and communication of information.
Information can be thought of as the amount there is to know about something.
For example, the information contained in a system is the amount one would expect to learn when measuring the system.
We quantify information in the units of bits, how many bits it takes to specify the information.
The more bits available, the more information we have, the larger the set of possibilities we can specify.

Physics can be formulated in the language of information theory \cite{Wheeler1989}.
In the previous chapter we introduced the concept that everything is modelled using fields and their excitations.
Now we describe a program that models all interactions between particles as the transfer of information.
This information may be regarding the properties of these particles or it may be to do with their motion.
As the physical world is fundamentally based on quantum physics, these interactions can mediate quantum information.
If we can understand how information is communicated using quantum systems we will be closer to understanding physics in terms of information transfer between particles.
We present in this chapter the tools required to study the communication of information using quantum systems.

In this thesis I will always refer to the sending observer as Alice and the receiving observer as Rob.
This is because this thesis is about relativistic quantum communication and we usually name the relativistic observer Rob instead of Bob.

\section{Physical form of information}
\label{sec:inf:encoding}
In this thesis we take the same operational view of information as presented by Timpson \cite{Timpson2008a}.
We take an information source, which produces states from an alphabet with certain relative probabilities.
The alphabet of an information source is the complete set of symbols the source is able to produce.
With classical binary sources, the alphabet simply consists of $\zero$ and $\one$.

The information source generates information by producing a particular sequence of physical states.
This can be sent using a communication protocol and reproduced exactly or approximately at a receiver by a sequence of states of the same permutation.
For a classical example, if we take the alphabet $\{a,b,c\}$ a sequence permutation may be ``$a,b,a,c$''.
This can be sent down a channel and reproduced by another sequence of permutation ``$a,b,a,c$'' made up of different physical states.
It is the same logical sequence however, so it contains the same information.

The simplest units of information are single instances of two level systems.
Classically, the alphabet is $\{\zero,\one\}$ and a single symbol from that alphabet is a bit.
A qubit is the quantum equivalent of a bit, and it is a two state quantum system in any superposition of those states.

The physical states used for communication of these sequences are often fields, as they allow easy transmission of information from one location to another.
In this thesis we will consider only bosonic fields.
The bosonic field most often used is the photon field, so we will use much of the nomenclature of optics.
In this thesis we are concerned with two methods of representing a qubit using an optical field.
They are the single rail method and the dual rail method, both presented here.

The single rail uses one field mode, representing the $\zero$ by a vacuum, and the $\one$ by a single particle state,
\begin{equation}
 \label{eq:inf:singlemode}
\Ket{\zero}^{(s)}=\vac \ \mbox{and} \ \Ket{\one}^{(s)}= \hat{a}^\dagger \vac.
\end{equation}
This is only possible when the energy cost of creating and annihilating an excitation is negligible.
For quantum information to be stored, there must sometimes be superpositions of the $\zero$ and $\one$ states.
This requires superpositions of the vacuum with the single particle state, which in certain situations has problems with conservation laws.

Dual rail uses two field modes with a single excitation in a superposition between the modes.
The information is stored in the location of this excitation, it is in the zero mode to represent a $\zero$ and in the one mode for a $\one$,
\begin{equation}
 \label{eq:inf:dualrail}
\Ket{\zero}^{(d)}=\hat{a}^\dagger_\zero\vac \ \mbox{and} \ \Ket{\one}^{(d)}=\hat{a}^\dagger_\one\vac.
\end{equation}
The dual rail method can be used with massive fields as there is always a single excitation, so creating superpositions of the $\zero$ and $\one$ modes is not a problem.
It is also possible with massless fields and is commonly used in optics with different polarizations representing the different required modes.

For communication in real life settings, Rob will want to reproduce the message that has been sent.
This is what defines successful transmission of the information.
When there is noise present in the system it is often the case that Rob will not receive a state exactly in one of the above encodings.
At this point Rob must know which encoding was intended and apply any error-correcting protocols to calculate which message was most likely to have been sent \cite{Nielsen2000}.
In this thesis we assume that this is possible and only look at the information content of Rob's state.
This information content, and it's correlation with the sender's state is what we use to calculate the information transferred.

\section{Classical information}
\label{sec:inf:classical}
We first study classical information because its behaviour is more intuitive.
As per the previous definition of information, we define classical information as the logical order of a sequence of classical states.
Communication of this information will be defined as the ability of a second party, Rob to reconstruct the type of sequence (the logical order) using his own classical states.

\subsection{Quantifying information using the Shannon entropy}
We quantify the information contained in a particular symbol using the logarithm of the probability that the symbol will occur,
\begin{equation}
  \label{eq:inf:information}
  i(x) = -\lb{p(x)}.
\end{equation}
It can be thought of as a measure of our surprise when that symbol occurs.
There is more information contained in the less frequent symbols.
This quantity of information is additive, which means that the amount of information contained in two independent symbols is simply the sum of the information of each individual symbol,
\begin{equation}
  \label{eq:inf:informationadditive}
  i(x,y) = i(x)+i(y).
\end{equation}

We are interested in the bulk properties of these information sequences treated in the large limit.
For this reason, the quantity we use is the expectation value of information, which is known as the entropy.
This is the expected number of bits required to transfer, store and reconstruct this logical order.
For a classical random variable $X$, with instances $x$ produced with probability $p(x)$, we use the Shannon entropy \cite{Shannon1949},
\begin{equation}
  \label{eq:inf:shannonentropy}
  H(X)=\sum_x - p(x) \lb{p(x)}.
\end{equation}

If a particular source is known, the probability of one outcome will be $1$ and the rest will be $0$.
We use the limiting cases of,
\begin{equation}
  \lim_{p\to \left\{0,1\right\}} \left[ p \lb{p}=0\right],
\end{equation}
so the entropy of a known source is zero.
The entropy is interpreted as the amount we do not know about a system.

\subsection{Communication}
Classical communication is easy to understand intuitively.
There is one resource, the classical channel, and it may have some level of noise or probability of error.
This can be accounted for using Shannon's noisy coding theorem \cite{Shannon1949}.

Any form of communication requires a sender and receiver, Alice and Rob.
They will have probability distributions associated with their systems (or sequence of systems).
Alice's probability distribution is related to what message she may want to send.
For example if she was using the English language, each letter in the alphabet has a probability of being used at any given point in a sentence.
Rob's probability distribution should be related to Alice's.
If Alice sends a ``k'', we would hope for a useful channel that Rob is most likely to receive an ``k''.
It is this correlation that we can quantify through the information capacity of a channel.

We should first formally define a communication channel.
A classical communication channel is a probabilistic transformation between Alice's input state (sequence of systems) resulting in Rob's output state.
For example, a simple symmetrical classical channel with a $10\%$ probability of bit-flip error can be represented as the following transformation matrix,
\begin{align}
  \label{eq:inf:simpleclassicalchannel}
  \begin{pmatrix}
          0.9&0.1\\
          0.1&0.9
        \end{pmatrix}
 ,
\end{align}
where both input and output use the $ (\zero, \one)$ basis.
This example is the case where if Alice sends a $\one$, Rob has a $10\%$ chance of receiving a $\zero$ and $90\%$ chance of receiving a $\one$ and vice versa.

\subsection{Quantifying the communication}
With the aim of quantifying the communication channel capacity between Alice and Rob we introduce the mutual information and conditional entropy.
These measure the correlations and relationship between the information contained in Alice's and Rob's states.

\begin{figure}[htp]
  \begin{center}
      \includegraphics[width=350pt]{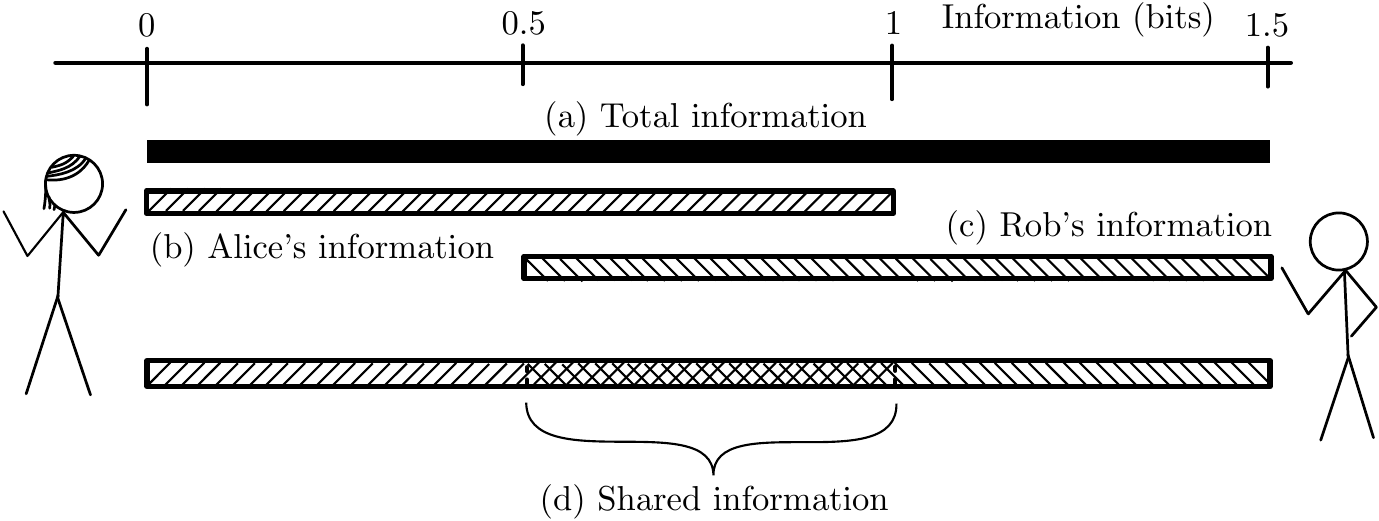}
  \caption[Visualising the mutual information]{The mutual information is constructed in this figure. We start with Alice and Rob having a total information of $1.5$bits between them (a). If Alice and Rob each have $1$bit of their own (b) and (c), they must share $0.5$bits of information (d). This $0.5$bits of shared information is the mutual information.}\label{fig:inf:mutualinfo}
  \end{center}
\end{figure}

The mutual information is a measure of the amount of information that Alice and Rob share.
From a communication perspective this is the amount of information Rob knows about Alice's state by looking at his own.
Taking Alice's random variable as $A$ and Rob's as $R$.
The mutual information is defined in terms of the Shannon entropy as,
\begin{equation}
  \label{eq:inf:classicalmutualinformation}
  I(A;R)=H(A)+H(R)-H(A,R).
\end{equation}
Where $H(A,R)$ is the joint Shannon entropy of the variables, which is where the Shannon entropy is calculated for the joint probability distribution of the variables.
This has been visualised in \figref{fig:inf:mutualinfo}, where we see the construction of mutual information.
We can take a simple example.
Let Alice have one bit of information and if asked separately Rob also has one bit of information.
If we can describe both of their bits with only one bit of information, then the information they have must be the same bit.
They share the one bit, their mutual information is one bit.

The conditional entropy is a measure of uncertainty about Alice's state, if one already knows Rob's state.
Again, it is defined in terms of the Shannon entropy as,
\begin{equation}
  \label{eq:inf:classicalconditionalentropy}
  H(A|R)=H(A,R)-H(R).
\end{equation}
The conditional entropy is interpreted as the amount of extra information Rob requires to fully specify Alice's state.
For example, Alice has two bits of information, while Rob has one bit.
Between them, they have two bits of information.
This requires the mutual information to be one bit, given the previous definition.
It also means that Rob would need Alice to send him one more bit of information for him to be able to describe Alice's full state.

The channel capacity is a measure of how much information it is possible to send using a particular physical channel.
More precisely, the capacity is generally expressed as a ratio,
\begin{equation}
  \label{eq:inf:channelcapacity}
  C = \frac{\text{information sent}}{\text{number of channel uses}}.
\end{equation}
This is often called the communication rate even though it is not related to time.
To transfer a message across a channel, it must be encoded into the alphabet that the channel uses.
This can be as simple as a direct mapping between the logical $\zero$ and $\one$ in the message to the representative $\zero$ and $\one$ in the channel.
The capacity of a channel is the maximum rate at which it can transfer information.
In a classical channel, this is simply the maximum mutual information over all possible source probability distributions.

This is formalized by proving in two parts a channel capacity theorem.
Firstly, there is the proof that the rate given is achievable.
If the rate is less than or equal to the capacity, there exists a set of codes that allow Alice to communicate reliably with Rob.
By reliably we mean that the probability of error tends to zero as the number of channel uses grows to infinity.
Secondly there is the proof that this rate is maximal.
That there does not exist a set of codes that allow a rate of communication greater than the channel capacity.
As a channel does not include any information about the encoding and decoding methods we must maximise over all possible encodings of the messages.

\section{Extending to quantum information}
\label{sec:inf:quantum}
The logical extension to quantum information requires some discussion.
It is clear that the classical states be replaced with quantum states.
Again, we treat the information as the potentially reproducible sequence of systems created by an information source.
In this case we have a quantum information source, producing quantum states from its alphabet according to a probability distribution.

The alphabet of quantum states can include any quantum state.
We must even include in the specification of these states entanglement with another hidden system, known as the purifying system.
Any mixed quantum state can be written as a pure, entangled quantum state between the system and its purifying system.
To regain the mixed state, it is only necessary to trace out the purifying system.
The alphabet for quantum information can contain quantum states that have entanglement with a purifying system.

\begin{figure}[tb]
  \begin{center}
      \includegraphics[width=272pt]{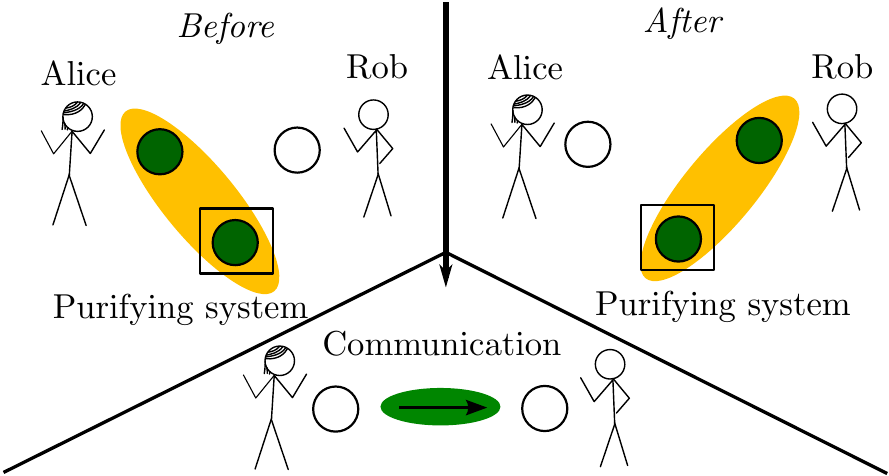}
  \caption[Schematic quantum communication]{Schematic diagram of quantum communication, representing quantum states with circles.
      Before the communication, Alice has some quantum information (circle filled green) that is entangled (yellow oval) with a purifying system (square and circle).
      Rob has no information in his quantum state (empty circle).
      Then some communication protocol is performed that does not involve the purifying system. This sends Alice's quantum information to Rob.
      After the communication, Rob has recovered the same information Alice had and also recovered the entanglement with the same purifying system.
      This is classed as successful communication.}
  \label{fig:inf:communication}
  \end{center}
\end{figure}

For quantum communication to be deemed successful, the receiver must reproduce the quantum states, and their entanglement with the purifying system.
The purifying system must be the same for both sender and receiver.
This requires a loss of entanglement between the sender and the purifying system and a gain of exactly the same entanglement between the receiver and the purifying system over the course of the communication, see \cref{fig:inf:communication}.

We define the simplest unit of quantum information to be the qubit, the amount of information contained in a two-state system.
However, we must make clear the distinction between specification information and accessible information.
Quantum states can be superpositions of the basis states, where the coefficients are continuous.
This may suggest that there is a continuous amount of information contained in a qubit.
This continuous information is the specification information, which is not particularly useful to talk about.
A qubit however measured, reveals at most one single bit of classical information, therefore we can say that the accessible information is not continuous.
We can only access one bit of information for every qubit \cite{Holevo1973}.

\subsection{Quantifying quantum information, the Von Neumann entropy}
\label{subsec:inf:vonneumann}
When using quantum states, the appropriate entropy measure is the von Neumann entropy, defined as,
\begin{equation}
  \label{eq:vonneumannentropy}
S\!\left(\rho\right) = -\Tr{\rho \lb{\rho}} = -\sum_i \left\{\lambda_i \lb{\lambda_i}\right\},
\end{equation}
where $\lambda_i$ are the eigenvalues of $\rho$.
This reduces to the Shannon entropy when the $\lambda_i$ are elements of a classical probability distribution.
It is the basic entropy measure of a single system, in terms of which all other measures are defined.
The units used by this entropy measure are qubits.

In the special case where $\rho$ is a pure state, it has one eigenvalue of $1$ and the rest are $0$.
This makes the Von Neumann entropy equal to zero for any pure state.
There is no true uncertainty in a pure state so it is expected that the entropy is zero.

\subsection{Multipartite states allowing communication}
As mentioned previously for classical communication, all communication requires a sender and a receiver.
These can both be described by a joint quantum state.
The joint state can be constructed in any way.
Conceptually the most useful construction for communication is where we take a tensor product between Alice's message state and Rob's initialization state.
We then apply the transformation of the channel, where depending on what Alice's state is, Rob's state will change, often with a random noise element included.
This allows us to easily see how Alice may send quantum information to Rob using the channel.
We can draw parallels between this multipartite state and the joint probability distribution in classical communication.

\subsection{Communication resources}
There are more methods of manipulating quantum information than there are for classical information.
In quantum communication we can have the quantum channel, whose aim is to directly send a qubit from Alice to Rob.
We can still use the classical channel too because quantum systems can model classical systems trivially.
Finally, we can also use quantum entanglement.
Entanglement is a fully quantum phenomenon with no classical analogue.
It is the correlation between quantum systems that goes beyond mere statistical correlations.

We can trade the three resources of classical communication, quantum communication and entanglement using various communication protocols.
These are sometimes written in resource inequalities as $[c\to c]$, the transfer of a classical bit, $[q\to q]$, the transfer of a quantum bit and $[qq]$ a shared entanglement bit.
We visualise a space where each resource is an axis, and the directions are creation or consumption of this resource.
Each protocol will have regions of achievable rate triples in this space, these regions define the possible communication abilities of the protocols.
An achievable rate triple, is a point in the space where it is possible to create that communication while consuming those resources using the protocol given.

\subsection{Distinguishability and fidelity}
\label{subsec:inf:fidelity}
The distinguishability is a useful concept in communication theory because it is the ease with which one can tell the two states apart.
If we send quantum information through a channel, we need to know that the received symbols are distinguishable.
If they are not, we have no way of knowing which state was sent, so there can be no information communicated.
Instead of directly calculating the distinguishability of quantum states, we use the fidelity, $F$, because it is related to the distinguishability, $D$, via the relation,
\begin{equation}
  \label{eq:inf:fidelitydistinguishability}
  D\equiv1-F.
\end{equation}

Fidelity is a measure of how closely two states resemble each other.
Jozsa took four fundamental axioms that any expression for the Fidelity of two mixed quantum states must obey.
These axioms are \cite{Jozsa1994},
\begin{enumerate}
  \item $0\le F(\rho_1,\rho_2) \le 1$ and $F(\rho_1,\rho_2)=1$ iff $\rho_1=\rho_2$.
  \item $F(\rho_1,\rho_2)=F(\rho_2,\rho_1)$.
  \item If $\rho_1$ is pure, $\rho_1=\Ketbra{\Psi_1|\Psi_1}\removetofixketbra $, then $F(\rho_1,\rho_2)=\Braket{\Psi_1|\rho_2|\Psi_1}$.
  \item $F(\rho_1,\rho_2)$ is invariant under unitary transformations on the state space.
\end{enumerate}
Ulhmann had already introduced a quantity that satisfied these axioms, \cite{Uhlmann1976}.
We now refer to this quantity as the fidelity, it is defined as,
\begin{equation}
 \label{eq:inf:fidelityfull}
 F = \left(\Tr{\sqrt{\sqrt{\rho_1}\rho_2\sqrt{\rho_1}}}\right)^2.
\end{equation}
Jozsa noted, \cite{Jozsa1994}, that this may be the only form satisfying all four axioms for general mixed quantum states.
Operationally it can be understood as the probability of mistaking one state for the other.

If the density matrices commute, the Fidelity takes the simpler form,
\begin{equation}
  \label{eq:inf:fidelitycommute}
  F = \left(\Tr{\sqrt{\rho_1\rho_2}}\right)^2.
\end{equation}
We will calculate this between the two types of received state to comment on how much the noise has mixed the logical states together.

The fidelity is used to calculate how successful certain communication protocols were.
We take quantum teleportation as an example.
Teleportation is a method of transferring quantum information without needing to send a physical qubit, which is often experimentally challenging.

We start with an unknown qubit in Alice's possession and a shared entanglement bit (ebit) between Alice and Rob.
Alice performs a bell measurement on her unknown qubit (to send) and her half of the ebit.
The results of this bell measurement she sends classically to Rob, this requires sending two classical bits of information.
Rob then performs two local single-qubit operations that each depend on one of the received classical bits.
He should end up with exactly the unknown qubit that Alice wanted to send.
This whole protocol is completed without ever finding out what the unknown qubit was because that would destroy it.
It also obeys the no cloning theorem by resetting Alice's qubits to a completely random state, one of the Bell states with equal probability.

This description of the teleportation protocol assumes a maximally entangled pair shared between Alice and Rob and perfect classical communication.
If the entanglement is not quite maximal, or the classical communication is not quite perfect there may be errors in Rob's received state.
For the measurement of success, we take the fidelity between Alice's initial unknown state, and Rob's received state.
This gives us a measure of how accurately the state was teleported between Alice and Rob.

\subsection{Differences in information measures}
\label{subsec:inf:quantuminfomeasures}
Classical information measures have equivalent quantum information measures.
The Shannon entropy is replaced with the Von Neumann entropy.
The measures are then valid for quantum information.
However, it is important that a few differences in the interpretation are pointed out.

The quantum mutual information is still a measure of the amount of correlations and information shared between the two systems.
It is defined in the same way as,
\begin{equation}
  \label{eq:inf:mutualinfo}
  I(A;R)_\sigma = S(\sigma_A)+S(\sigma_R)-S(\sigma_{AR}),
\end{equation}
where $S(\sigma)$ is the von Neumann entropy.

Quantum correlations can take a different form to classical statistical correlations.
The mutual information measures all correlations between Alice's and Rob's states.
A system can have positive mutual information even if it is in a pure state where the total information is zero.
This is in contrast to the classical mutual information which must always be less than the information content of each state and the total information.
Quantum entanglement allows for stronger correlations between the two qubits than is allowed classically with purely probabilistic correlations.
The maximum mutual information that two qubits can have is 2 bits rather than two classical bits having 1 bit of mutual information.

We can use the mutual information to measure information transferred over a channel.
When we are looking at the amount of classical information shared over a quantum channel we need to look at the classical mutual information between Alice's information source and Rob's measurement results.
This will of course depend on the exact observable Rob measures.
To overcome this ambiguity, we maximise the mutual information over all possible observables that Rob can measure.
This quantity is known as the accessible information.
It is not known currently how to calculate the accessible information, however, it is upper bounded by the Holevo information \cite{Schumacher1997,Holevo1998}.

The Holevo information is defined over an ensemble of quantum states \cite{Wilde2011,Holevo1973},
\begin{equation}
  \label{eq:inf:ensembleforholevo}
  \varepsilon \equiv \{p_X(x), \sigma_{(x,R)}\},
\end{equation}
where the random variable $X$, is Alice's information source.
Alice then gives the density operator $\sigma_{(x,R)}$ to Rob.
Rob has no information about $x$, so he specifies the quantum state as the expected value of $\sigma_{R}=\sum_x p_X(x)\sigma_{(x,R)}$.
The Holevo information is given by the quantity,
\begin{equation}
  \label{eq:inf:holevoinfo}
  \chi(\varepsilon)\equiv S(\sigma_{R})-\sum_x p_X(x)S(\sigma_{(x,R)}).
\end{equation}
It can be shown that, for a classical quantum state of the form, 
\begin{equation}
  \label{eq:inf:classicalquantumstate}
  \sigma_{AR} = \sum_x p_A(x) \ket[A]{x}\bra{x} \otimes \sigma_{R}(x),
\end{equation}
the Holevo information is equivalent to the quantum mutual information \cite{Wilde2011}.
In this thesis, when calculating the classical capacity of a quantum channel, we always use classical quantum states of the form in \cref{eq:inf:classicalquantumstate}.
For the classical channel capacity, we will always calculate the quantum mutual information, which, being equivalent to the Holevo information, provides us with an upper bound on the classical capacity.

Sending classical information over a quantum channel is perhaps the simplest protocol \cite{Schumacher1997}.
In this situation, the required message is encoded into a classically probabilistic ensemble of quantum states,
\begin{equation}
  \label{eq:inf:classicalinfoprobabilitystate}
  \sigma_A = \sum_x p(x) \Ket{x}\!\Bra[A]{x},
\end{equation}
where $x \in \{\zero,\one\}$ is the representation of the classical binary logical bit.
The quantum states used may or may not be orthogonal, in this thesis we only consider orthogonal states because this maximises the information transferred.
This ensemble can be represented by a density matrix,
\begin{equation}
  \label{eq:inf:classicalinfodensitystate}
  \sigma_A = \modsq{\alpha} \Ket{\zero}\!\Bra[A]{\zero} + \modsq{\beta} \Ket{\one}\!\Bra[A]{\one},
\end{equation}
where $\modsq{\alpha}=p(\zero)$ and $\modsq{\beta}=p(\one)$.
Alice sends this state down the channel to Rob who receives something that may have some error or noise introduced.
To calculate how much of Alice's information Rob has received we use the Holevo information.
Assuming for the example that there is perfect transmission of the quantum state, we can calculate this Holevo information as follows.
The joint quantum state can be written as the density matrix,
\begin{equation}
  \label{eq:inf:sigmaARholevoexample}
  \sigma_{AR}=\begin{pmatrix}
              \modsq{\alpha}&0&0&0\\
              0&0&0&0\\
              0&0&0&0\\
              0&0&0&\modsq{\beta}
            \end{pmatrix}.
\end{equation}
Taking the partial trace over $R$ will recover $\sigma_A$ and due to the symmetry taking the partial trace over $A$ will result in $\sigma_R=\sigma_A$.
As these matrices are all diagonal, it is trivial to calculate their Von Neumann entropies,
\begin{equation}
  \label{eq:inf:entropyholevoexample}
  S(\sigma_A)=S(\sigma_R)=S(\sigma_{AR})=-\modsq{\alpha} \lb{\modsq{\alpha}}-\modsq{\beta} \lb{\modsq{\beta}}.
\end{equation}
From this, the Holevo information calculation cancels down to just the following,
\begin{equation}
  \label{eq:inf:holevoinfexample}
  I(A;R)_\sigma=-\modsq{\alpha} \lb{\modsq{\alpha}}-\modsq{\beta} \lb{\modsq{\beta}}.
\end{equation}
This is now equal to the information content of the original message, as expected because there was no noise present.

There is a complimentary way of looking at this situation, and that is to look at how much information Rob is missing about Alice's state.
This is Rob's uncertainty about Alice, known as the conditional entropy.
Just as in the classical case, the conditional entropy is a measure of the amount of extra information Rob needs to fully specify Alice's state given that his own state is known \cite{Horodecki2005}.
It is defined as a difference between the total entropy of the composite system and the entropy of Rob's local system,
\begin{equation}
  \label{eq:CEdefinitionold}
  S(A|R) = S(AR)-S(R).
\end{equation}
Perhaps a more enlightening definition, is in terms of the mutual information,
\begin{equation}
  \label{eq:inf:CEintermsofMI}
  S(A|R) = S(A) - I(A;R).
\end{equation}
Here we can clearly see that it's the amount of information contained in Alice's state, minus the amount of information Rob already has about Alice's state.
To illustrate the conditional entropy we introduce the state merging protocol.

\subsection{The state merging protocol and coherent information}
\label{subsec:inf:statemerging}
Initially the protocol considers a pure state distributed between Alice, Rob and a purifying system.
This is stricter than considering mixed states that are shared between Alice and Rob, because the entanglement with the purifying system must also be preserved.
We must also assume that Alice and Rob have spare quantum information registers to store extra information.
They may also be required to start with some shared entanglement.
The protocol must end with the same pure state but with Alice's part stored on Rob's extra registers instead.
This is the merging of the state from being distributed to being all in one location (Rob's quantum information registers).

The state merging protocol is a name given to any process that performs the above transformation using only LOCC operations.
The LOCC operations are Local Operators and Classical Communication, only local quantum operations are allowed, combined with any amount of classical communication.
This is a common restriction used when dealing with quantum communication because LOCC operations cannot affect the intrinsic nature of any entanglement.
This definition was first stated by Horodecki et al. \cite{Horodecki2007a} as an extension to the Slepian-Wolf theorem for reproducing classical information with certain prior knowledge \cite{Slepian1973}. 

Horodecki et al. showed that the entanglement cost of this protocol is equal to the conditional entropy \cite{Horodecki2007a}.
The interesting part of the protocol is that the conditional entropy can be negative.
Classically, conditional entropy can never be negative because $S(AR) \ge S(R)$.
The best that is achievable classically is that Rob fully learns Alice's state from looking at his own.
However, for states containing quantum information the conditional entropy can be negative, this happens when the states are entangled \cite{Vollbrecht2002}.
When these states are entangled, and the conditional entropy is negative, the state merging protocol costs a negative amount of entanglement.
This is the same as saying the state merging protocol generates entanglement between the parties.

The negative of conditional entropy is such an important quantity that it is given a name, the quantum coherent information,
\begin{equation}
  \label{eq:inf:coherentinfodefinition}
  I(A\rangle R)_\rho =- S(A|R)_\rho.
\end{equation}
The coherent information, is an equivalent measure to conditional entropy, but it describes more clearly how much extra information Alice and Rob can share by using the state merging protocol.
It is the amount of entanglement gained between Alice and Rob.
Once the entanglement is created it can be used for communication via teleportation provided that classical communication is possible \cite{Horodecki2005}.

\subsection{Quantum communication channel capacity}
\label{subsec:inf:quantumchannelcapacity}
As we stated at the beginning of this chapter, the information task that corresponds to quantum communication must include preserving quantum correlations with a hidden system.
These correlations may be entanglement, and Rob must recover the states that Alice sent, including the correlations.
We can include quantum correlations that have been set up through the use of the channel as part of the information transferred.

Any quantum channel can be described as a unitary transformation from a sender's Hilbert space, to the tensor product of Hilbert spaces of the receiver and the environment, followed by a trace over the environment.
This defines a complementary channel, the one from the sender to the environment.
A channel is degradable if a map exists between the receiver and the environment that transforms the channel into it's complimentary channel.

The regularized coherent information plays a similar role in quantum channel capacities to the mutual information in classical capacities.
The channel capacity is the maximum of all achievable rates.
It has already been proven (Chapter 23 of \cite{Wilde2011}) that this is given by the maximum regularized coherent information over all possible encoding methods.
Regularization is not necessary for the degradable class of quantum channels.
The channels we study in this thesis have not been proven to be degradable.
The coherent information without any regularization provides us with an achievable rate, and therefore a lower bound on the capacity.

It is beyond the scope of this thesis to regularize the coherent information.
We do however, know that the coherent information is itself a measure of the amount of entanglement created by the state merging protocol.
This has been interpreted as a quantum channel capacity \cite{Horodecki2005}, even though it has not been proven for all channels.
Here we take the stricter view and just look at the entanglement that it is possible to generate.
For a full explanation of the state merging protocol, see \cref{subsec:inf:statemerging}.

The channel capacity must be maximised over any encoding scheme.
We do not study all encoding schemes in this thesis.
This leads us to define a term `restricted capacity' to represent the maximum information transferred per use of a particular channel in a particular encoding method.
We use the word restricted because we restrict ourselves to a particular encoding method rather than optimising over everything.
The word capacity is used here even though we do not prove that these are upper bounds on the information transferred.
We have merely numerically optimised the information transfer to find the maximum.
Due to the infinite dimensionality of most of the states we work with here, it is not possible to find an analytical expression for the coherent information.
We truncated the dimensionality of the relevant Fock bases at a level appropriate to have a small enough error.

We have discussed all three quantum communication resources, namely classical communication, quantum communication and entanglement.
Previously we mentioned that they can be manipulated using communication protocols.
Each protocol will use up some resources, and produce others.
A trivial example is sending a qubit down a quantum channel to communicate quantum information.
We use 1 qubit of the quantum channel, and it gives us one qubit of quantum information communicated.

To summarise the protocols mentioned earlier, using resource conversions.
Classical communication is denoted by $C$, quantum communication by $Q$ and entanglement by $E$.
We write them as a triple, always in the order $(C,Q,E)$, the units are always bits, qubits and ebits.
Sending classical information over a quantum channel is possible, we just take the qubits in the computational basis and send them.
This uses the quantum communication resource to achieve classical communication.
The resource conversion is,
\begin{equation}
  \label{eq:inf:classicaloverquantumresources}
  (0_C,1_Q,0_E)\to(1_C,0_Q,0_E).
\end{equation}
If we use dense coding, as was introduced by Bennett and Weisner \cite{Bennett1992}, we can use an ebit to assist the communication.
This allows us to send two bits of classical information with one ebit and one qubit of quantum communication.
\begin{equation}
  \label{eq:inf:densecodingresources}
  (0_C,1_Q,1_E)\to(2_C,0_Q,0_E).
\end{equation}
The teleportation protocol allows us to send a qubit using only classical communication and entanglement.
The resource conversion is,
\begin{equation}
  \label{eq:inf:teleportationresources}
  (2_C,0_Q,1_E)\to(0_C,1_Q,0_E).
\end{equation}

If one represents used resources as negative, and gained resources as positive each resource conversion can be mapped to a point in a three dimensional space.
These rate triples are ideal maximums for the protocols.
When channel noise is taken into account, it is useful to study the regions of achievable rate triples, which is outside the scope of this thesis.

\subsection{Parameter estimation}
\label{subsec:inf:fisherinformation}
There is another type of information task, parameter estimation, that we perform in this thesis.
Parameter estimation is where a continuous parameter is encoded in a state, and we measure the state to estimate the value of the parameter.
We assume this state is in the possession of Alice to begin with.
Alice then performs a communication protocol with Rob.
Finally we require either Rob alone, or both together to measure their states to determine what the value of the parameter is.

Every measurement of a continuous parameter is subject to an uncertainty and we would like to calculate that uncertainty.
The uncertainty describes the range of values in which we expect the true value of the parameter to be.
Generally the uncertainty is quantified as the standard deviation, however, for ease of notation we talk about the variance.
The variance on an estimation of the parameter is lower bounded by the Cram\'er-Rao bound \cite{Braunstein1996} which states,
\begin{equation}
  \label{eq:inf:cramerraobound}
  \left\langle (\Delta \theta)^2 \right\rangle \ge \frac{1}{N \mathscr{F}(\theta)},
\end{equation}
where $\mathscr{F}(\theta)$ is the Fisher information and $N$ is the number of measurements.

The Fisher information can be thought of as the expected amount we learn about the parameter in each measurement.
It is calculated from the density matrix of the state,
\begin{equation}
  \label{eq:inf:fischerinfo}
  \mathscr{F}(\theta)=\Tr{{\rho'} \mathscr{L}_\rho({\rho'})}, \quad \rho' = \frac{\ud \rho}{\ud \theta},
\end{equation}
where $\mathscr{L}$ is the lowering operator given by,
\begin{equation}
  \label{eq:inf:loweringoperator}
  \mathscr{L}_A(B)=\sum_{jk} \frac{2B_{jk}}{(\lambda_A)_j+(\lambda_A)_k} \Ketbra{j|k} \junktofixketbra,
\end{equation}
where $\lambda_A$ are the eigenvalues of $A$, making $A$ diagonal in the $\Ketbra{j|k}\junktofixketbra$ basis.

In the case of calculating the Fisher information for a parametrized density matrix that is diagonal we can vastly simplify the lowering operator.
In this thesis, the density matrices we study are often diagonal due to the partial traces required in many of the transformations.
The density matrix being diagonal means that the differentiated density matrix $\rho'$ will also be diagonal in the same basis.
Therefore we can simplify the lowering operator to,
\begin{equation}
  \label{eq:inf:simplelowering}
  \mathscr{L}_\rho(\rho')=\sum_{j} \frac{\rho'_{jj}}{\rho_{jj}} \Ketbra{j|j}\junktofixketbra.
\end{equation}
This in turn allows simpler calculation of the Fisher information,
\begin{equation}
  \label{eq:inf:simplefisher}
  \mathscr{F}(\theta) = \Tr{\frac{(\rho'_d)^2}{\rho_d}},
\end{equation}
where $\rho_d$ reminds us that $\rho$ must be diagonal, and similarly for $\rho'_d$.

We can use the Fisher information to find out how much information can be extracted by Rob from the states that Alice sends him.
This can tell us more about the information preserving capabilities of the various channels we use for Alice to communicate with Rob.

\section{Full example}
\label{sec:inf:simpleexample}
We find it helpful to first go through the calculation for ideal communication.
This allows us to set out the structure of the calculations and how we model them.
We will use the single rail encoding method and demonstrate an example encoding and decoding.

Alice and Rob each have their own quantum systems, and the communication channel will be found from the mutual entropy between these systems.
The communication channel itself will be a field, the state of which Alice creates, and Rob detects.
After propagation, we denote Alice's state with $\ket[A]{x}$, and Rob's state with $\ket[R]{x}$, the vacuum field we will denote with $\vac$.

Alice will send an arbitrary qubit, so she first prepares her system in the state of that qubit.
The full initial state will then be,
\begin{equation}
\label{eq:idealinitial}
\left(\alpha\Ket[A]{\zero} + \beta\Ket[A]{\one}\right)\otimes\vac\otimes\ket[R]{D},
\end{equation}
where $\ket[R]{D}$ is the state of Rob's detector before it detects anything.

Then Alice's machine will create an excitation in the appropriate mode of the field.
This will be given by the transformations,
\begin{subequations}
\begin{align}
  \label{eq:idealalicecreate}
  \Ket[A]{\zero}\otimes\vac &\to \Ket[A]{\zero}\otimes \vac,\\
  \Ket[A]{\one}\otimes\vac &\to \Ket[A]{\one}\otimes \ket{1},
\end{align}
\end{subequations}
where $\vac$ and $\ket{1}$ are the states of exactly zero and one photon in the field respectively.

At this stage the full state is,
\begin{equation}
\label{eq:idealfieldcreated}
\left(\alpha\Ket[A]{\zero}\otimes\vac + \beta\Ket[A]{\one}\otimes\ket{1}\right)\otimes\ket[R]{D}.
\end{equation}

We apply the propagation to the field states, this is where all the physics comes in.
Any effects due to relative motion or spacetime curvature will happen in this section.
For the ideal communication we use the identity operator, so nothing happens.
Formally this is given by the transformations,
\begin{subequations}
\begin{align}
  \label{eq:idealfieldtransformation}
  \vac &\to \Ket[t]{\varnothing} = \identity \vac,\\
  \ket{1} &\to \Ket[t]{1} = \identity \ket{1}.
\end{align}
\end{subequations}

After the transformation the full state becomes,
\begin{equation}
\label{eq:idealfieldtransformed}
\left(\alpha\Ket[A]{\zero}\otimes\Ket[t]{\varnothing} + \beta\Ket[A]{\one}\otimes\Ket[t]{1}\right)\otimes\ket[R]{D}.
\end{equation}

Rob receives these transformed fields, and detects them.
His ideal detector exactly transfers the $\vac$ and $\ket{1}$ states to Rob's own system's zero and one states.
This is formally given by,
\begin{subequations}
\begin{align}
  \label{eq:idealbobdetect}
  \vac\otimes\ket[R]{D} &\to \vac\otimes \Ket[R]{\zero},\\
  \ket{1}\otimes\ket[R]{D} &\to \vac\otimes \Ket[R]{\one},
\end{align}
\end{subequations}

The full state at the end of the process is,
\begin{equation}
\label{eq:idealfielddetected}
\Ket{\psi}=\alpha\Ket[A]{\zero}\otimes\vac\otimes\Ket[R]{\zero} + \beta\Ket[A]{\one}\otimes\vac\otimes\Ket[R]{\one}.
\end{equation}

We can trace out the field vacuum, as it is not relevant, without any loss, as it is completely separable from Alice and Rob's combined state.
This leaves us with a state involving Alice and Rob's systems.
We need to write this as a density matrix in the logical basis,
\begin{equation}
\label{eq:idealrhoARmatrix}
\rho_{AR} = \begin{pmatrix}
              \modsq{\alpha}&0&0&\alpha\beta^*\\
              0&0&0&0\\
              0&0&0&0\\
              \alpha^*\beta&0&0&\modsq{\beta}
            \end{pmatrix}
\end{equation}

We can find the individual density matrices for Alice and Rob by partial trace,
\begin{subequations}
\begin{align}
  \label{eq:idealrhoAmatrixRmatrix}
  \rho_A = \Tr[R]{\rho_{AR}} &=
  \begin{pmatrix}
    \modsq{\alpha}&0\\
    0&\modsq{\beta}
  \end{pmatrix}\\
  \rho_R = \Tr[A]{\rho_{AR}} &=
  \begin{pmatrix}
    \modsq{\alpha}&0\\
    0&\modsq{\beta}
  \end{pmatrix}
\end{align}
\end{subequations}
Each of these matrices is diagonal already.
The overall state is a pure state, so has zero entropy.
They are also both the same, so the mutual entropy is double the entropy of one of them.
If we use $p=\modsq{\alpha}$, the mutual entropy is then given by,
\begin{equation}
  \label{eq:idealmutualentropy}
  S\left(\rho_A;\rho_R\right)= 2\left(-p\lb{p} - (1-p) \lb{1-p}\right).
\end{equation}
This is exactly as expected, and is maximal at $p=\frac12$ as expected.

\chapter{Review}
\label{ch:review}

Quantum information theory has developed out of attempts to generalize information theory to include the quantum world.
In \cref{ch:information} we introduced the concept of quantum information.
We saw that many concepts could be adapted but there were also some totally new ones.
Relativistic quantum information theory studies how special and general relativity affects quantum information.
Some special relativistic effects have been reviewed by Peres and Terno, \cite{Terno2005,Peres2003}, in this thesis we will mostly focus on general relativistic effects.

Relativistic quantum information theory is a young field.
Researchers investigating relativistic quantum information are motivated by the potential practical use and the fundamental understanding.
There are computing advantages using quantum information.
There are examples that show quantum algorithms improve on the best classical algorithms for solving certain difficult problems \cite{Shor1997,Deutsch1992,Grover1996}, summarised by Nielsen and Chuang \cite{Nielsen2000}.
These improved resources would be very useful for solving problems in the sciences in general, and quantum physics specifically.

One of the key tasks that becomes easy using quantum computers is factorisation \cite{Shor1997}.
This creates a practical problem in that the current technique for private communication relies on the difficulty of factorisation.
Quantum information theory also provides the solution because there are quantum channels that can transmit private information between parties.
These quantum channels do not rely on the difficulty of any numerical problem, rather they use the fundamental laws of quantum physics to maintain privacy.
The private information that is communicated can be either classical or quantum in nature.
Many real world situations involving communication need to take into account general relativistic effects, for example, when satellites are involved.
We must understand these effects on the quantum channels.

There is a natural human desire to understand more about our universe, which drives us to investigate deeper.
We want to find a fundamental theory describing everything around us.
We would like to unify quantum theory with general relativity, this quantum gravity would form a fundamental theory of the universe.
There are many candidate theories of quantum gravity, however, the only way to distinguish the better ones is by comparing to experimental results.
We need to understand more about these and other fundamental theories and the way they relate to each other to design appropriate experiments.
Much of the research that I review in this chapter works towards a greater understanding of the relationship between general relativity and quantum physics, particularly entanglement and quantum communication.

One of the big unanswered questions in the overlap between quantum information and general relativity is known as the black hole information paradox.
The paradox comes from the apparent conflict between unitary quantum evolution and evaporating black holes.
Quantum physics requires the time evolution of closed systems to be unitary, preserving all information.
Black holes can absorb matter and information, then evaporate the energy as thermal radiation.
This results in information loss, and non-unitary evolution.
As an extreme example, take a universe in a pure state, with a black hole.
In this universe there are some particles in pure state that get absorbed by the black hole.
The black hole eventually emits the equivalent energy as thermal radiation, totally erasing the information contained in the particles.
The thermal radiation is maximally mixed, so the universe would have to undergo non-unitary evolution.
Researchers are trying to understand more about this situation and related ones by studying entanglement and communication in relativistic situations.
Some recent work on the topic has led to possible solutions of this paradox \cite{Braunstein2013}.

\section{Entanglement}
\label{sec:rev:entanglement}
Quantum entanglement is the name given to non-classical correlations between quantum states.
These correlations are stronger than purely statistical correlations.
One can demonstrate this strength by the violation of Bell's inequalities \cite{Bell1964}.
The details of entanglement are reviewed by both Horodecki \emph{et al.} \cite{Horodecki2007}, and Plenio and Virmani \cite{Plenio2007}.
The special relativistic effects on entanglement was the topic of Friis' thesis \cite{Friis2010}.
General relativistic effects on entanglement was the topic of Mart\'i{}n-Mart\'i{}nez's Ph.D. thesis \cite{Martin-Martinez2011}.
In this thesis, we restrict the scope to relativistic effects.
One of the relativistic effects that has been studied is the Unruh-Hawking effect.

\subsection{Unruh-Hawking effect}
The Unruh-Hawking effect is where particles are created in field modes due to either acceleration of the observer detecting the field modes, or the curvature of spacetime where the field exists \cite{Hawking1975,Davies1975}.
These excitations behave as if they are a thermal bath of radiation with a particular temperature.
This temperature is referred to as the Hawking temperature or Unruh temperature, depending on whether it refers to the field near a black hole, or an accelerated observer.
A full study of the quantum field theory involved in the Unruh effect was published by Takagi \cite{Takagi1986}.

The exact nature of this detected radiation has been the subject of much discussion.
It is common to calculate it as a consequence of the lack of knowledge about half of the system.
This results in a mixed quantum state corresponding to a thermal spectrum, and an associated von Neumann entropy.
However, the standard calculations require the observer to continue accelerating forever.
It has been suggested that this modelling technique is not causal.
A real effect cannot depend on a future event, that the observer continues to accelerate.

This led to the introduction of a causal Unruh effect \cite{Schlicht2004}.
In this description, the observer is asymptotically at rest at time $t\to \pm \infty$.
This removes the horizon completely.
It is found that, during the acceleration, the observer continues to detect thermal radiation.
This suggests that the thermal radiation is not a consequence of lack of knowledge about half of the field.

A constantly accelerated observer can also exist where there are reflecting boundary conditions cutting through the Minkowski space, such that the full field is known.
This removes the lack of knowledge as the entire field is accessible.
Thermal radiation is also detected in this situation.
The detector and field exist in a pure state, meaning that the von Neumann entropy is zero \cite{Rovelli2012}.

Temperature is measured as a distribution over energy eigenstates, determined by the Hamiltonian operator.
For an accelerated detector, however, instead of measuring the Hamiltonian operator, the detector measures the boost operator because this generates translations in the detector's proper time.
The field state is not in an eigenstate of this observable, because this operator does not commute with the Hamiltonian operator.
It is suggested that instead of the von Neumann entropy, the appropriate entropy is the Shannon entropy resulting from the quantum uncertainty of measuring an observable when the field is not in an eigenstate of that observable \cite{Rovelli2012}.
This may lead to a better fundamental understanding of the Bekenstein black hole entropy through a better interpretation of the entropic quantities associated with the Unruh effect.

\subsection{Entanglement generation}
The Unruh-Hawking effect is normally seen as noise in the task that is being attempted.
However, the Unruh-Hawking effect can be used to generate entanglement.
Depending on the initial state of the field, entanglement between an inertial and an accelerated observer can actually be created by this effect \cite{Montero2011c}.
One can even create entanglement between causally disconnected observers under certain conditions by having them each interact with the vacuum \cite{Reznik2003,Lin2010}.
This entanglement does not persist, and is only created for specific ranges of times and separations of the causally disconnected observers.
Curvature can increase the amount of entanglement that is possible to extract from the vacuum using causally disconnected detectors \cite{Cliche2011}.

It is possible to create entanglement with field transformations.
This is demonstrated in cavity modes, by using non-uniform acceleration.
The entanglement generation can be increased by using single mode squeezing before the acceleration starts \cite{Friis2012a}.
Combining two cavities, one inertial and one accelerated, can generate entanglement between them.
This entanglement generation requires the correct initial state \cite{Friis2012}.
Simply boosting to a relativistic velocity can also generate entanglement \cite{Palge2012}.

Quantum computing uses entanglement in its algorithms.
Cluster state quantum computing relies on the creation of a very specific entangled state \cite{Nielsen2006}.
This state is then measured with single qubit measurements, which depend on previous outcomes to perform the computation.
Understanding how entanglement is generated can help with the design of quantum computers that use entanglement.

\subsection{Effects of motion}
The spin degrees of freedom are often used in quantum information processes because of the relative ease their use involves, both in the theoretical calculation of the details and practically manipulating them.
There have been spin based implementations of quantum computing \cite{Kane1998,Kloeffel2013}.
It is natural therefore, to study the effects of motion on spin entanglement.

Relative inertial motion always transforms the states unitarily.
There can be no loss of entanglement under boosts \cite{Alsing2002}.
However, when observers are moving relative to each other they must alter the angle of their spin measurements to be able to detect the entanglement that is present \cite{Terashima2002}.
This is due to the rotational nature of Lorentz boosts when applied to spins.
It does not maintain the entanglement perfectly, however, because the boost operation on spin alone is not unitary.
It is only when spin is viewed jointly with momentum that the complete transformation becomes unitary \cite{Gingrich2002}.

The spin rotation effect is similar in general relativistic situations.
The observers must correct for the rotations by altering their measurement bases.
However, this is challenging because the correction depends strongly on the observer's exact location \cite{Terashima2004}.
For inertial observers free falling into a Kerr-Newman black hole, the spin basis changes.
It is only possible to calculate the required corrections up to the horizon \cite{Said2010}.

\subsection{Effects of acceleration}
Inertial motion induces unitary transformations on the field modes.
These transformations are local to each mode and therefore keep the total entanglement invariant.
Conversely, acceleration degrades the entanglement through the Unruh effect.
Understanding how entanglement is degraded, or even perceived to be degraded is important in designing any experiment or device that uses entanglement.
Many communication protocols use entanglement.
We need to know if and how entanglement may be altered when the communicating parties are subject to acceleration.

For Dirac fields, this degradation approaches a finite non-zero value for infinite acceleration \cite{Alsing2006}.
Information tasks that require entanglement in a Dirac field are still possible as one observer approaches infinite acceleration.
This effect is not due to the limited modes available to fermions.
The entanglement survival is still present in fields that have an infinite number of accessible modes \cite{Martin-Martinez2009}.
However, not all fermionic fields have entanglement that survives to infinite acceleration, some do vanish in this limit \cite{Montero2011}.

The entanglement may be perceived to be degraded but this is not a complete erasure.
The Unruh effect can create entanglement between the causally disconnected observers Rob and AntiRob.
If Alice initially shares some entanglement with Rob, then the Unruh effect will just redistribute the entanglement as multipartite entanglement, sharing some with AntiRob \cite{Adesso2007}.

Cavity mode entanglement is also degraded by the Unruh effect.
When a cavity with entangled modes is accelerated for a finite but long time, the entanglement is degraded \cite{Bruschi2012}.
The acceleration does not need to last forever, like most models of the Unruh effect.
It is enough to start both observers as inertial, and gradually increase one observer's acceleration.
As expected, this results in a time dependent entanglement degradation \cite{Mann2009}.

Montero and Mart\'i{}n-Mart\'i{}nez developed a generalized formalism to study arbitrary spin fields in accelerated frames \cite{Montero2011a}.
They found that for the two examples studied, one bosonic and one fermionic, there was no significant difference in the infinite acceleration limit.

\subsection{Black hole effects}
Black holes behave in a very similar way to acceleration at sufficient proximity, as discussed in \cref{subsec:fields:rindlerapprox}.
Entanglement decreases between an inertial observer and those that hover near a black hole.
Bosonic modes retain no entanglement in the infinite acceleration limit \cite{Fuentes-Schuller2005}.

The Hawking radiation is the source of noise in the same way that Unruh radiation is for accelerated frames.
The effect of the Hawking temperature has been studied and found to degrade the entanglement \cite{Pan2008}.
The teleportation fidelity of scalar fields also decreases with the Hawking temperature.

For observers in the presence of a horizon, the correlations, both classical and quantum, become redistributed.
The bipartite entanglement between modes outside the horizon is converted into multipartite entanglement with modes behind the horizon \cite{Adesso2009}.
This is a redistribution of entanglement that follows certain conservation laws for fermionic fields.
Bosonic fields do not have an equivalent entanglement conservation law.
Both fermionic and bosonic fields obey a mutual information conservation law, which includes all correlations both classical and quantum \cite{Martin-Martinez2010b}.

The equivalence principle states that it is not possible to detect the difference between an accelerated frame and a curved spacetime, using only localized measurements.
An accelerated observer in Minkowski spacetime and a static observer hovering above a black hole both measure the same proper acceleration.
The responses of their detectors to fields in the state of Rindler vacuum and Schwarzschild vacuum respectively is zero, by definition of these vacuum states.
However, when comparing the responses of their detectors to the more physically realistic Minkowski vacuum and Unruh vacuum, the observer near a black hole will measure a higher temperature.
This violates the equivalence principle.
The principle is restored on the horizon in the infinite acceleration limit because both temperatures become infinite \cite{Singleton2011}.

Black holes are not eternal static objects.
We believe that they are created through a dynamic gravitational collapse process.
The process of collapse itself can create entanglement \cite{Martin-Martinez2010}.
The entanglement created is between the future and past, between the infalling radiation and future Hawking radiation.
Fermions are more sensitive to this entanglement generation than bosons.
The dynamical collapse was studied for non-trivial initial states of the field by Mart\'i{}n-Mart\'i{}nez \emph{et al.} \cite{Martin-Martinez2012b}.
This paper also introduces a formalism to study these dynamical systems.

The redistribution of entanglement may be an important part of understanding the black hole information paradox.
Fundamentally understanding how entanglement is affected by black holes, could shed some light into the evaporation process.
This may reveal what happens to the information that is apparently lost.

\subsection{Spacetime structure}
We know that the curvature of spacetime affects entanglement.
Instead of trying to protect the entanglement for use in communication, it is possible to use these effects to measure the spacetime itself.
A detector coupled to a field in an expanding (de Sitter) vacuum will observe thermal radiation.
It is by placing two spatially separated comoving detectors that the difference between an expanding universe and a thermal bath can be distinguished.
For certain configuration parameters, the detectors in a thermal bath become entangled, while the detectors in expanding spacetime do not \cite{Steeg2009}.

It is possible to go a step further.
A two dimensional expanding Freedman Robertson Walker universe with a scalar field was analysed by Ball \emph{et al.} \cite{Ball2006}.
Starting from a state where the modes of this field are separable, the entanglement entropy is used to calculate the expansion parameters of the universe.
In a similar set up, it is found that fermionic fields encode more information about the expansion than bosonic fields do \cite{Fuentes2010}.
This may help us measure and understand the fundamental structure of our own universe and local spacetime.

\subsection{Timelike entanglement}
The vacuum is a delocalized state that can exhibit entanglement between remote regions of the field.
An analytical framework to probe the spacetime structure of vacuum entanglement was developed \cite{Dragan2011}.
The entanglement exists between spacetime regions that are causally disconnected.

For a massless field, any signal travels along the light cone.
This means that the future and past regions inside a particular light cone are independent systems.
The vacuum state in these two regions is non-separable resulting in timelike entanglement, similar to the spacelike entanglement already discussed.
Some experiments have been proposed to investigate a detector switched on at a particular time that may register a thermal signal in response to this entanglement \cite{Olson2011}.

The extraction of this entanglement is also possible, with careful timing of switching on and off of the detectors.
If two detectors at the same spatial location interact with the field, at their specific times, they can become entangled at a particular time \cite{Olson2012}.

Two photons that do not exist at the same time can also be entangled \cite{Wiegner2011}.
This is due to two entangled atoms emitting the photons at different times, with the first photon being detected before the second is emitted.
This setup has been theoretically analysed and relies on the lack of spatial and temporal location knowledge of the photons.
They have been shown to violate Bell's inequalities and therefore exhibit entanglement.

Storage of quantum information is currently technologically challenging due to the decoherence of the qubits through interactions with their environment.
Timelike entanglement suggests a possible way to create a quantum information memory, where qubits can be `stored' using this entanglement and the storage of classical information.
This could benefit quantum computing greatly in certain situations where qubits must be stored for extended periods of time.

\section{Communication}
\label{sec:rev:communication}
The non-local correlations of entanglement make it an ideal resource for communication.
The study of quantum communication is sometimes called Quantum Shannon Theory.
It uses similar entropic quantities and brings together classical and quantum communication.
A good introduction to this topic is presented by Wilde in the book Quantum Information Theory \cite{Wilde2011}.

Quantum channels have a richer structure than classical channels.
Two zero-capacity quantum channels can be combined in a way to make the overall channel non-zero \cite{Smith2009}.
This implies that the channel capacity is not the only relevant quantity when studying these channels.

Communication can be described by the means of protocols.
Various procedures showing how to transfer information by using the communication resources.
These protocols can all be derived from more general protocols.
There are two main families, those derived from the mother protocol and those from the father protocol \cite{Hsieh2008}.
These families can be related by a transformation, when the mother protocol takes a certain form \cite{Abeyesinghe2009}.
To derive a child protocol such as state merging, all that is required is to add simple transformations before or after the main protocol.
A father protocol exists for broadcast channels that can be used to calculate capacity regions \cite{Dupuis2011}.

\subsection{Fundamental limitations of quantum information transfer}
There are some fundamental limitations to quantum information processing that are worth mentioning.
These are limitations that are not present in classical information.
The limitations must be investigated, to understand what is and is not possible to do with quantum information.
Quantum computing is not simply an extension of classical computing, but a fundamentally different process.

The no cloning theorem says that it is not possible to perfectly copy an unknown quantum state.
This follows directly from the principles of linearity and superposition \cite{Wootters1982}.
The no signalling theorem says that it is impossible to communicate any information using an entangled state alone.
To do so would enable faster than light communication, violating causality.
Any measurement that Alice makes on her half of the state, does not change Rob's measurement results, see Section 5.1 of \cite{Barnett2009}.

Combining these theorems leads to the no-summoning theorem.
Summoning is where Alice gives Rob a localized state, with the requirement that Rob give it back instantaneously at some unspecified time and place in the future.
Rob does not know where or when the state will be summoned.
Classically this is possible, however, it is not possible in quantum mechanics \cite{Kent2012}.
It is possible classically because Rob is able to clone the classical state and broadcast it to every point in the future.
Rob cannot clone a quantum state, he also cannot instantaneously request and receive it at the summons location without faster than light communication.

As well as these fundamental limitations to what can be achieved with quantum information, there are limitations based on the energy transfer.
All information transfer and storage requires energy, and there are limits to communication given a specific signal energy \cite{Bekenstein1990}.

\subsection{Communication with accelerated observers}
As seen in the entanglement section, there are a lot of interesting effects when one considers an inertial and an accelerated observer.
These effects persist into the study of quantum communication.
Fundamentally understanding communication in this simplest situation of non-inertial observers is a good starting point to attempting to understand communication in general relativistic situations.
It may also prove useful when designing communication systems between accelerating observers.

Early speculation on relativistic quantum communication involved causally disconnected counter-accelerating observers communicating using the vacuum state.
These observers obtain entanglement through the Unruh effect.
No true communication can occur as they are causally disconnected.
However, they are able to perform some protocols, for example the shared coin toss \cite{vanEnk2003}.
The causality restrictions limit the parties to only be able to jointly create identical classical information.

A more feasible setup is where the sending party, Alice, is inertial, and the receiving party, Rob, is accelerated.
Particles are created in the field modes to which Rob is sensitive due to the Unruh effect.
Kok and Braunstein developed a unified description of the transformations and interferometry required to effectively study these states \cite{Kok2006}.

The difference in entanglement between fermionic fields and bosonic fields does not translate to an improvement in quantum channel capacity.
When an inertial observer communicates with an accelerated observer, both the fermionic and bosonic channels' quantum capacities drop to zero in the limit of infinite acceleration.
The quantum channel capacity for this particular set of fermionic channels is zero even though there is entanglement that survives in the infinite acceleration limit.
This suggests that entanglement measures should not be used na\"ively as a direct substitute for the quantum channel capacity \cite{Bradler2010b}.

\subsection{Trade-off capacities of the Unruh channel}
The channel between an inertial and an accelerated observer is known as the Unruh channel, due to the noise being caused by the Unruh effect.
The trade-off capacity regions of the Unruh channel have been found in a few examples.
The trade-off capacity region is the region of the three dimensional space representing the consumption or creation of communication resources.
This was discussed for a few perfect examples in \cref{subsec:inf:quantumchannelcapacity}.
The single points of optimization for the noiseless protocols described previously extend to regions of achievable rate triples.
These have been plotted for the Unruh channel amongst others \cite{Jochym-OConnor2011,Wilde2012b}.
The Unruh channel is part of a larger class of channels called the Hadamard channels for which the capacity region is possible to calculate \cite{Bradler2010a}.

\subsection{Effects of localization}
Localization is a challenge for relativistic quantum information theory.
The mathematics generally becomes very difficult once the fields become localized.
Most people use plane waves or the accelerated equivalent, Unruh modes, to calculate capacities.
There are a number of techniques used to localize the fields involved in a communication protocol.

When one considers a localized beam, for example in the photon field, diffraction effects must be taken into account.
These diffraction effects mean that it is not possible for a detector to be placed perpendicular to the direction of propagation because the wavefront is curved.
This means that there can be no perfect photon qubits and no perfectly orthogonal states when using localized detectors.
However, an effective reduced density matrix formalism was introduced to deal with these states, which takes into account the particular detection scheme \cite{Peres2003a}.

One can localize the observers by using Unruh-DeWitt detectors, which are pointlike two-level systems coupled to a quantum field.
This channel capacity was calculated for various spacetime separations.
As expected, both the quantum and classical capacities were zero for spacelike separations \cite{Cliche2010}.

Unruh-DeWitt detectors are omnidirectional, which is not a very good model for communication between two localized observers since the communication must follow the path between them.
A refinement of the localization techniques is to model directional communication assuming that this is in the same direction as the acceleration.
It is then possible to use the paraxial wave approximation with a directed Gaussian field mode.
Rob can use a broadband detector to measure states that Alice sends.
This model has been used to calculate an example of continuous variable quantum key distribution \cite{Downes2012}.

Localization is extremely important in constructing models of physically realisable communication.
Absolutely everything that could communicate is localised to some degree.
We need to develop the field by using localised observers for the protocols.
This would enable us to make stronger claims to what is and isn't possible.

\subsection{Private communication}
As stated before, an important use of quantum communication is to establish private communication channels.
Relativity plays a part directly in the privacy of communication.
When an eavesdropper, Eve, is accelerated, the extra noise they observe allows two inertial observers to communicate privately.
This private capacity depends on the magnitude of Eve's acceleration \cite{Bradler2007,Bradler2009,Bradler2010}.

Relativistic quantum communication allows another form of privacy, which is spacetime dependence.
It is impossible classically to construct a channel where the receiver of the information must occupy a particular location in spacetime.
However, with the use of entanglement, real-time communication is achievable with a channel that can only be received from a specific location.
This ensures privacy, in that it is possible to know exactly where your communication partner is located \cite{Malaney2010}.

\subsection{Teleportation}
Teleportation is the most important quantum communication protocol because it involves all the aspects of quantum communication that set it apart from classical communication.
Practically, it is particularly important because it provides universal quantum computation when combined with single qubit gates \cite{Gottesman1999}.
This concept of using teleportation to accomplish quantum computation can be extended into a full model for quantum computing.
The model may involve performing a specific sequence of single qubit measurements on a particular type of state called cluster states \cite{Nielsen2006}.
The information in the computation is teleported from one qubit to the next in the cluster state when the previous qubit is measured.

Teleportation in a relativistic situation suffers from the same type of noise that standard quantum communication does.
When one of the observers is accelerated, the teleportation fidelity is reduced \cite{Alsing2003}.

Teleportation can be performed between two different observers that are separated in spacetime.
To obey the no-signalling theorem, Rob cannot extract any information about the teleported state until he receives the classical information from Alice about the result of the Bell measurement and performs the required rotations according to the teleportation protocol.
This means that Alice can teleport a state to Rob, at any point in her future light cone.
This is particularly useful when combined with the principle of timelike entanglement \cite{Olson2012}.
If Alice creates the entanglement, then performs her part of the teleportation protocol, storing the classical bits rather than sending them.
At a later time but at the same point in space, Rob can create his part of the timelike entanglement, and perform the rotation using the classical bits of information.
This effectively means that Rob has recalled the quantum state so timelike entanglement can be used as a quantum memory.
The storage of classical information is much easier than quantum information, so if timelike entanglement can be created, it can be used to store quantum information.

\subsection{Conclusion}
The field of relativistic quantum information may be young, but there are already many interesting developments.
It is also clear that there is still so much more to be discovered.
In this review, I have focussed on relativistic effects and phenomena.
The field could be interpreted as being much broader than I have presented it.
It relates to many other areas of physics, and is continuously evolving.
There are technological implications of the field that may have far reaching effects.
These implications are mainly found in the speed up that quantum computing provides.
Private communication using quantum channels would also replace the current technology.
Through studying this field we will have a better understanding of the fundamental nature of information in our universe.

\chapter[Communication near black holes]{Communication between agents near black holes}
\label{ch:blackholes}
Black holes are an interesting setting in which to test any relativistic effects.
The Schwarzschild metric is well known \cite{Schwarzschild1916}, and many consequences have been studied in this setting as discussed in \cref{ch:review}.
It is already known that quantum fields near black holes gain excitations in the form of Hawking radiation \cite{Hawking1975,Davies1975}.
This provides us with an interesting topic to see how the black hole and it's associated noise affects both classical and quantum information in a quantum channel.
This chapter is based on a paper written in collaboration with Pieter Kok and Carsten van de Bruck \cite{Hosler2012}.

\section{Communication setup}
\label{sec:bh:setup}
\begin{figure}[thp]
\begin{center}
  \frame{\includegraphics[scale=0.6]{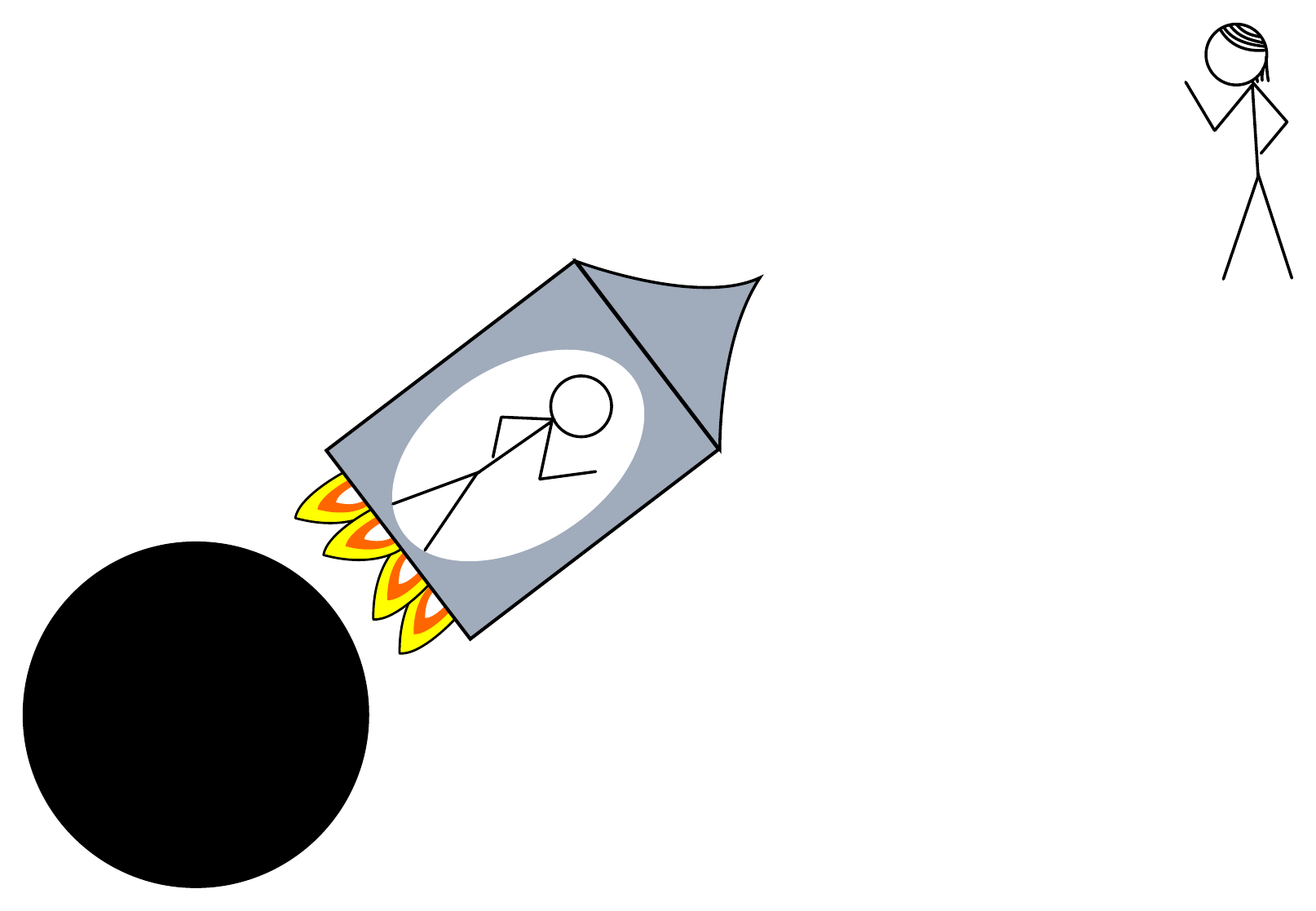}}
  \caption[Alice with Rob near a black hole]{This figure depicts Alice freefalling far away from the black hole. It also shows Rob, maintaining his distance by accelerating away from the black hole.\label{fig:bh:alicerobbh}}
\end{center}
\end{figure}

We have two observers in this situation.
Alice is an inertial observer free-falling into a Schwarzschild black hole, while Rob hovers at a fixed distance from the horizon.
This situation is depicted in \cref{fig:bh:alicerobbh}.
Alice wants to send a message to Rob.

We compare classical and quantum communication between Alice and Rob using two encoding methods of bosonic fields.
These encoding methods are single rail and dual rail, discussed in \cref{sec:inf:encoding}.
As Rob hovers above the horizon he describes the field with a different set of modes to those of Alice.
Here we take Rob to be close to the horizon and so we can approximate his modes with Rindler modes.
We also use the Single Wedge Mapping as explained in \cref{ch:SMA}.

Alice creates the appropriate field excitations, in her own frame for the message she is trying to send.
We transform the excitations that Rob will measure into his non-inertial frame of reference.
The transformation between Alice and Rob's modes is discussed in \cref{sec:fields:transformation}.
We end up with the field state as shared between Alice and Rob, $\rho_{AR}$.
This allows calculation of the information transferred using the entropic quantities discussed in \cref{subsec:inf:quantuminfomeasures}.

We start by giving the full state of the field as seen by the observers Alice and Rob in \cref{sec:bh:fieldstate}
Then we quantify the distinguishability between classical states that Alice sends in \cref{sec:bh:distinguishability}.
This allows us to get an idea of the maximum information transfer of the classical encodings.
Then we calculate the Holevo information and conditional entropy of all these encoding methods in \cref{sec:bh:channels}.
We also calculate the coherent information in \cref{sec:bh:channels}.
Finally we conclude the study on black hole communication in \cref{sec:bh:conclusion}.

\section{Resulting field state}
\label{sec:bh:fieldstate}
It is useful to explicitly give the state of the field for each of the encoding methods of this channel.
We start by stating the logical states Alice sends to Rob.
These are, \hyperref[eq:bh:classicallogical]{(a)}, classical bits and, \hyperref[eq:bh:quantumlogical]{(b)}, quantum qubits respectively,
\begin{subequations}
  \label{eq:bh:logicalmessage}
\begin{align}
  \label{eq:bh:classicallogical}
  \sigma_A(\mbox{general bit}) =& \modsq{\alpha} \Ket[A]{\zero}\Bra{\zero} + \modsq{\beta} \Ket[A]{\one}\Bra{\one},\\
  \label{eq:bh:quantumlogical}
  \rho_A(\mbox{general qubit}) =& \left(\alpha \Ket[A]{\zero} + \beta \Ket[A]{\one}\right) \otimes \left(\alpha^* \! \Bra[A]{\zero} + \beta^* \! \Bra[A]{\one}\right).
\end{align}
\end{subequations}

After the transformation discussed in \cref{sec:fields:transformation}, we end up with a full field state that is entangled between Alice, Rob and AntiRob.
As we do not care about AntiRob in this chapter, we trace out his part of the field.
Leaving us with all the information we require to calculate all the entropic quantities.
To obtain the separate field states of Alice and Rob individually, a partial trace is performed over the other observer's basis.
The full Alice and Rob field state in the single rail encoding is,
\begin{subequations}
  \label{eq:bh:singlerailfieldstate}
\begin{align}
  \label{eq:bh:singlerailclassical}
  \sigma_{AR}^{(s)} = \sum_{n=0}^\infty
                \Bigg[&\Ket[A]{\zero}\Bra{\zero}\otimes \Ket[R]{n}\Bra{n} \frac{\modsq{\alpha}}{\cosh^2 (r)} \tanh^{2n} (r) \nonumber \\
                     +&\Ket[A]{\one}\Bra{\one}\otimes \Ket[R]{n}\Bra{n} \frac{n \modsq{\beta}}{\cosh^4 (r)} \tanh^{2(n-1)}(r)\Bigg],
\end{align}
\begin{align}
  \label{eq:bh:singlerailquantum}
  \rho_{AR}^{(s)} = \sum_{n=0}^\infty
                \Bigg[&\Ket[A]{\zero}\Bra{\zero}\otimes \Ket[R]{n}\Bra{n} \modsq{\alpha} \frac{\tanh^{2n} (r)}{\cosh^2 (r)} \nonumber \\
                     +&\Ket[A]{\zero}\Bra{\one}\otimes \Ket[R]{n}\Bra{n+1} \alpha\beta^* \sqrt{n+1}\frac{\tanh^{2n} (r)}{\cosh^3 (r)} \nonumber \\
                     +&\Ket[A]{\one}\Bra{\zero}\otimes \Ket[R]{n+1}\Bra{n} \alpha^*\beta \sqrt{n+1}\frac{\tanh^{2n} (r)}{\cosh^3 (r)} \nonumber \\
                     +&\Ket[A]{\one}\Bra{\one}\otimes \Ket[R]{n+1}\Bra{n+1} \modsq{\beta} (n+1) \frac{\tanh^{2n} (r)}{\cosh^4 (r)} \Bigg].
\end{align}
\end{subequations}
It is helpful to note that both of the above density matrices are block diagonal, with each value of $n$ being one of the blocks.
This allows easy calculation of the partial traces and the diagonalisation.

In the dual rail encoding there are two field modes that Rob has access to, making the expression much longer.
To shorten it again, we use extra sums.
Now we have $n$, which is the extra noise in Rob's zero mode, $R_0$, $m$, which is the extra noise in Rob's one mode, $R_1$.
We also have $p$, the ket part of Alice's logical message and $q$, the bra part of Alice's logical message.
For the classical communication, $p=q$ at all times, simplifying the expression further.
Putting this all together, results in a field state given by,
\begin{subequations}
  \label{eq:bh:dualrailfieldstate}
\begin{align}
 \label{eq:bh:classicaldualrailstate}
  \sigma_{AR}^{(d)} = \sum_{n=0}^\infty &\sum_{m=0}^\infty \sum_{p=0}^1 |\alpha|^{2-2p}|\beta|^{2p} \frac{\tanh^{2n+2m}(r)}{\cosh^6(r)} \nonumber\\
       \times& (n+1)^{1-p} (m+1)^{p} \Ket[A]{p}\Bra{p}\nonumber\\
       \otimes& \Ket[{R_0}]{n+(1-p)}\Bra{n+(1-p)} \otimes \Ket[{R_1}]{m+p}\Bra{m+p},\\
 \label{eq:bh:quantumdualrailstate}
  \rho_{AR}^{(d)} = \sum_{n=0}^\infty &\sum_{m=0}^\infty \sum_{p=0}^1 \sum_{q=0}^1 \alpha^{1-p}\beta^p(\alpha^*)^{1-q}(\beta^*)^{q}\frac{\tanh^{2n+2m}(r)}{\cosh^6(r)} \nonumber\\
       \times& (n+1)^{\frac{2-p-q}{2}} (m+1)^{\frac{p+q}{2}} \Ket[A]{p}\Bra{q}\nonumber\\
       \otimes& \Ket[{R_0}]{n+(1-p)}\Bra{n+(1-q)} \otimes \Ket[{R_1}]{m+p}\Bra{m+q}.
\end{align}
\end{subequations}

To take the partial traces and diagonalise the classical dual rail state, \myeqref{eq:bh:classicaldualrailstate} is trivial as it is already diagonal.
The quantum dual rail state, \myeqref{eq:bh:quantumdualrailstate}, is not already diagonal.
The partial trace over $A$ is taken by setting $q=p$ and summing over $p$.
We already know that Alice's partial state should be \myeqref{eq:bh:logicalmessage}, however, we still take the partial trace over $R$ to verify this.

Diagonalising this is possible once you rearrange the basis order so the matrix becomes block diagonal.
Each block can be found by fixing the value of $n$ and $m$ and letting $p$ and $q$ vary.
This gives a $2\times2$ matrix as the block.
Each different set of values for $n$ and $m$ gives each different block.
Leaving $n$ and $m$ general we can solve the diagonalisation problem on a general block, allowing us to find the general eigenvalues dependant on $n$ and $m$.
We can then sum over these eigenvalues by summing over $n$ and $m$ to calculate any entropic quantity.

\section{Classical distinguishability}
\label{sec:bh:distinguishability}

If the receiver, Rob, is unable to tell the difference between the logical $\zero$ and the logical $\one$ that have been sent, then he cannot extract any information from the message.
Hence, it is useful to calculate the distinguishability between these states of the field, $\sigma_R(\zero)$ and $\sigma_R(\one)$, for both single and dual rail classical encoding methods.
To quantify how the channel reduces the distinguishability between the logical states sent by Alice, we use the probability of mistaking one for the other when performing a measurement.
When performing any measurement, the probability of getting the wrong result is given by the fidelity between the two possible states \cite{Jozsa1994},
\begin{equation}
 \label{eq:bh:fidelityfull}
 F = \left(\Tr{\sqrt{\sqrt{\sigma_R(\zero)}\sigma_R(\one)\sqrt{\sigma_R(\zero)}}}\right)^2.
\end{equation}
If the density matrices commute, the Fidelity takes the simpler form,
\begin{equation}
  \label{eq:bh:fidelitycommute}
  F = \left(\Tr{\sqrt{\sigma_R(\zero)\sigma_R(\one)}}\right)^2.
\end{equation}
This is discussed in full in \secref{subsec:inf:fidelity}

This is calculated using the states to which Rob has access.
In the single rail case this is given by \myeqref{eq:bh:singlerailclassical}, taking the logical $\zero$ term only for $\sigma_R(\zero)$ and logical $\one$ for $\sigma_R(\one)$.
These states are diagonal in the same basis, and therefore commute, so the fidelity is calculated using \myeqref{eq:bh:fidelitycommute},
\begin{equation}
 \label{eq:bh:singlerailfidelity}
F^{(s)} = \left(\sum_{n=0}^\infty \frac{\tanh^{2n-1}r}{\cosh^3r}\sqrt{n}\right)^2.
\end{equation}
This function is a convergent infinite sum that gets closer to a geometrical series as $n\to\infty$.

In the dual rail case we need to use the dual rail equivalent, which involves taking each term where Alice sends a pure logical $\zero$ ($p=0$) or $\one$ ($p=1$) from \myeqref{eq:bh:classicaldualrailstate}.
Similar to the single rail case, these are already both diagonal in the same basis, so we can calculate fidelity in the same way resulting in,
\begin{equation}
  \label{eq:bh:dualrailfidelity}
  F^{(d)} = \left(\sum_{n=0}^\infty \sum_{m=0}^\infty \frac{\tanh^{2n + 2m -2}r}{\cosh^6r}\sqrt{nm}\right)^2,
\end{equation}
which is a double sum.

Using Mathematica \cite{Mathematica} we were able to find analytical solutions for the Fidelities.
These are, for the single rail method,
\begin{equation}
  \label{eq:bh:fidelitysingleanalytical}
  F^{(s)}=\left[\frac{\Li_{-\frac12}(\tanh^2r)}{\sinh r \cosh^2 r} \right]^2,
\end{equation}
and for the dual rail method,
\begin{equation}
  \label{eq:bh:fidelitydualanalytical}
  F^{(d)}=\left[\frac{\Li_{-\frac12}(\tanh^2r)}{\sinh r \cosh^2 r} \right]^4,
\end{equation}
where $\Li_n(z)$ is the polylog function defined as,
\begin{equation}
  \label{eq:bh:polylogdefinition}
  \Li_n(z)=\sum_{k=1}^\infty \frac{z^k}{k^n}.
\end{equation}

We notice the Fidelity for the dual rail is exactly the square of the single rail Fidelity.
The dual rail method is therefore consistently more distinguishable than the single rail method.
For this purely classical transmission of just a known zero, and a known one the dual rail method has no links between each rail.
It can be thought of in this situation as two parallel single rails giving the square of the Fidelity.

Fundamentally, with the restriction to single excitations in each mode, the single rail method fully utilizes it's mode.
The dual rail however, has the possibility of using four fundamental states, that of $\ket{\varnothing,\varnothing}$, $\ket{\varnothing,1}$, $\ket{1,\varnothing}$ and $\ket{1,1}$.
In the dual rail method as used in optics and here, we restrict this to only the $\ket{\varnothing,1}$ and $\ket{1,\varnothing}$ states.
There is no problem with energy conservation in composite states using just these as there is always only ever one excitation.

Sometimes, the dual rail method is treated as two uses of the channel, we do not use this interpretation here.
In this thesis we treat the dual rail as equivalent to the single rail and not as two parallel single rail channels.
This is because we take the same view as optics in that the difficulty in realizing the communication comes not from making the modes available but from each usage of the modes that are set up.

\begin{figure}[tb]
  \begin{center}
      \includegraphics[width=272pt]{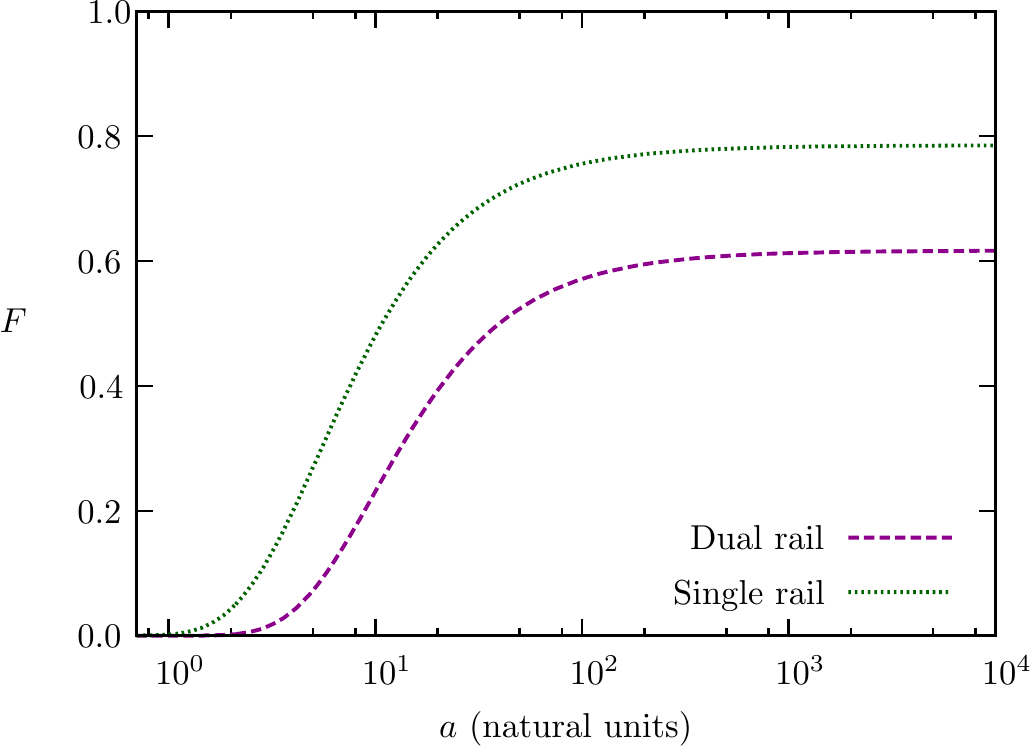}
  \caption[Fidelity between received zero and one]{Fidelity between the logical zero and one states received by Rob as a function of the acceleration $a$ for single and dual rail bits.}  \label{fig:bh:fidelity-of-zero-vs-one}
  \end{center}
\end{figure}

The squeezing parameter $r$ is given by,
\begin{equation}
  \label{eq:bh:r-func-of-a}
  \tanh r = \exp(-\omega\pi/a),
\end{equation}
where $\omega$ is the Rindler frequency of the mode being used.
This allows us to directly relate the fidelity back to Rob's proper acceleration $a$.
Using as an example $\omega=1$, the results are plotted in \figref{fig:bh:fidelity-of-zero-vs-one}.

For small $a$ the system approaches a perfect channel, indicated by a fidelity that is exponentially close to zero between the logical states.
As $a$ increases, the probability of Rob getting an incorrect measurement result increases, reducing the channel channel capacity.
Note that when $a\to\infty$ we find $F\to F_\infty<1$.
This is due to the persistent difference of a single photon between the two states that Rob receives.
Consequently, the classical channel capacity should never drop to zero.

\section{Communication channels}
\label{sec:bh:channels}

In this section we calculate and compare the Holevo information, conditional entropy and coherent information of single and dual rail classical and quantum communication.
Dual rail communication is physically symmetrical in the logical $\zero$ and logical $\one$ modes, which means that the maximum Holevo information is obtained when $\modsq{\alpha} = \modsq{\beta} =\frac12$.
For single rail encoding the situation is slightly more complicated so we do not assign values to $\alpha$ and $\beta$ yet.

The Holevo information, discussed in \cref{subsec:inf:quantuminfomeasures}, is equivalent to,
\begin{equation}
  \label{eq:bh:holevoinformation}
  I(A;R)_\sigma = S(\sigma_A)+S(\sigma_R)-S(\sigma_{AR}),
\end{equation}
where $S(\rho)$ is the von Neumann entropy, with respect to the classical quantum state,
\begin{equation}
  \label{eq:bh:classicalquantumstate}
  \sigma_{AR} = \sum_x p_A(x) |x\rangle_A\langle x| \otimes \sigma_{R_x},
\end{equation}
where $x \in \{0,1\}$ and $\sigma_{R_x}$ is the state Rob receives when Alice sends the logical value $x$.
All of the following calculations were done using partial traces of \cref{eq:bh:singlerailfieldstate,eq:bh:dualrailfieldstate}.

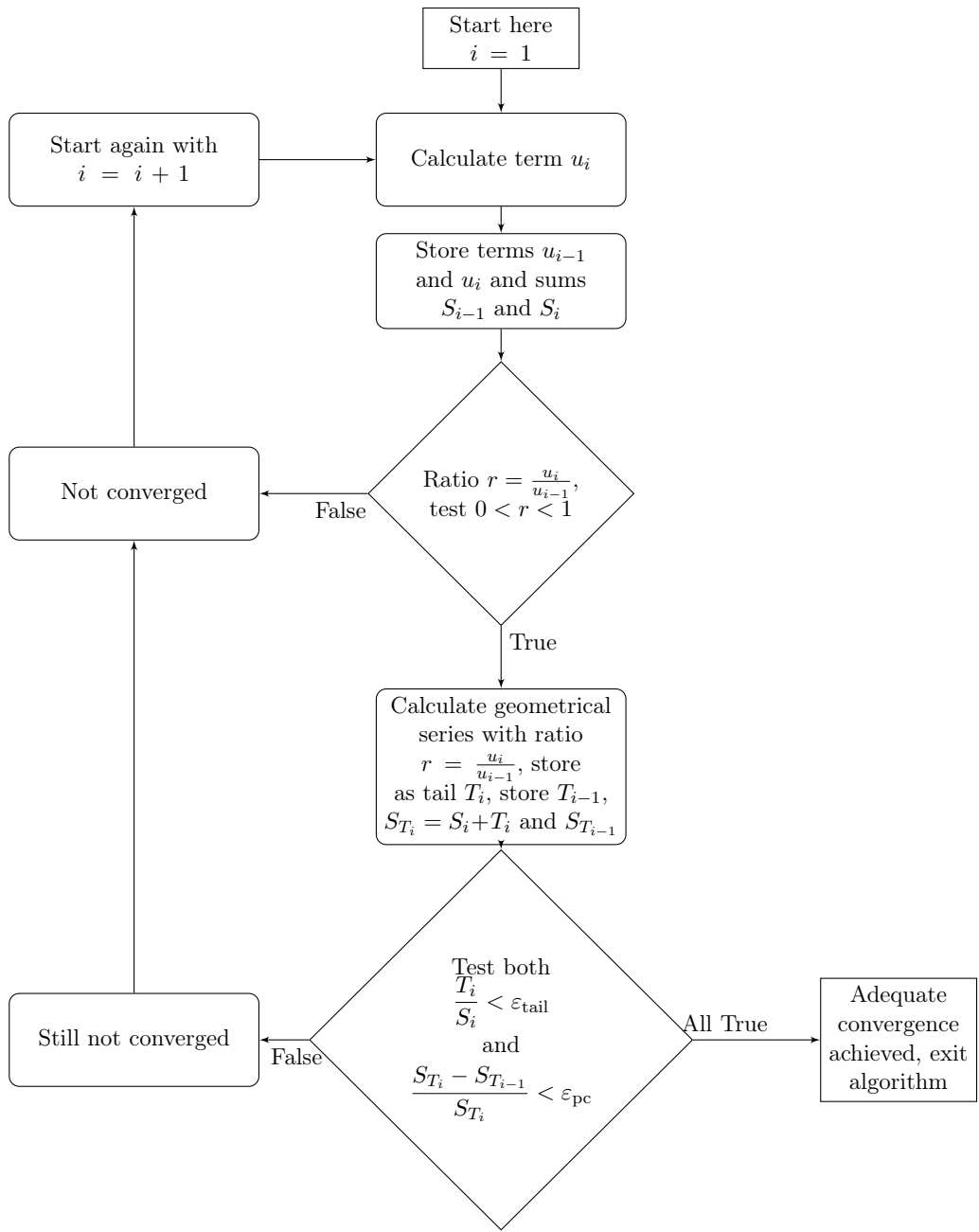
\begin{figure}[htp]
 \begin{center}
\begin{tikzpicture}[node distance=2cm, auto, scale=0.85, transform shape]
  \node[endpoint] (start) {Start here $i=1$};
  \node[block, below of=start, node distance=2cm] (calculateith) {Calculate term $u_i$};
  \node[block, below of=calculateith, node distance=2cm] (storeterms) {Store terms $u_{i-1}$ and $u_i$ and sums $S_{i-1}$ and $S_{i}$};
  \node[decision, below of=storeterms, node distance=3.5cm] (ratiodecide) {Ratio $r=\frac{u_i}{u_{i-1}}$, test $0 < r < 1$};
  \node[block, left of=calculateith, node distance=6cm] (newi) {Start again with $i=i+1$};
  \node[block, left of=ratiodecide, node distance=6cm] (notconverged) {Not converged};
  \node[block, below of=ratiodecide, node distance=4.5cm] (calculater) {Calculate geometrical series with ratio $r=\frac{u_i}{u_{i-1}}$, store as tail $T_i$, store $T_{i-1}$, $S_{T_i}=S_i+T_i$ and $S_{T_{i-1}}$};
  \node[decision, below of=calculater, node distance=4.5cm] (testtail) {Test both $\dfrac{T_i}{S_i}<\varepsilon_{\text{tail}}$\\ \vspace{0.1cm} and \\ \vspace{0.2cm} $\dfrac{S_{T_i}-S_{T_{i-1}}}{S_{T_i}}<\varepsilon_\text{pc}$};
  \node[block, left of=testtail, node distance = 6cm] (stillnotconverged) {Still not converged};
  \node[endpoint, right of=testtail, node distance=6.5cm] (finish) {Adequate convergence achieved, exit algorithm};
  \path [line] (start) -- (calculateith);
  \path [line] (calculateith) -- (storeterms);
  \path [line] (storeterms) -- (ratiodecide);
  \path [line] (ratiodecide) -- node[decision answer] {True} (calculater);
  \path [line] (ratiodecide) -- node[decision answer] {False} (notconverged);
  \path [line] (notconverged) -- (newi);
  \path [line] (newi) -- (calculateith);
  \path [line] (calculater) -- (testtail);
  \path [line] (testtail) -- node[decision answer] {False} (stillnotconverged);
  \path [line] (testtail) -- node[decision answer] {All True} (finish);
  \path [line] (stillnotconverged) -- (notconverged);
\end{tikzpicture}
\end{center}
 \caption[Convergence test algorithm]{\label{fig:bh:convergence} Flow chart depicting convergence test algorithm. The error tolerances are given by $\varepsilon_{\text{tail}}$ and $\varepsilon_{\text{pc}}$.}
\end{figure}

\begin{figure}[ht]
  \begin{center}
      \includegraphics[width=272pt]{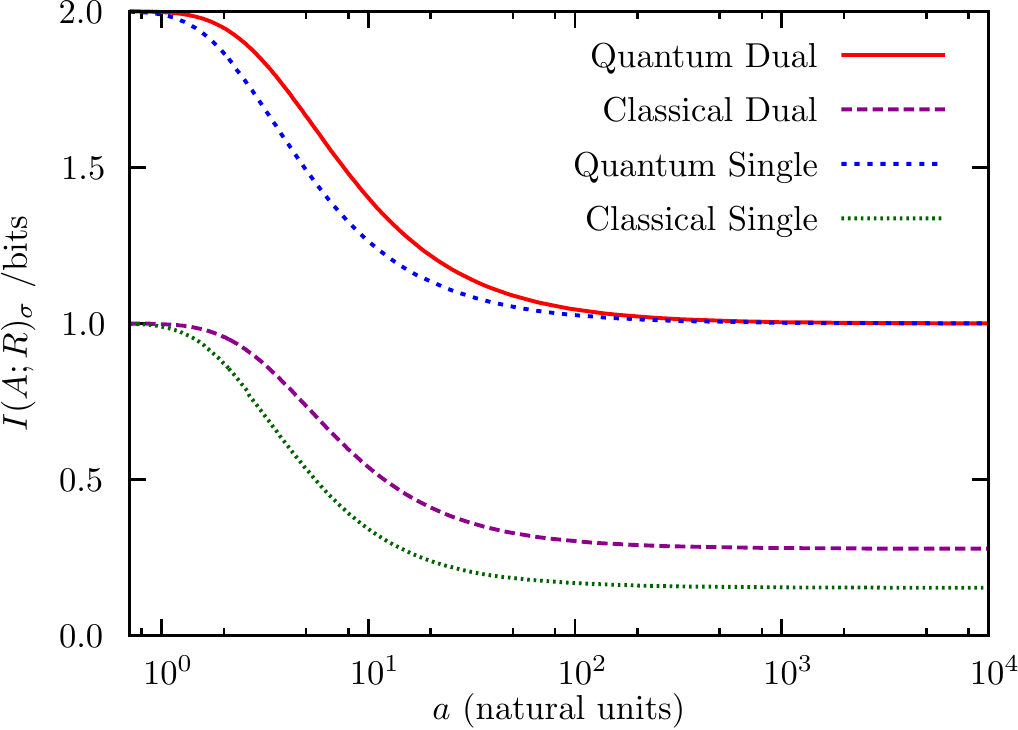}
  \caption[Holevo information]{Holevo information as a function of the  acceleration $a$ when $\modsq{\alpha} = \modsq{\beta} =\frac12$.}
  \label{fig:bh:ME-compare-single-dual-mixed-full}
  \end{center}
\end{figure}

We calculate the Holevo information for the classical case in the single rail encoding, $\sigma_{AR}^{(s)}$, as,
\begin{align}
\label{eq:bh:MIclassicalsingle}
S\left(A;R\right)_\sigma^{(s)} = & -\modsq{\alpha} \lb{\modsq{\alpha}} - \modsq{\beta}\lb{\modsq{\beta}} \\
                              & - \sum_{n=0}^\infty \Bigg[ \frac{\modsq{\alpha}}{\cosh^2r} \tanh^{2n}r \lb{1+\frac{n \modsq{\beta}}{\modsq{\alpha}\sinh^2r}} \nonumber \\
                              &\qquad \quad + \frac{n \modsq{\beta}}{\cosh^2r\sinh^2r}\tanh^{2n}r \lb{1+\frac{\modsq{\alpha}\sinh^2r}{n \modsq{\beta}}} \Bigg] ,\nonumber
\end{align}
The Holevo information for the classical case in the dual rail encoding, $\sigma_{AR}^{(d)}$, having substituted the optimal by symmetry parameter values $\modsq{\alpha} = \modsq{\beta} =\frac12$, is calculated as,
\begin{align}
  \label{eq:bh:MIclassicaldual}
S\left(A;R\right)_\sigma^{(d)}= 1 - \sum_{p=0}^\infty  \frac{\tanh^{2p}r}{\cosh^6r} \sum_{q=0}^p (q+1)\lb{1+\frac{p-q}{q+1}},
\end{align}
For simplicity when plotting we use the parameters $\modsq{\alpha} = \modsq{\beta} =\frac12$ for the single rail encoding as well as the dual rail encoding.
Even though these parameter values are not optimal they provide representative behaviour of all information measures considered here and allow a direct comparison with dual rail encoding.
These series were numerically calculated using the convergence test algorithm detailed in \figref{fig:bh:convergence}.
The Holevo information for each encoding are plotted together in \figref{fig:bh:ME-compare-single-dual-mixed-full}.
For completeness the figure also includes the mutual information of the quantum encodings.

\begin{figure}[ht]
  \begin{center}
      \includegraphics[width=272pt]{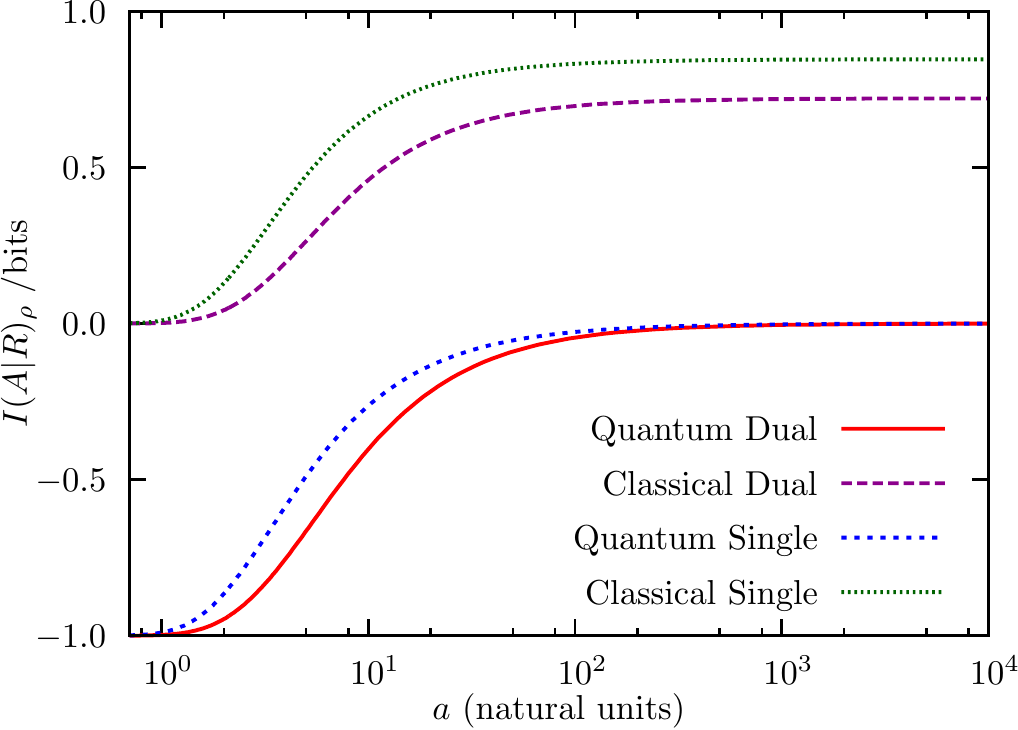}
  \caption[Conditional entropy]{Conditional Entropy as a function of the acceleration $a$ when $\modsq{\alpha} = \modsq{\beta} =\frac12$.}
  \label{fig:bh:conditionalentropy}
  \end{center}
\end{figure}

We calculate the conditional entropy for the quantum case in the single rail encoding, $\rho_{AR}^{(s)}$, as,
\begin{align}
  \label{eq:bh:CEquantumsingle}
S\left(A|R\right)_\rho^{(s)} = & - \sum_{n=0}^\infty \left[ \frac{\modsq{\alpha}}{\cosh^2r} \tanh^{2n}r
                                        \lb{\tanh^2r \left(\frac{\modsq{\alpha}\cosh^2r + \modsq{\beta}(n+1)}
                                            {\modsq{\alpha}\sinh^2r + \modsq{\beta}n}\right)}\right. \nonumber \\
                                 & + \left. 2n \tanh^{2n}r \frac{\modsq{\beta}}{\cosh^2r} \left( \frac{n+1}{\cosh^2r}-\frac{n}{\sinh^2r}\right) \lb{\tanh r} \right. \nonumber \\
                                 & + \frac{\modsq{\beta}}{\cosh^2r} \tanh^{2n}r \frac{n+1}{\cosh^2r} \lb{\frac{\modsq{\alpha}}{\cosh^2r}+\frac{\modsq{\beta}(n+1)}{\cosh^4r} } \nonumber \\
                                  & - \left. \frac{\modsq{\beta}n}{\sinh^2r\cosh^2r} \tanh^{2n}r\lb{\frac{\modsq{\alpha}}{\cosh^2r} + \frac{\modsq{\beta}n}{\sinh^2r\cosh^2r}}  \right].
\end{align}
The conditional entropy for the dual rail quantum case, $\rho_{AR}^{(s)}$, using the optimal by symmetry parameter values $\modsq{\alpha} = \modsq{\beta} =\frac12$, is given by
\begin{align}
  \label{eq:bh:CEquantumdual}
S\left(A|R\right)_\rho^{(d)}= - \sum_{p=0}^\infty \frac{\tanh^{2p}r}{\cosh^6r} \lb{\frac{p+2}{p+1}} \sum_{q=0}^p (q+1).
\end{align}
Using the same parameter values for the plot of the single rail case, we obtain the conditional entropy shown in \figref{fig:bh:conditionalentropy} again using numerical calculations and the convergence test described in \figref{fig:bh:convergence}.

The Holevo information graph, \figref{fig:bh:ME-compare-single-dual-mixed-full}, starts at one for both classical communication methods, showing that Rob learns the full bit of Alice's state when there is no squeezing.
We also find that as the squeezing increases, the information Alice shares with Rob decreases.

For quantum communication, we see that the conditional entropy, \cref{fig:bh:conditionalentropy}, starts at $-1$ in the absence of acceleration ($a=0$).
This means that Alice and Rob share one qubit of entanglement, which allows them to share a qubit in the future using additional classical communication.
The conditional entropy tends to zero when $a\to\infty$.
Note that we have not taken relativistic time dilation into account, which will affect the \emph{rate} of each communication channel in equal measure.

\begin{figure}[tb]
  \begin{center}
      \includegraphics[width=272pt]{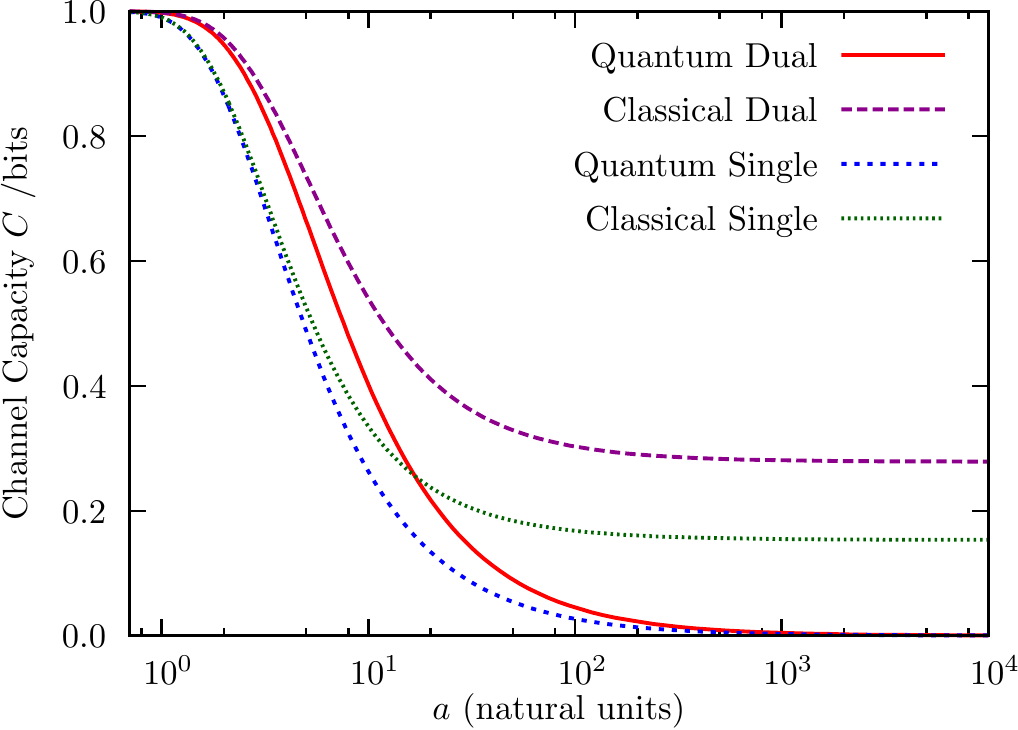}
  \caption[Restricted channel capacities]{The restricted channel capacity $C$ as a function of the acceleration $a$ when $\modsq{\alpha} = \modsq{\beta} =\frac12$.
          The figure does not take into account relativistic time dilation, which reduces all communication \emph{rates} to zero equally.
          While the classical restricted channel capacity reaches a finite value at the Schwarzschild horizon, the quantum restricted channel capacity (also measured in bits) drops to zero.
          Dual rail encoding outperforms single rail encoding in both classical and quantum communication.}
  \label{fig:bh:compare-capacities}
  \end{center}
\end{figure}

Next, we compare the maximum Holevo and coherent information, representing the restricted channel capacity within these encodings.
See \cref{subsec:inf:quantumchannelcapacity} for a detailed discussion of our use of the term `restricted capacity'.
The restricted classical capacity measures the amount of recoverable classical bits, while the restricted quantum capacity in this situation measures the entanglement that is generated using the state merging protocol, see \cref{subsec:inf:statemerging}.
This quantum communication method requires unlimited classical communication due to the state merging protocol.
The comparison of all four restricted capacities is shown in \cref{fig:bh:compare-capacities}.
When Rob is hovering far away from a black hole, there is little Unruh-Hawking noise ($a\simeq 0$), and all encoding methods result in communication equal to one bit.
Hovering closer to the black hole increases the noise and consequently reduces the restricted capacity.
Both classical communication restricted capacities tend towards finite values in the limit of large $a$, indicating that even when the communication \emph{rate} drops to zero due to time dilation, each channel use that reaches Rob carries a finite amount of information.
This is explained by the fidelity between the logical bit states sent by Alice, which never reaches one when $a\to\infty$ (see \cref{fig:bh:fidelity-of-zero-vs-one}).

On the other hand, the quantum restricted capacity falls off exponentially with $r$ as Rob hovers closer to the Schwarzschild horizon ($C \propto \exp(-\gamma r)$ for large $r$, with $\gamma\approx 2$ for the dual rail quantum channel).
Relating this back to the acceleration and distance from the horizon, we find that the quantum restricted capacity fully drops to zero for infinite acceleration, or at the horizon.
This is a fundamental difference between classical and quantum communication.
Classical communication relies on the occupation number, where, even at infinite acceleration, there is always a difference.
Quantum communication however relies on coherence, which gets completely washed out at infinite acceleration.

\section{Conclusion}
\label{sec:bh:conclusion}

In conclusion, we studied classical and quantum restricted communication channel capacities between an inertial observer Alice and an accelerated observer Rob, using both single and dual rail encoding of a bosonic field.
We found that the quantum restricted channel tends to zero with increasing acceleration (i.e., approaching the horizon).
The classical restricted channel remained finite arbitrarily close to the Schwarzschild horizon, indicating that statistical correlations still exist between Alice and Rob even in the limit of infinite acceleration, whereas quantum correlations are fully removed.
In both cases the channel degradation is due to Unruh-Hawking noise, and in both cases we ignored the time dilation that affects the rate of all communication channels equally.
In both classical and quantum communication, the dual rail encoding outperformed the single rail in this setting.

The quantum correlations seem to be more susceptible to the type of noise introduced by spacetime curvature.
Further studies that use quantum communication in this type of situation must take these effects into account.
The classical statistical correlations survive, suggesting that they are fundamentally stronger and more robust.
The distinguishability results are highly relevant to the classical communication.
Quantum communication does require distinguishable basis states, so it is important to know that the bases are always distinguishable by a finite amount.

There are many situations currently that require communication whilst taking relativistic effects into account.
With the introduction of more satellite based experiments and space exploration this will only increase.
Any quantum communication technology that utilizes these quantum correlations will need to be aware of the limitations.
The fact that quantum communication is not possible at the horizon of a black hole should not pose any huge problem for current technology but future designs should take it into account.
Applications where the quantum communication is highly important will perhaps need to modify the external situation so that it is not subject to large gravitational fields.

The channels studied here are restricted to using either the single rail or dual rail encodings.
One effect of this is that the communication basis is firmly fixed as the number basis for both encoding schemes.
We store the information in the amplitude and phase between these basis states, so it is feasible that a different basis may improve on the situation.
The limitation to number basis may prohibit some technique where quantum correlations survive in the limit of the horizon.
We could also use some of the ideas from quantum metrology and use some variations on the commonly used states.
Extending the basis encoding states to multi-excitation states, perhaps NOON states may assist in improving the channel capacities.
This is a topic of future research.

\chapter{Parameter estimation over a quantum channel}
\label{ch:measure}
We now take our study of black holes to a new area, that of parameter estimation.
Parameter estimation is the study of precision measurements, the amount of information about a physical parameter it is possible to extract with measurement.
Any physical measurement of a continuous parameter can be thought of in two stages: an encoding stage and a measurement stage.
First we encode the parameter onto the state of a system (in our case we are looking at qubit systems).
Then we measure the system to try and find the value of the parameter.
In this chapter we study this parameter estimation system, where the measurement is performed by Rob hovering close to the black hole horizon.
This chapter is partly based on a paper written in collaboration with Pieter Kok \cite{Hosler2013}.

\section{Quantum Fisher information}
\label{sec:meas:fisherinfo}
The quantum Fisher information was originally derived by Braunstein and Caves \cite{Braunstein1994} from considerations on optimally resolving nearby quantum states.
They defined a Riemannian metric using a generalization of the statistical distance to quantum states.
The following derivation has been adapted from \cite{Kok2010} to find the Cram\'er-Rao bound and the Fisher information.

\subsection{Cram\'e{}r-Rao bound and Fisher information}
To find the parameter, we must make a measurement on the encoded state.
This measurement will have a measurement outcome $x$ which will in general be different to the parameter $\theta$.
The measurement outcomes will be related to the parameter by a probability distribution (or probability density distribution) $p(x|\theta)$.
Once we have the measurement outcome, we construct an estimator $T(x)$ for $\theta$ with an expected value,
\begin{align}
  \label{eq:meas:estimator}
  \Braket{T}_\theta= \int \ud x\, p(x|\theta) T(x).
\end{align}
If the expected value of the estimator is equal to $\theta$, the estimator is unbiased.

We want to quantify the expected error on the parameter $\theta$ so to begin with, we take the error on the estimator $\Delta T=T(x)-\Braket{T}_\theta$.
The expected error on the estimator for $N$ independent measurements is always zero from the definition and can be written,
\begin{align}
  \label{eq:meas:expectederrorestimator}
  \int \ud x^N p(x_1|\theta) \cdots p(x_N|\theta) \Delta T = 0.
\end{align}
We start to relate back to $\theta$ by taking the derivative with respect to $\theta$, using the chain rule on $\Delta T$ and collecting terms together,
\begin{align}
  \label{eq:meas:derivTerror}
  \int \ud x^N p(x_1|\theta) \cdots p(x_N|\theta)\left(\sum_{i=1}^N \frac{1}{p(x_i|\theta)} \frac{\partial p(x_i|\theta)}{\partial \theta}\right) \Delta T = \Braket{\frac{\ud \Braket{T}_\theta}{\ud \theta}}.
\end{align}

A study of the size of errors usually involves the variance rather than the expected value.
This is because the expected value of the error is zero and the expected value of the variance gives a good measure for the size of the error.
We can square the equation to give us the variance, and apply the Cauchy-Schwarz inequality,
\begin{align}
  \label{eq:meas:cauchy-schwarz}
  \modsq{\Braket{f,g}}\le \Braket{f,f} \Braket{g,g},
\end{align}
where $\Braket{a,b}$ is the inner product.
In this case the inner product is defined as,
\begin{align}
  \label{eq:meas:innerproductforprobabilities}
  \Braket{a,b}=\int \ud x^N p(x_1)\cdots p(x_N) a(x^N) b(x^N),
\end{align}
where $a(x^N)$ and $b(x^N)$ are quantities potentially depending on all values of $x$.

Taking the substitutions that are useful for \cref{eq:meas:derivTerror},
\begin{subequations}
\begin{align}
  \label{eq:meas:substitutions}
  f=& \sum_{i=1}^N \frac{1}{p(x_i|\theta)} \frac{\partial p(x_i|\theta)}{\partial \theta}, \\
  g=& \Delta T,
\end{align}
\end{subequations}
we find an inequality expressed in terms of the variance, $(\Delta T)^2$,
\begin{align}
  \label{eq:meas:inequalityforvariance}
  \Braket{\sum_{i=1}^N \frac{1}{p(x_i|\theta)} \frac{\partial p(x_i|\theta)}{\partial \theta},\sum_{i=1}^N \frac{1}{p(x_i|\theta)} \frac{\partial p(x_i|\theta)}{\partial \theta}} \Braket{\Delta T,\Delta T} \nonumber \\
        \ge \Braket{\sum_{i=1}^N \frac{1}{p(x_i|\theta)} \frac{\partial p(x_i|\theta)}{\partial \theta},\Delta T}= \modsq{\Braket{\frac{\ud \Braket{T}_\theta}{\ud \theta}}}.
\end{align}

The first inner product can be simplified because each $x_i$ is independent, so the expected value over $N$ of them is equal to $N$ times the expected value for one.
We define this expected value for a single measurement to be the classical Fisher information,
\begin{align}
  \label{eq:meas:fisherinfodefinition}
  \mathscr{F}(\theta)\equiv \int \ud x \frac{1}{p(x|\theta)}\left( \frac{\partial p(x|\theta)}{\partial \theta}\right)^2.
\end{align}
This can be used to simplify \cref{eq:meas:inequalityforvariance} greatly,
\begin{align}
  \label{eq:meas:simpleinequality}
  N \mathscr{F}(\theta) \Braket{(\Delta T)^2}_\theta \ge \modsq{\frac{\ud \Braket{T}_\theta}{\ud \theta}}.
\end{align}

Finally this needs to be related back to the expected variance in $\theta$.
The relationship between the variances for $T$ and $\theta$ is,
\begin{align}
  \Braket{(\Delta T)^2}_\theta = \modsq{\frac{\ud \Braket{T}_\theta}{\ud \theta}} \left(\Braket{(\Delta \theta)^2}_\theta-\Braket{\Delta \theta}_\theta^2 \right).
\end{align}
For an unbiased estimator, giving unbiased values for $\theta$, we can eliminate $\Braket{\Delta \theta}_\theta^2$ immediately.
This gives us the Cram\'e{}r-Rao bound,
\begin{align}
  \label{eq:meas:cramerraubound}
  \Braket{(\Delta \theta)^2}_\theta \ge \frac{1}{N \mathscr{F}(\theta)}.
\end{align}
The Cram\'e{}r-Rao bound is a minimum uncertainty that any possible encoding and measurement scheme can achieve for the measurement of a continuous parameter.
In the next section we derive a quantum version of the Fisher information in the way it relates to the distances between density operators.
We find that the information extractable is represented by the quantum Fisher information.

\subsection{Distance between density operators}
\label{subsec:meas:distancebetweendensityoperators}
Any parametrized state can be represented as a path through the state space of the system with the parameter varying along the path.
The ease of determining the parameter is related to how easy it is to distinguish the states of the system.
This geometrical formulation was originally calculated by Braunstein \cite{Braunstein1996a}, the reproduction here follows the notation of Kok and Lovett \cite{Kok2010}.
It is helpful first to define a distance between quantum states.
A distance function $s(a,b)$ must obey the following axioms,
\begin{enumerate}
  \label{eq:meas:distanceaxioms}
  \item $s(a,b)\ge0$.
  \item $s(a,b) = 0$ iff $a=b$.
  \item $s(a,b) = s(b,a)$.
  \item $s(a,c) \ge s(a,b) + s(b,c)$.
\end{enumerate}

The state space of the system is a form of generalized vector space.
We can include a distance function on the vector space.
The vector space can be extended to a metric space if we define the distance infinitesimally using the metric,
\begin{align}
  \label{eq:meas:metric}
  \ud s^2 = d_{\mu\nu} \ud x^\mu \ud x^\nu.
\end{align}

It is helpful to point out at this stage that the vectors in this vector space are the states of the system represented by density matrices, $\rho$.
The derivatives $\ud \rho$ form the tangent vectors. 
The relevant dual space to this vector space are zero expectation Hermitian observables $\hat{A}_0=\hat{A}-\Braket{A}$.

We can use the metric and its inverse to convert between a vector and its dual.
In this case, the natural quadratic form for the duals is the correlation.
The correlation is also an observable and therefore must be Hermitian.
We therefore use the expectation value of the anti-commutator.
The quadratic form between duals involves the inverse metric,
\begin{align}
  \label{eq:meas:metriconduals}
  A_\mu B_\nu d^{\mu\nu} = \frac12 \Braket{\left\{\hat{A},\hat{B} \right\}} = \frac12 \Tr{\left\{\hat{A},\hat{B}\right\}\rho}.
\end{align}
This inverse metric can also be interpreted as a raising operator for the indices.

We can write this correlation in a different form by using the cyclic property of the trace, and having the inverse metric act as a raising operator on only one of the Hermitian operators
\begin{align}
  \label{eq:meas:raisinguseincorrelation}
  \frac12 \Tr{\left\{\hat{A},\hat{B}\right\}\rho} = \frac12 \Tr{\hat{A}\left\{\rho,\hat{B}\right\}} \equiv \Tr{\hat{A} \,\mathscr{R}_\rho(\hat{B})},
\end{align}
where the raising superoperator $\mathscr{R}_\rho(\hat{B})$ is defined as
\begin{align}
  \label{eq:meas:raisingdefinition}
  \mathscr{R}_\rho(\hat{B}) = \frac12 \left\{\rho,\hat{B}\right\}.
\end{align}
Due to the fact that this operator appears inside a trace, we can write the density operator $\rho$ in its diagonal basis.

For the distances we actually need the lowering operator form of the metric $\mathscr{L}_\rho(\hat{B})$ which can be calculated as the inverse of the raising operator.
Using the diagonal basis where $\rho = \sum_j p_j \Ketbra{j|j}\junktofixketbra$ this takes the form \cite{Braunstein1994}
\begin{align}
  \label{eq:meas:loweringdefinition}
  \mathscr{L}_\rho(\hat{B})= \mathscr{R}_\rho^{-1}(\hat{B}) = \sum_{jk} \frac{2 B_{jk}}{p_j + p_k} \Ketbra{j|k}\junktofixketbra.
\end{align}

From this we can define the infinitesimal distance between $\rho$ and $\rho+\ud\rho$ using this form of the metric as
\begin{align}
  \label{eq:meas:metricdistance}
  \ud s_\rho^2 = d_{\mu\nu}\ud x^{\mu} \ud x^{\nu} = \Tr{\ud \rho \mathscr{L}_\rho (\ud \rho)}.
\end{align}
This form can relate straight back to the Fisher information by dividing by $\ud \theta^2$ \cite{Braunstein1994}
\begin{align}
  \label{eq:meas:Fisherasdistance}
  \mathscr{F}(\theta)= \left(\frac{\ud s_\rho}{\ud \theta}\right)^2 = \Tr{\rho' \mathscr{L}_\rho(\rho')} \mbox{ where } \rho'=\frac{\ud\rho}{\ud \theta}.
\end{align}

From considerations on the Heisenberg equation of motion, the Fisher information is bounded above by the variance of the generator of translations in $\theta$
\begin{align}
  \label{eq:meas:Fisherupperbound}
  \mathscr{F}(\theta)\le \frac{4}{\hbar^2} \left(\delta \hat{K}\right)^2,
\end{align}
where $\hat{K}$ is the generator of translations in $\theta$.
This bound gives a physical limitation to the precision of any quantum measurement and leads to a derivation of the Heisenberg uncertainty principle \cite{Braunstein1996}.

\section{Estimation of relative amplitude}
\label{sec:meas:setup}
We have the same two observers as in the previous chapter.
Alice is an inertial observer, free-falling into a Schwarzschild black hole.
Rob is an accelerated observer, hovering a small distance above the horizon.
As Rob is hovering close enough to the horizon, we approximate Rob as a Rindler observer where his proper acceleration is equal in both the Schwarzschild and Rindler spacetimes.
We will use Rindler modes for Rob, and therefore it is useful for us to use Unruh modes for Alice.
We also make use of the Single Wedge Mapping (SWM, discussed in \cref{ch:SMA}).

The task here is for Rob to attempt to determine the value of a continuous parameter $\theta$ encoded onto a state in the possession of Alice.
Alice must first communicate the parametrized state with Rob.
This communication is represented by the transformation that was discussed in \cref{sec:fields:transformation}.
They must then estimate the parameter, either by Rob measuring the state alone, or by Alice and Rob performing a joint measurement on their respective subsystems.
We compare the amount of information they learn about the parameter using each of the possible ways of performing this task.

The parameter $\theta$ is either encoded onto a quantum state or a classical-quantum state of Alice's qubit.
These are given in the logical basis of the qubit onto which we encode it as,
\begin{subequations}
\begin{align}
  \label{eq:meas:quantumparameterised}
  \ket{\psi(\theta)}=& \cos\theta\ket{\zero}+\sin\theta\ket{\one}, \\
  \label{eq:meas:classicalparameterised}
  \sigma(\theta)=& \cos^2 \theta \Ketbra{\zero|\zero}\junktofixketbra + \sin^2\theta \Ketbra{\one|\one}\junktofixketbra.
\end{align}
\end{subequations}
In communicating the state to Rob, Alice can use either of the two encoding schemes discussed in \cref{sec:inf:encoding}, single rail or dual rail.
The communication can create entanglement between Alice and Rob's respective subsystems.

To summarise, there are three options in this task.
One is whether the parameter is encoded into a quantum state or a classical-quantum state of the qubit.
Another is whether Alice uses single rail, or dual rail encoding to communicate this qubit to Rob.
Finally, Rob may either perform a measurement by himself, or Alice and Rob perform a joint measurement on both of their subsystems.

\section{Mathematical analysis of amplitude estimation}
\label{sec:meas:mathsanalysis}
The quantum states held by Alice and Rob are almost identical to those in the previous chapter.
The only difference, is that we have now parametrized the coefficients in the qubit.
We use the states given in \cref{eq:bh:singlerailfieldstate,eq:bh:dualrailfieldstate} with the replacements
\begin{subequations}
\begin{align}
  \label{eq:meas:alphabetareplacements}
  \alpha \to& \cos\theta, \\
  \beta \to& \sin\theta.
\end{align}
\end{subequations}

\subsection{Single rail encoding}
\label{subsec:meas:singlerailamplitudemaths}
First, let us consider the single rail encoding method.
Taking the classical-quantum state \cref{eq:bh:singlerailclassical}, we separate the task into two cases:
Rob performing the measurement alone, and both Alice and Rob performing a joint measurement.
The state required for the joint measurement, where both Alice and Rob measure the single rail classical state to determine the value of $\theta$, is exactly \cref{eq:bh:singlerailclassical}.
This is already diagonal, so we can use the simplified form of the Fisher information, \cref{eq:inf:simplefisher}.
We first need the eigenvalues, two for each value of $n$
\begin{subequations}
\begin{align}
  \label{eq:meas:singleclassicalAReigenvalues}
  (\lambda_{1})_n =& \cos^2\theta \frac{ \tanh^{2n} (r)}{\cosh^2 (r)}, \\
  (\lambda_{2})_n =& n \sin^2\theta \frac{ \tanh^{2(n-1)}(r)}{\cosh^4 (r)}.
\end{align}
\end{subequations}
Differentiating these with respect to $\theta$ gives the following for the eigenvalues of $\sigma'$
\begin{subequations}
\begin{align}
  \label{eq:meas:SCARrhoprimeeigenvalues}
  (\lambda'_{1})_n =& - 2 \cos\theta \sin\theta \frac{\tanh^{2n} (r)}{\cosh^2 (r)}, \\
  (\lambda'_{2})_n =&  2 n \sin\theta \cos\theta \frac{\tanh^{2(n-1)}(r)}{\cosh^4 (r)}.
\end{align}
\end{subequations}
Putting this together in the equation to calculate Fisher information gives us
\begin{align}
  \label{eq:meas:SCARfisherinfo}
  \mathscr{F}(\theta)= \sum_n \left( 4 \sin^2\theta \frac{\tanh^{2n}(r)}{\cosh^2(r)}  + 4 n \cos^2\theta \frac{\tanh^{2(n-1)}(r)}{\cosh^4 (r)} \right)= 4.
\end{align}
It is worth noting that the result is not dependent on either $r$ or $\theta$.
It is in fact equal to the Fisher information of the classical-quantum state \cref{eq:meas:classicalparameterised} as directly measured by Alice, before any communication.
This is because in classical communication, the sender always keeps a perfect copy of the state they have sent.
If Alice has a perfect copy, then the optimal measurement where Alice is allowed to assist Rob, is just the measurement that Alice would perform herself.

We now look at the situation where Rob must measure his own state to determine $\theta$ without any help from Alice.
Alice's subsystem is treated as an external, unknowable system and therefore must be traced out.
Rob's state once Alice has been traced out is the same no matter whether the logical state contained quantum or classical information to begin with.
This partial trace leaves Rob with the system in a state,
\begin{align}
  \label{eq:meas:SCRrho}
  \rho_{R}^{(s)}(\theta)=\sigma_{R}^{(s)}(\theta) = \sum_{n=0}^\infty \Ket[R]{n}\Bra{n} \left[\cos^2\theta\frac{\tanh^{2n} (r)}{\cosh^2 (r)} +n \sin^2\theta\frac{ \tanh^{2(n-1)}(r)}{\cosh^4 (r)}\right].
\end{align}

It is again diagonal, allowing us to use the simplified Fisher information \cref{eq:inf:simplefisher}.
We can just read off the eigenvalues from the state to perform the calculation.
Taking the derivative with respect to $\theta$ is very similar to the previous case,
\begin{align}
  \label{eq:meas:SCRrhoprime}
  \rho_{R}^{\prime(s)}(\theta) = \sum_{n=0}^\infty \Ket[R]{n}\Bra{n} \left[- 2 \sin\theta \cos\theta\frac{\tanh^{2n} (r)}{\cosh^2 (r)} +2 n \sin\theta \cos\theta\frac{ \tanh^{2(n-1)}(r)}{\cosh^4 (r)}\right].
\end{align}
Each eigenvalue is now the sum of what were the eigenvalues in the previous calculation.
When we square the eigenvalues of $\sigma'$ we end up with cross terms that drastically change the form of the Fisher information
\begin{align}
  \label{eq:meas:SCRfisherinfo}
  \mathscr{F}(\theta)=\sum_{n=0}^\infty \left( \frac{4 \sin^2\theta \cos^2\theta \tanh^{2n+2}(r) (n \csch^2(r)-1)^2}{n \sin^2\theta + \sinh^2(r)\cos^2\theta}\right).
\end{align}
This sum was calculated using Mathematica \cite{Mathematica}, which came up with an analytical solution involving the generalised hypergeometric function and the Lerch transcendent function.
This solution is given in \cref{ch:measurefullmaths}.

It is worth remembering that these infinite sums will always converge.
The power series part of the sum will always become more important than any linear part.
The function $\tanh(r)<1$ for finite $r$ and only tends towards a value of $1$ for $r\to\infty$.

The last remaining setup for the single rail encoding is that of a quantum state being sent and both Alice and Rob measuring it.
The act of sending the quantum state actually creates entanglement between Alice and Rob due to the no-cloning theorem.
We see what effect this has on the Fisher information by performing the full calculation.

We start with a general $2\times2$ block of the density matrix
\begin{align}
  \left(\rho_{AR}^{(s)}(\theta)\right)_n =
  \begin{pmatrix}
    \cos^2\theta \frac{\tanh^{2n}(r)}{\cosh^2(r)}&\sqrt{n+1}\sin\theta\cos\theta\frac{\tanh^{2n}(r)}{\cosh^3(r)} \\
    \sqrt{n+1}\sin\theta\cos\theta \frac{\tanh^{2n}(r)}{\cosh^3(r)}&(n+1)\sin^2\theta\frac{\tanh^{2n}(r)}{\cosh^4(r)}
  \end{pmatrix}.
\end{align}
We find the relevant eigenvalues and eigenvectors.
The eigenvalues written as a column vector are
\begin{align}
  \left(\lambda_{AR}^{(s)}(\theta)\right)_n=
  \begin{pmatrix}
   0 \\
   \frac14 \left(3+2n+\cosh(2r)+\cos(2\theta)(-1-2n+\cosh(2r))\right)\frac{\tanh^{2n}(r)}{\cosh^4(r)}
  \end{pmatrix}.
\end{align}
The eigenvectors written as columns of the transformation matrix to the basis where $\left(\rho_{AR}^{(s)}(\theta)\right)_n$ is diagonal are
\begin{align}
  T_d^{(s)}=
  \begin{pmatrix}
    -\sqrt{n+1}\sech(r)\tan\theta & \frac{\cosh(r)\cot\theta}{\sqrt{n+1}} \\
    1&1
  \end{pmatrix}.
\end{align}

We now take the derivative with respect to $\theta$ and then transform to the same basis with $T_d^{(s)}$.
This leaves $\left(\rho_{AR}^{\prime(s)}(\theta)\right)_n$ as
\begin{align}
  \left(\rho_{AR}^{\prime(s)}(\theta)\right)_n =
  \begin{pmatrix}
   0 & \cot\theta \frac{\tanh^{2n}(r)}{\cosh^2(r)} \\
   (n+1)\tan\theta \frac{\tanh^{2n}(r)}{\cosh^4(r)} & \sin\theta\cos\theta (1+2n-\cosh(2r)) \frac{\tanh^{2n}(r)}{\cosh^4(r)}
  \end{pmatrix}.
\end{align}

Having these matrices both in the basis of diagonal $\left(\rho_{AR}^{(s)}(\theta)\right)_n$ makes it easy to apply the lowering superoperator.
We can then calculate the Fisher information by multiplying $\left(\rho_{AR}^{\prime(s)}(\theta)\right)_n$ and the lowered $\left(\rho_{AR}^{\prime(s)}(\theta)\right)_n$, tracing and summing over $n$.
This leads to the following
\begin{align}
  \mathscr{F}(\theta)= \sum_{n=0}^\infty \left(2n+3 + (2n+1)\cos(2\theta)+2\sin^2\theta\cosh(2r)\right) \frac{\tanh^{2n}(r)}{\cosh^4(r)} = 4.
\end{align}
This is the maximum information possible to extract from even the noiseless state.

The determination of $\theta$ using the classically encoded state is optimal.
Therefore, any entanglement between Alice and Rob can just be discarded by tracing out Rob's subsystem.
This leaves Alice with a partially or fully decohered version of her original state.
However, for the determination of $\theta$ the coherence is irrelevant as the optimal measurement is classical anyway.

\subsection{Dual rail encoding}
The dual rail is somewhat different because there are two modes with noise and we need to sum over both modes.
This results in a double sum.
We found the first sum possible to calculate using Mathematica \cite{Mathematica}, but the second sum had to be calculated numerically.

We first look at the situation where the parameter was encoded classically onto the dual rail qubit.
This is the state given by \cref{eq:bh:classicaldualrailstate}.
Following the argumentation of the previous section, if both Alice and Rob perform the measurement we should have a Fisher information of $4$.
This is again because Alice will always keep a perfect copy of the state.
We must check that we indeed find $\mathscr{F}(\theta)=4$ in this case.

As the state is already diagonal, we can trivially extract the eigenvalues, indexed by three labels $n$, $m$ and $p$,
\begin{align}
   \left(\lambda_{AR}^{(d)}(\theta)\right)_{nmp}=\cos^{2-2p}\theta \sin^{2p}\theta \frac{\tanh^{2n+2m}(r)}{\cosh^6(r)} (n+1)^{1-p} (m+1)^p,
\end{align}
where the indices $n,m \in \{0,\mathbb{Z}^+\}$ and $p\in \{0,1\}$.
It is helpful to write out the different $p$ eigenvalues separately to perform the differentiation
\begin{subequations}
\begin{align}
   \left(\lambda_{AR}^{\prime(d)}(\theta)\right)_{nm0}=&-2\cos\theta \sin\theta \frac{\tanh^{2n+2m}(r)}{\cosh^6(r)} (n+1), \\
   \left(\lambda_{AR}^{\prime(d)}(\theta)\right)_{nm1}=&2\sin\theta \cos\theta \frac{\tanh^{2n+2m}(r)}{\cosh^6(r)} (m+1).
\end{align}
\end{subequations}

Substituting these into the expression for the Fisher information and writing the sum over $p$ explicitly, we find
\begin{align}
  \mathscr{F}(\theta)= \sum_{n,m=0}^\infty \frac{\tanh^{2n+2m}(r)}{\cosh^6(r)} 4\left( (m+1)\cos^2\theta + (n+1)\sin^2\theta\right)=4.
\end{align}
Once the double sum is completed, this does indeed give a result of $4$, as expected.

In a similar way to the single rail, both the quantum and classical encodings become equal and diagonal if you trace out Alice.
We write the eigenvalues as
\begin{align}
  \left(\lambda_R^{\prime(d)}(\theta)\right)_{nm} = \frac{\tanh^{2n+2m-2}}{\cosh^6(r)} \left(n \cos^2\theta + m \sin^2\theta\right).
\end{align}
Differentiating and putting it into the simple expression for Fisher information, \cref{eq:inf:simplefisher}, results in
\begin{align}
  \mathscr{F}(\theta)= \sum_{n,m=0}^\infty \frac{\tanh^{2m+2n-2}(r)}{\cosh^6(r)} \frac{(m-n)^2 \sin^2(2\theta)}{n\cos^2\theta + m\sin^2\theta}.
\end{align}
Mathematica was able to provide an analytical solution for the sum over $m$.
If $n$ was summed first instead, the result would be very similar and of equal complexity.
The result is given in \cref{ch:measurefullmaths}, and the sum over $n$ was performed numerically in Matlab \cite{Matlab}.

The quantum encoding sent over the dual rail is different when Alice also participates in the measurement.
This is due to the entanglement that is created between the two parties.
The state is not diagonal but it is block diagonal, so we follow a similar process to that of the equivalent single rail Fisher information.

We start with a general block,
\begin{align}
\left(\rho_{AR}^{(d)}(\theta)\right)_n= \frac{\tanh^{2n+2m}(r)}{\cosh^6(r)}
  \begin{pmatrix}
    (n+1) \cos^2\theta &\sqrt{n+1}\sqrt{m+1} \sin\theta\cos\theta \\
    \sqrt{n+1}\sqrt{m+1} \sin\theta\cos\theta & (m+1) \sin^2\theta
  \end{pmatrix}.
\end{align}
The eigenvalues can be found and written as a column vector,
\begin{align}
  \begin{pmatrix}
    0 \\
    -\frac12 \left(-2-m-n+(m-n)\cos(2\theta)\right)\frac{\tanh^{2(m+n)}(r)}{\cosh^6(r)}
  \end{pmatrix}
\end{align}
We can also find the eigenvectors and write them as columns of the matrix that transforms into the basis where $\rho$ is diagonal,
\begin{align}
  T_d^{(d)}=
  \begin{pmatrix}
    -\frac{\sqrt{m+1}\tan\theta}{\sqrt{n+1}}&\frac{\sqrt{n+1}\cot\theta}{\sqrt{m+1}}\\
    1&1
  \end{pmatrix}.
\end{align}

\begin{figure}[tb]
  \begin{center}
      \includegraphics[width=272pt]{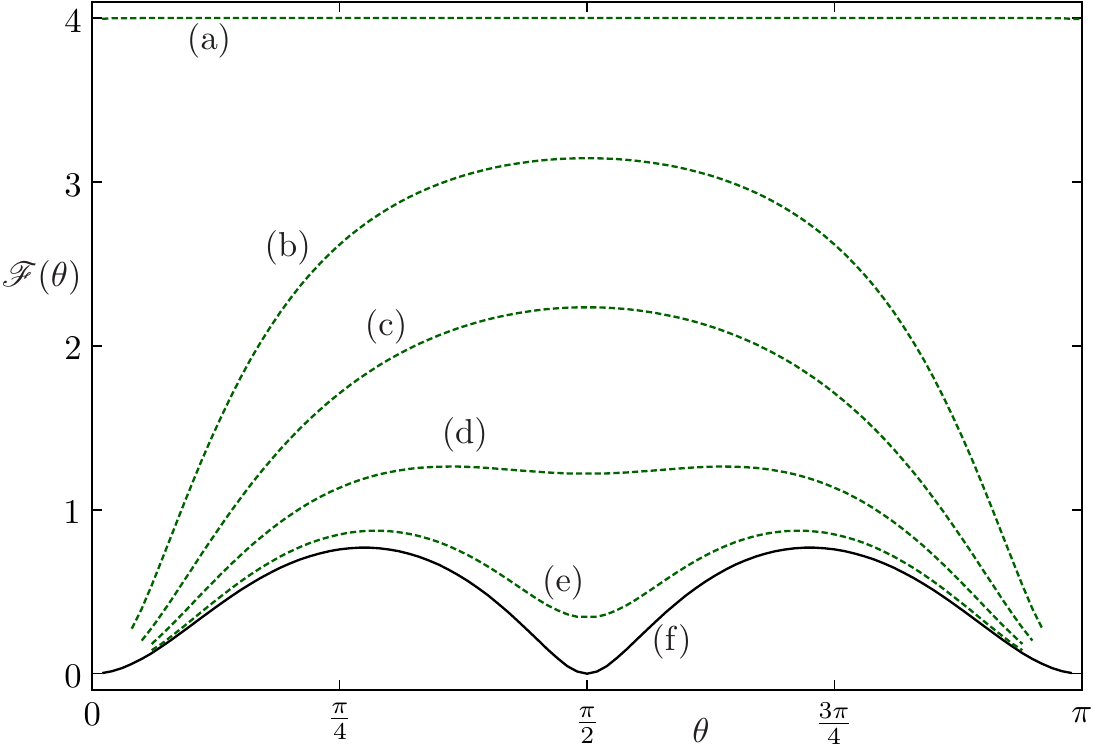}
  \caption[Single rail quantum Fisher information against $\theta$]{The Fisher information $\mathscr{F}(\theta)$ for Rob measuring the parameter $\theta$ when Alice has sent him the state using the single rail, plotted as a function of $\theta$ for various values of $r$.
          For line (a) $r=0.001$, which is close enough to zero to be the starting point.
          Lines (b), (c), (d), and (e) have $r=0.5$, $r=0.8$, $r=1.2$, and $r=1.9$ respectively.
          Line (f) has $r=2.8$ which we are taking to be the converged value within the approximate thickness of the line.}
  \label{fig:meas:single-theta}
  \end{center}
\end{figure}

We can now differentiate with respect to $\theta$ and transform into the same basis,
\begin{align}
\left(\rho_{AR}^{\prime(d)}(\theta)\right)_n=
  \begin{pmatrix}
    0& (n+1)\cot\theta\frac{\tanh^{2(n+m)}(r)}{\cosh^6(r)} \\
    (m+1)\tan\theta\frac{\tanh^{2(n+m)}(r)}{\cosh^6(r)}& (m-n)\sin(2\theta)\frac{\tanh^{2(n+m)}(r)}{\cosh^6(r)}
  \end{pmatrix}.
\end{align}
This makes it possible to apply the lowering operator.
We can then perform the multiplication and trace for each block, finally summing every block to find the Fisher information,
\begin{align}
  \mathscr{F}(\theta)=\sum_{n,m=0}^\infty 2 \left(2+m+n+(m-n)\cos(2\theta)\right)\frac{\tanh^{2(n+m)}(r)}{\cosh^6(r)} = 4.
\end{align}
This is again the maximum information possible to extract from the noiseless state.
The same reasoning applies to the dual rail as to the single rail.
If Alice is included in the measurement, she can always ignore Rob's state and measure her own.
The decoherence does not matter in the type of measurement she needs to do to find $\theta$.
Alice is therefore capable of performing an optimal measurement regardless of any communication noise with Rob.

\section{Results for amplitude estimation}
\label{sec:meas:results}
We now plot the results of Rob measuring the parameter alone since these are the nontrivial results.
As discussed in the previous section, if Alice assists in the measurement, she can always achieve the Fisher information of $4$.
Rob, however, is affected by the noise generated in the communication channel due to his acceleration.

\begin{figure}[tb]
  \begin{center}
      \includegraphics[width=272pt]{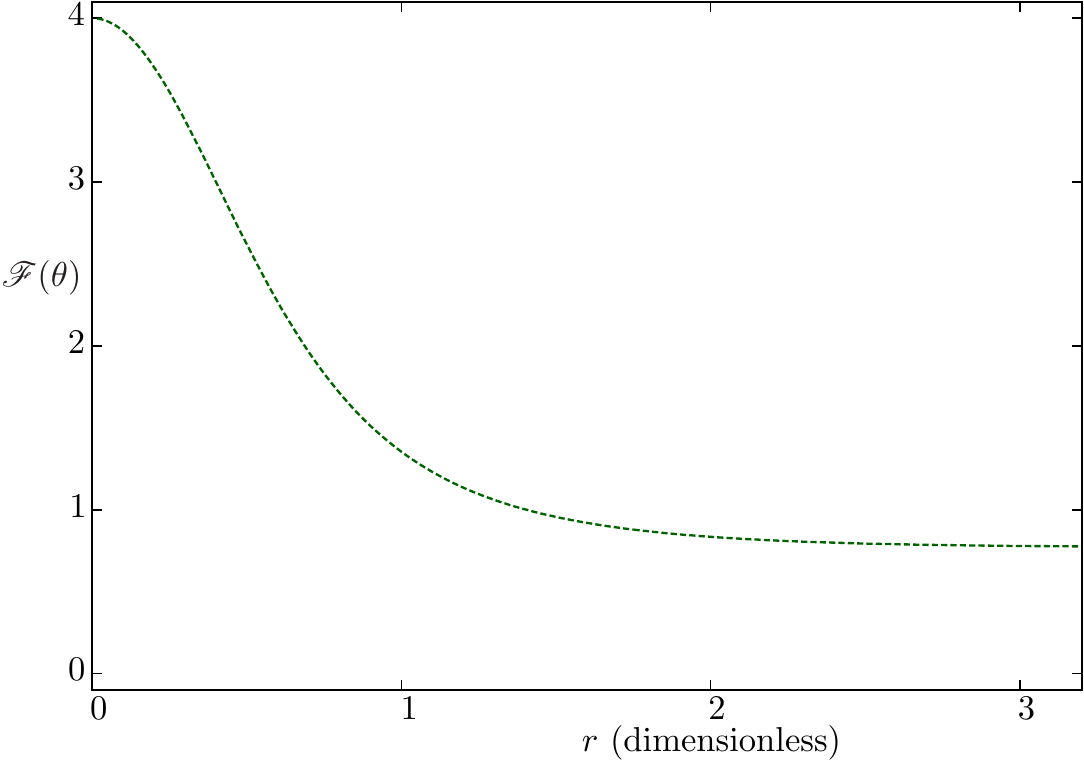}
  \caption[Single rail quantum Fisher information against $r$]{The Fisher information $\mathscr{F}(\theta)$ for Rob measuring the parameter $\theta$ when Alice has sent him the state using the single rail, plotted as a function of $r$ for a representative value of $\theta=\frac{\pi}{4}$.}
  \label{fig:meas:single-r}
  \end{center}
\end{figure}

We briefly noted in \cref{ch:blackholes} that there was an asymmetry between the $\ket{\zero}$ and $\ket{\one}$ states when using the single rail.
This becomes very pronounced when the parameter we are trying to measure oscillates the amplitude between the $\ket{\zero}$ and $\ket{\one}$ states.
The Fisher information is plotted as a function of the parameter $\theta$, for various values of $r$ in \cref{fig:meas:single-theta}.
The ends of the lines are where the numerical algorithm did not achieve the required accuracy and so have been left out of the plot.
The black solid line is the limiting line as $r$ gets large.

For low accelerations (small $r$) we see a huge difference in the sensitivity of the measurement when $\theta$ is at different values.
When $\theta\approx0$ or $\theta\approx\pi$, the state is almost $\ket{\zero}$, we see a minimum Fisher information.
If $\theta\approx\frac{\pi}{2}$, the state is almost $\ket{\one}$ and we see a maximum.
When the accelerations get larger (around $r>1$), the sensitivity decreases for both $\ket{\zero}$ and $\ket{\one}$.
The maximum sensitivity starts to move towards the points where they have equal amplitude.

\begin{figure}[tb]
  \begin{center}
      \includegraphics[width=272pt]{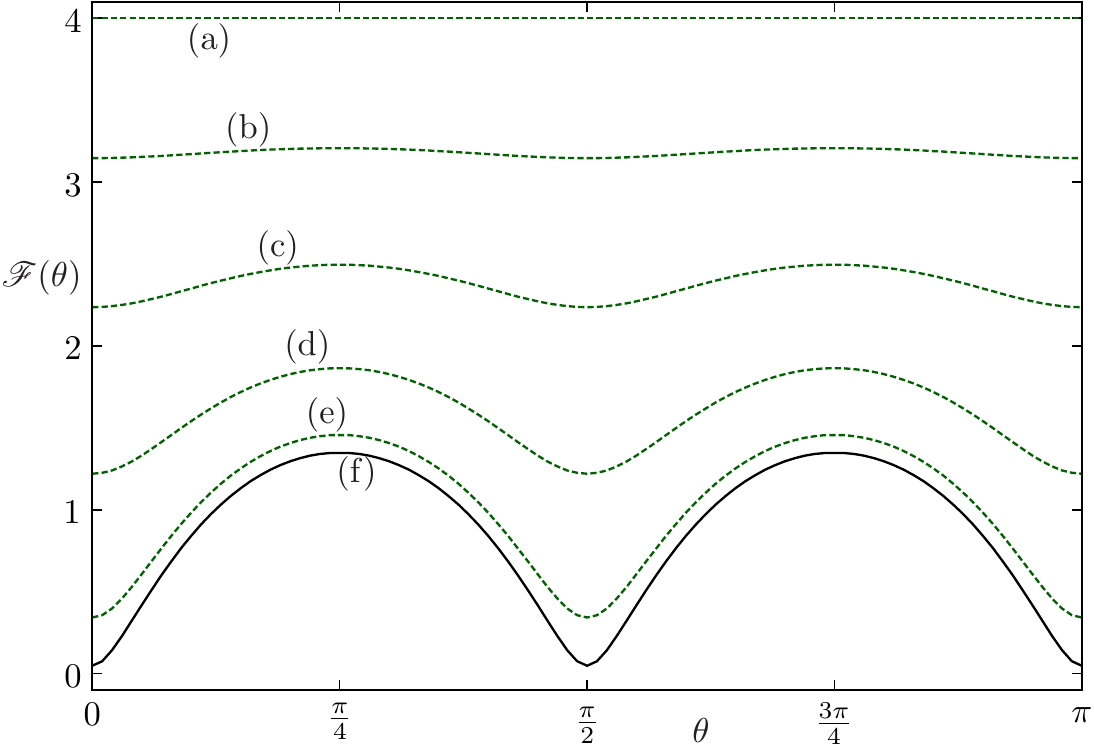}
  \caption[Dual rail quantum Fisher information against $\theta$]{The Fisher information $\mathscr{F}(\theta)$ for Rob measuring the parameter $\theta$ when Alice has sent him the state using the dual rail, plotted as a function of $\theta$ for various values of $r$.
           For line (a) $r=0.001$, which is close enough to zero to be the starting point.
           Lines (b), (c), (d), and (e) have $r=0.5$, $r=0.8$, $r=1.2$, and $r=1.9$ respectively.
           Line (f) has $r=2.9$ which we are taking to be the converged value within the approximate thickness of the line.}
  \label{fig:meas:dual-theta}
  \end{center}
\end{figure}

\begin{figure}[tb]
  \begin{center}
      \includegraphics[width=272pt]{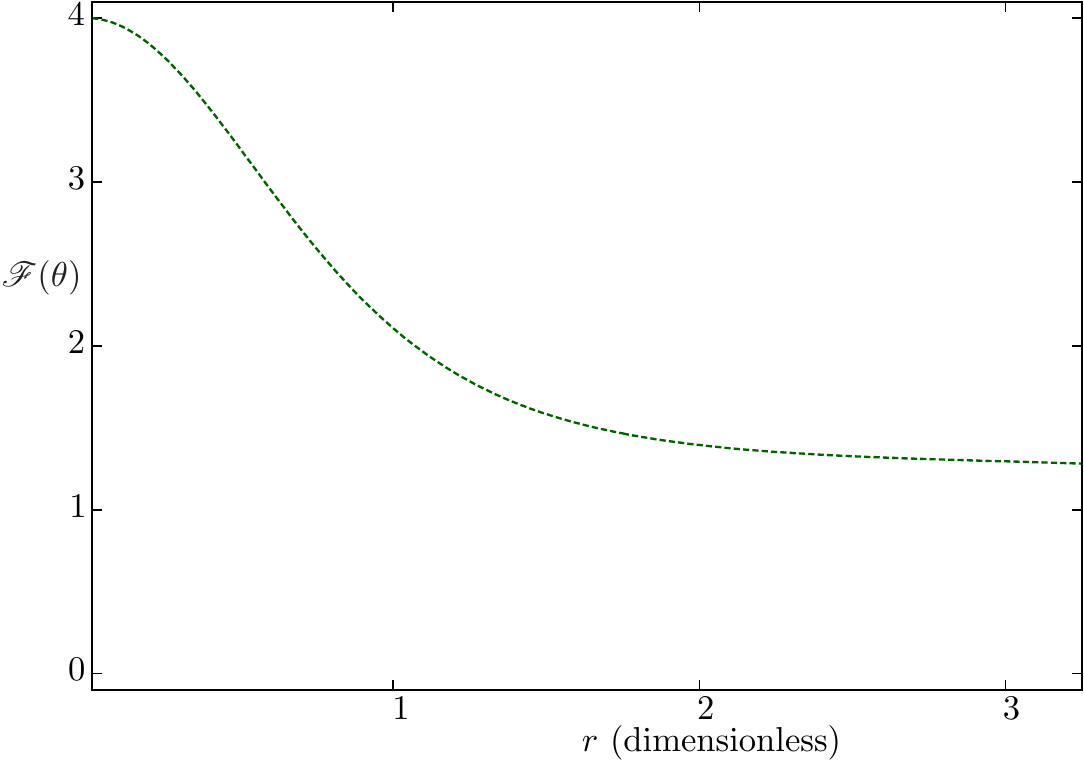}
  \caption[Dual rail quantum Fisher information against $r$]{The Fisher information $\mathscr{F}(\theta)$ for Rob measuring the parameter $\theta$ when Alice has sent him the state using the dual rail, plotted as a function of $r$ for a representative value of $\theta=0.65$.}
  \label{fig:meas:dual-r}
  \end{center}
\end{figure}

At low accelerations, the probability of measuring a noise excitation in the mode is very low.
Therefore, if $\theta=\frac\pi2$ there is definitely at least one photon in the mode.
If $\theta=0$ there are most likely to be no photons in the mode, but there is a possibility of having one photon.
This gives you a higher probability of estimating the parameter correctly if $\theta=\frac\pi2$ because you will measure at least one photon and guess the correct value of $\theta$.
If $\theta=0$, you have a chance of getting the estimate wrong if you do measure a noise excitation instead of the vacuum.

As the acceleration gets larger there is a transition into the region where the best sensitivity comes from the most mixed states, those where $\theta\approx\frac\pi4$ and $\theta\approx\frac{3\pi}{4}$.
If the state is near a logical basis state, it becomes difficult to estimate $\theta$ due to the large amount of noise photons flooding the signal.
If there are approximately equal amplitudes of each basis state, the correlations between them become important.
The shift between these two behavioural regions occurs near to the acceleration for which the expected number of noise excitations is one.

We have plotted the Fisher information over $r$ for a typical value of $\theta=\frac{\pi}{2}$ in \cref{fig:meas:single-r}.
This allows us to see how the acceleration affects the measurement sensitivity.
We expect, from our calculations in \cref{ch:blackholes} and the fact that the parameter estimation is a classical information task, that the sensitivity will drop off from its noiseless maximum but tend towards a finite value for large accelerations.
This is indeed reflected in the graph.

When communicating via the dual rail, there is perfect symmetry between the $\ket{\zero}$ and $\ket{\one}$ states.
This is clear to see in \cref{fig:meas:dual-theta}, where we have plotted the Fisher information over $\theta$ for various values of $r$.
The solid black line is the limiting line as $r$ gets large.

We can see from \cref{fig:meas:dual-theta} that again the communication noise has a variable effect depending on the parameter $\theta$.
There is less sensitivity in resolving $\theta$ when the state is closer to a computational basis state.
The greatest sensitivity is when the amplitudes are equal in the computational basis.

The noise in each mode is independent of the other mode.
This means that correlating the two modes tends to reduce the noise.
The highest sensitivity comes from when the state is correlated, and the measurement can detect these correlations.
If the state is not correlated between the two modes (one of the logical basis states), it becomes an attempt to measure which mode has a higher average excitation.

To see the form of the loss of sensitivity due to acceleration, we have plotted the Fisher information over $r$ for a representative value of $\theta=0.65$ in \cref{fig:meas:dual-r}.
This value was chosen because the computational time to calculate for $\theta=\frac{\pi}{4}$ was prohibitively large.
This is very similar to the form of the classical communication reduced channel capacity seen in \cref{ch:blackholes}.
We see the Fisher information decrease from it's maximal value of $4$ as $r$ increases but converge to a finite value that is larger than the convergent value of the single rail.
This reflects the conclusion that the dual rail can carry more information that the single rail.

\section{Using NOON states to estimate relative phase} 
We now study the similar situation where an inertial observer, Alice, communicates information about a parameter with a constantly accelerated observer, Rob.
Alice prepares a state with $\theta=0$, which undergoes a unitary evolution that imparts the phase $\theta$ onto the state.
If encoded using the dual rail, the state is,
\begin{equation}
  \label{eq:dualrailstate}
  \ket[A]{\psi}^{(d)}= \ket{N,\varnothing} + e^{i N \theta}\ket{\varnothing,N}.
\end{equation}
This dual rail state is known as the NOON state in quantum interferometry, originally introduced by Bollinger \emph{et al.} \cite{Bollinger1996}.
It encodes the parameter onto the relative phase.
Equivalently, the state encoded using the single rail is,
\begin{equation}
  \label{eq:singlerailstate}
  \ket[A]{\psi}^{(s)}= \ket{\varnothing} + e^{i N \theta}\ket{N}.
\end{equation}
This requires a fully quantum channel because if it became classical at any point, all phase information would be lost.

The task is for Rob to measure the field and obtain an estimate for the parameter $\theta$.
We need to find out precisely how well he is able to perform this task.
Rob's detector will be subject to the relativistic noise present in the form of the Unruh-Hawking effect \cite{Hawking1975,Unruh1976,Takagi1986}.

\begin{figure}[tb]
  \begin{center}
      \includegraphics[width=272pt]{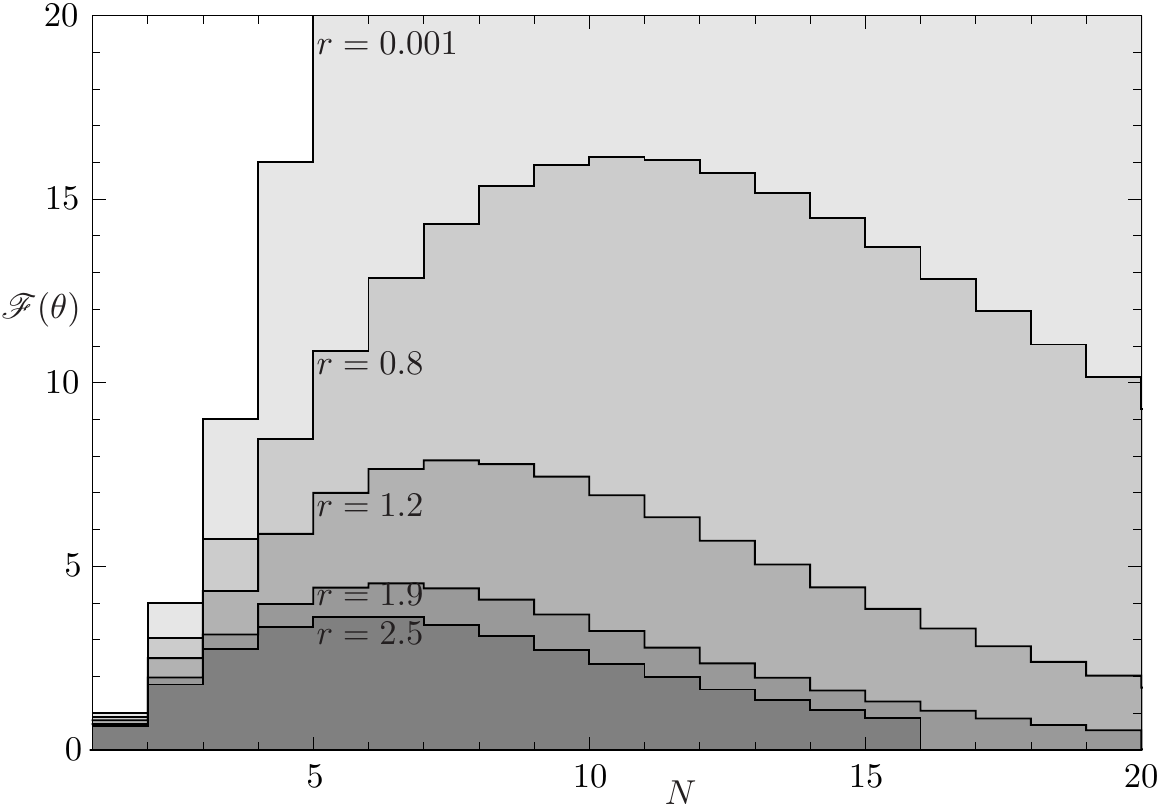}
  \caption[Single rail NOON state quantum Fisher information against $N$]{The quantum Fisher information $\mathscr{F}(\theta)$ plotted as a function of $N$ for Rob measuring the parameter $\theta$ when Alice has sent him the state using the single rail for various values of $r$.}
  \label{fig:single-noon-N}
  \end{center}
\end{figure}

\begin{figure}[tb]
  \begin{center}
      \includegraphics[width=272pt]{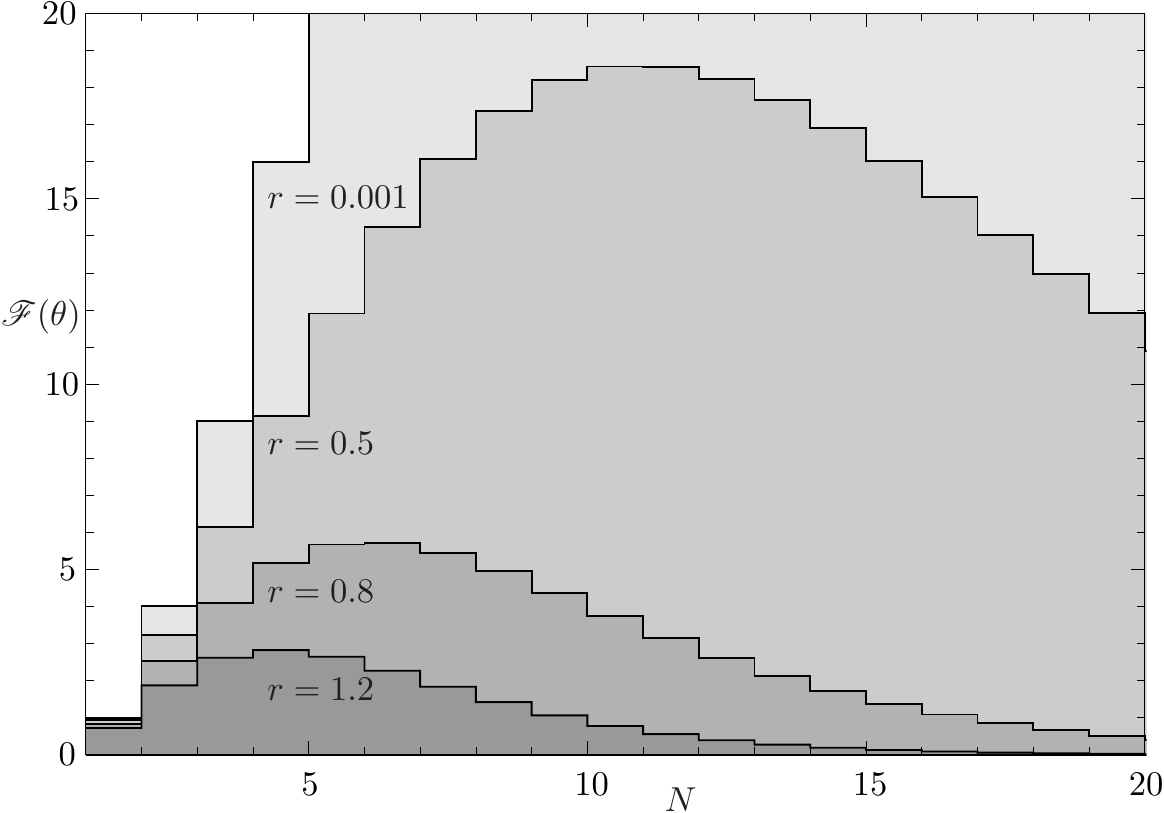}
  \caption[Dual rail NOON state quantum Fisher information against $N$]{The quantum Fisher information $\mathscr{F}(\theta)$ plotted as a function of $N$ for Rob measuring the parameter $\theta$ when Alice has sent him the state using the dual rail for various values of $r$.}
  \label{fig:dual-noon-N}
  \end{center}
\end{figure}

\begin{figure}[tb]
  \begin{center}
      \includegraphics[width=272pt]{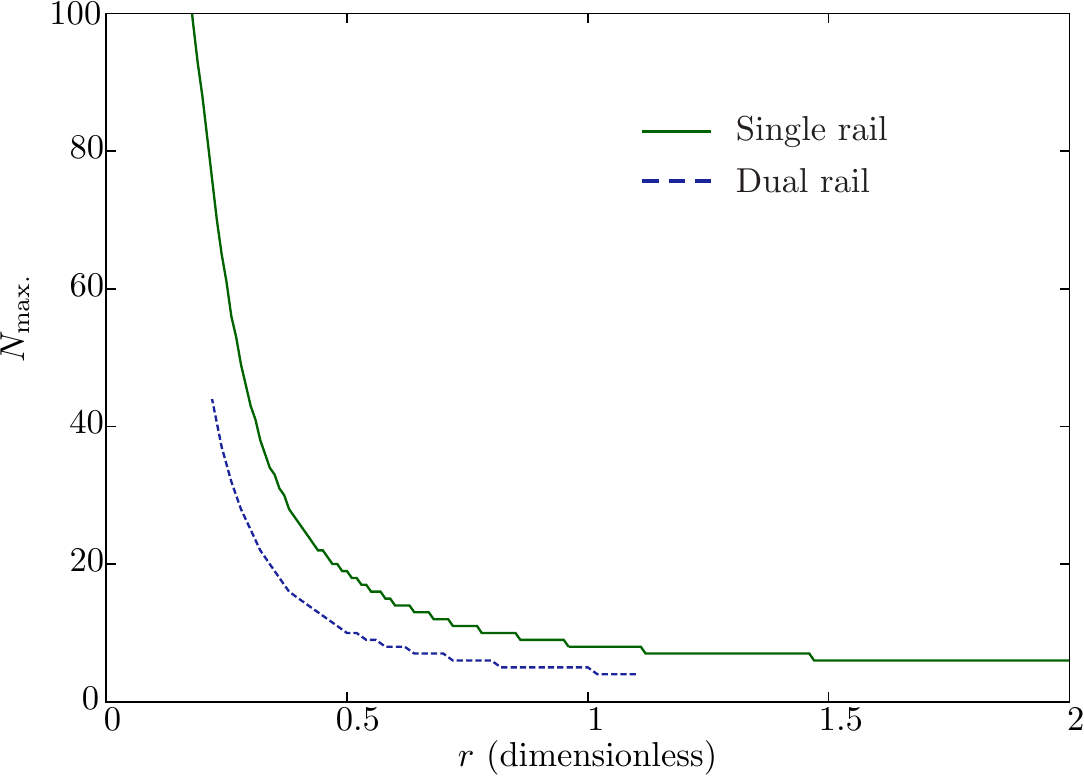}
  \caption[Value of $N$ for maximum quantum Fisher information against $r$]{The value of $N$ resulting in a maximum quantum Fisher information $\mathscr{F}(\theta)$ for Rob measuring the parameter $\theta$ when Alice has sent him the state using the single rail plotted as a function of $r$ for both the single rail and dual rail. The dual rail has a shorter line because the computational resources required prohibited further calculation in either direction.}
  \label{fig:single-noon-maxN}
  \end{center}
\end{figure}

Once Alice has created the state, she no longer has access to it.
Rob must then measure the field to determine the parameter.
To find out what Rob can measure, we must describe the field in terms of his mode functions \cref{sec:fields:transformation}.
As in \cref{ch:blackholes}, we use the Single Wedge Mapping, where Alice creates Unruh modes that directly map to Rindler modes in Rob's Rindler wedge, see \cref{ch:SMA}.
When we apply the squeezing transformation given in \cref{eq:fields:twomodesqueezing} to a mode with N excitations present $\ket[A]{\psi}=\ket{N}$, the transformed state in both Rob and anti-Rob's bases is,
\begin{equation}
  \label{eq:transformedN}
  \ket[R\bar{R}]{\psi} = \sum_{p=0}^\infty \frac{(ir)^p}{|r|^p} \frac{\tanh^p |r|}{\cosh^{N+1}|r|} \begin{pmatrix}p+N \\ p\end{pmatrix}^\frac12 \ket[R]{N+p} \otimes \ket[\bar{R}]{p}.
\end{equation}
To transform the full states, we construct Alice's density matrix from the states with zero and $N$ excitations, then transform each term according to \cref{eq:transformedN}.
Finally, we trace out anti-Rob leaving Rob with a mixed state represented by an infinite matrix in the number basis.

To calculate the quantum Fisher information numerically we need a finite matrix.
The coefficients of the matrix decay as the number of excitations increases.
We can therefore use a truncation of the matrix to get the result within a specified precision.
We numerically evaluated the matrices, for a size $k$, and calculated the quantum Fisher information.
To check that the truncated size was sufficient, we calculated the quantum Fisher information for matrices of size $k+1$.
We continued to increase $k$ until the quantum Fisher information changed less than the prescribed precision.

Using this truncated matrix, $\rho$, we calculated the quantum Fisher information using the algorithm as follows.
\begin{enumerate}
  \item Find eigenvalues and eigenvectors of $\rho$.
  \item Differentiate $\rho$ with respect to $\theta$ to find $\rho'$.
  \item Use the eigenvectors of $\rho$ to transform both $\rho$ and $\rho'$ into the basis where $\rho$ is diagonal.
  \item Perform the lowering operator according to \cref{eq:inf:loweringoperator}.
  \item Calculate the quantum Fisher information using the trace of the matrix product of $\rho'$ and the lowered $\rho'$.
\end{enumerate}
The quantum Fisher information includes $\theta$ in its calculation, however the final result does not depend on $\theta$.
We set $\theta=0.65$ throughout this section.
The quantum Fisher information does depend on both $r$, the squeezing parameter, and $N$, the number of excitations Alice created in the states.

\begin{figure}[tb]
  \begin{center}
      \includegraphics[width=272pt]{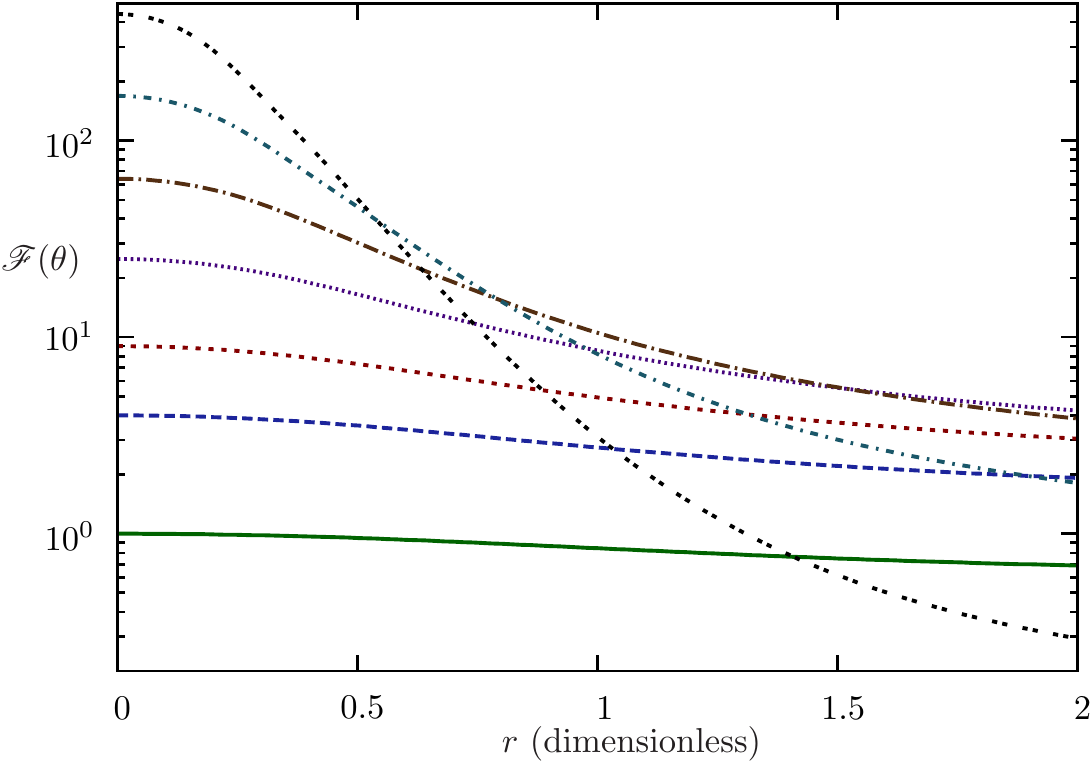}
  \caption[Single rail NOON state quantum Fisher information against $r$]{The quantum Fisher information $\mathscr{F}(\theta)$ plotted as a function of $r$ for Rob measuring the parameter $\theta$ when Alice has sent him the state using the single rail. The values of $N$ for each line are $1$, $2$, $3$, $5$, $8$, $13$ and $21$, corresponding to increasing Fisher information for $r=0$.}
  \label{fig:single-noon-r}
  \end{center}
\end{figure}

\begin{figure}[tb]
  \begin{center}
      \includegraphics[width=272pt]{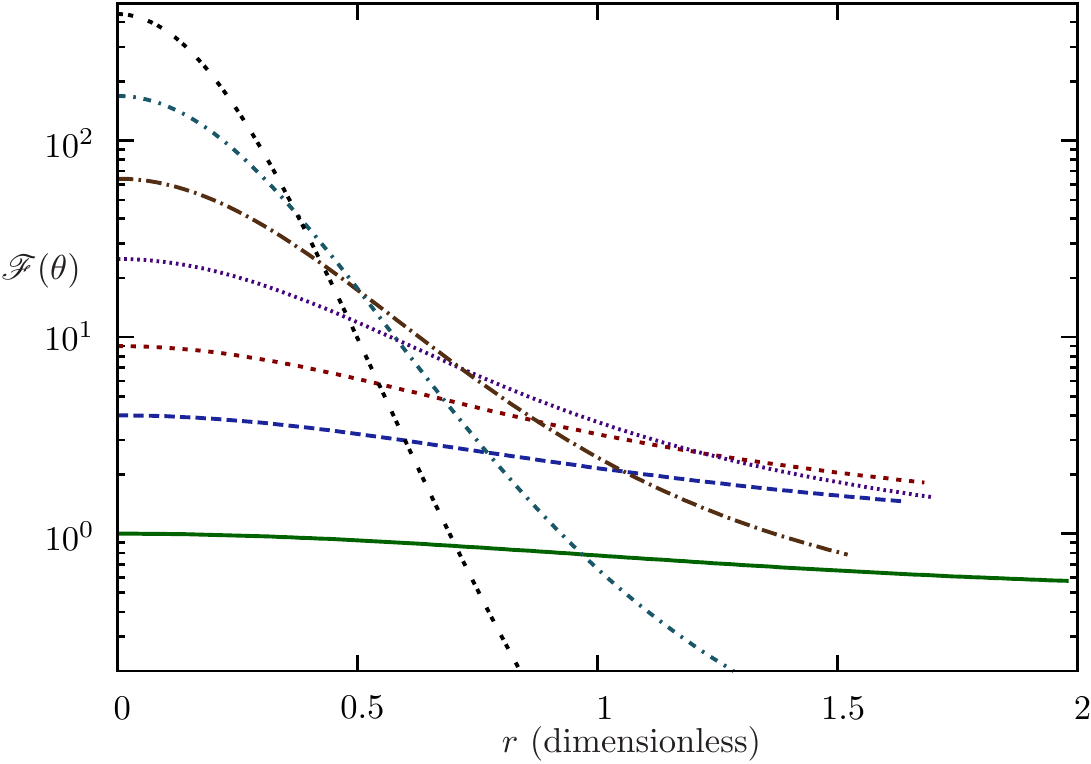}
  \caption[Dual rail NOON state quantum Fisher information against $r$]{The quantum Fisher information $\mathscr{F}(\theta)$ plotted as a function of $r$ for Rob measuring the parameter $\theta$ when Alice has sent him the state using the dual rail. The values of $N$ for each line are $1$, $2$, $3$, $5$, $8$, $13$ and $21$, corresponding to increasing Fisher information for $r=0$.}
  \label{fig:dual-noon-r}
  \end{center}
\end{figure}

\section{Results for phase estimation}
The quantum Fisher information is shown against $N$ for various values of the parameters, in \cref{fig:single-noon-N} for single rail encoding, and in \cref{fig:dual-noon-N} for dual rail encoding.
The channel is similar to an amplifying channel in that the more excitations that are present initially the more excitations are created by the channel.
Because the excitations created by the channel are thermal, they contribute to a higher noise level.
If there is a higher number of initial excitations there is more noise.
We find an exponential decay in the quantum Fisher information with respect to, $N$, the initial number of excitations, for a fixed acceleration.
This results in an optimal value of $N$ to use for each particular noise level.
Higher $N$ states sent via the dual rail are more susceptible to the noise than the single rail equivalents.
This is due to the fact that both the logical zero and logical one contain $N$ excitations, whereas in the single rail, the zero contains no excitations.
This results in a smaller expected number of excitations for the single rail, and hence less noise.

In \cref{ch:blackholes,ch:alicerobantirob}, we found the dual rail channel to perform better than the single rail for quantum communication between inertial Alice and accelerated Rob.
Here we find the opposite effect, that the single rail channel performs better.
These studies may use similar states, however, they are measuring a different quantity.
The information in the states studied previously was contained in the relative amplitude of the logical zero and one basis states.
In this study, the information about $\theta$ is contained entirely in the relative phase.
Whilst perhaps surprising, it is not impossible that different encoding methods are favoured for these situations.

The optimal value of $N$ is shown against $r$ in \cref{fig:single-noon-maxN} for the single rail.
We expect it to be asymptotic as $r\to 0$ because at $r=0$, $\mathscr{F}(\theta)=N^2$.
The fact that it drops at all is perhaps surprising because it is known for noiseless situations that NOON states are optimal \cite{Bollinger1996}.
Using NOON states in a noiseless situation results in a Fisher information that scales with $N^2$.

For zero noise, the quantum Fisher information rises with $N^2$.
When there is noise present, there is an exponential decay reducing the quantum Fisher information for larger $N$.
To match this physical situation, we fitted this for each value of $r$ with a model of the form,
\begin{equation}
  \label{eq:fitmodel}
  \mathscr{F}(\theta)= N^2 e^{- a(r) N + b(r)}.
\end{equation}
We found that for each $r$, this functional form fitted very well.
For large $r$, the coefficient $a(r)$ tended towards a linear function of $r$.
This function has a gradient of $(41.6 \pm 0.3) \times 10^{-3}$, when calculated using data for which $2.08\le r \le 3.10$.
It would be possible to use the linear dependence of the fitting parameter $a(r)$ to find the maximum value of $N$ for large $r$.
However, the interesting behaviour happens near $r=1$, because this is where the expected number of noise excitations is approximately $1$.

The quantum Fisher information is shown as the noise increases for various values of $N$ in \cref{fig:single-noon-r,fig:dual-noon-r} for the single and dual rails respectively.
Due to the large computational requirements of calculating these graphs, some of the lines stop before the end of the graph.
The quantum Fisher information decreases with increasing noise for each value of N, which is expected behaviour.
It is clear from the graphs that the single rail performs better than the dual rail for all values of $r$.

\section{Conclusion}
\label{sec:meas:conclusion}

In this chapter we studied the parameter estimation capabilities of Alice and Rob once the state encoding the parameter had passed through a quantum communication channel.
We looked at different types of states, ones similar to states used in \cref{ch:blackholes}, and NOON states.

\subsection{Amplitude estimation}
The protocol used in the first part was where Alice encoded the parameter onto her state, then created entanglement with Rob followed by a measurement performed by one or both of them to determine the parameter $\theta$.
If Alice is involved in the measurement, she can always perform an optimal measurement as expected.
This is true regardless of what communication happened with Rob, the optimal measurement is the one where Alice discards any measurement from Rob and just uses her own subsystem.
This allows her the full Fisher information of $\mathscr{F}(\theta)=4$ every time.

For Rob to measure alone, after performing this channel, we first discard Alice's subsystem, giving Rob a mixed state.
When Rob measures this mixed state, the Fisher information degrades with increased channel noise.
Rob's sensitivity to measuring $\theta$ also becomes dependent on the value of $\theta$ itself.
Certain state configurations are better at protecting the information they contain from the noise present in the communication channel.
This is due to the fact that we have used $\theta$ as the amplitude on a number basis encoding.
The dual rail has a much higher Fisher information than the single rail, which we also expect, given the reduced classical communication capacity.
However, the extractable information is highly dependant on both the value of $\theta$ and the choice of single or dual rail encoding.
This suggests that the techniques used are very sensitive to particular conditions.
Any application would need to have prior information of what the value of $\theta$ may be to choose the correct encoding.

The degradation of information with acceleration follows the same pattern as with classical communication in \cref{ch:blackholes}.
We expect this similarity because the communication protocols used are almost identical.
In both cases, we take a state encoded onto a logical basis given by the single or dual rail encodings, where the information is present in the amplitudes of the basis states.
We then create entanglement between Alice and Rob using this state, followed by Rob performing any operations or measurements on his subsystem.
The Fisher information measures the accuracy to which a continuous variable can be determined.
The reduced classical channel capacity that we were studying in \cref{ch:blackholes} is a measure of the amount of discrete information extractable from the message.
The entanglement-generating communication channel between an inertial observer and an observer hovering near a black hole degrades both continuous variable information in a similar way than it degrades discrete information.

\subsection{Phase estimation using NOON states}
The previous study was limited to single excitation states, and to a communication protocol that created entanglement between Alice and Rob.
In the second part of this chapter we extend the work to NOON states, using a communication protocol where Alice simply sends the state to Rob.
There is no entanglement generated so Alice does not need to be traced out, which preserves phase information in Rob's subsystem.

The channel used is amplifying in nature; the noise is dependent on the number of excitations already present.
Because of this $N$-dependent noise, there exists an optimum $N$ for maximum information transfer.
The $N$ dependence of the quantum Fisher information has a simple form which we fitted numerically.
The coefficients of the fit are dependent on $r$, the squeezing parameter, as expected.
We found that the decay coefficient had asymptotically linear dependence on $r$.
This could allow extrapolation to any noise levels.
The decay of the quantum Fisher information with noise was monotonic regardless of any possible optimisation over NOON states, as expected.

We find in this case that the single rail outperforms the dual rail.
Whilst further research would be required to ascertain exactly why, the main differences are as follows.
In this study, we used the parameter encoded onto the phase rather than the amplitude.
The single and dual rail encodings may be better suited to preserving these different types of information.
The communication protocol was different, Alice simply sent the state to Rob rather than generating an entangled version of the state.

\subsection{Implications}

The study of metrology is important to direct technological techniques for measuring physical parameters.
Any measurement protocols using these types of fields under accelerated conditions would need to use this knowledge to tailor their encodings specifically to the parameters they are measuring.
We have studied the effect on extractable information about a continuous variable under accelerated conditions.
Continuous variable quantum computing may benefit from knowing what encodings of the variables maintain the best information in these situations.

Further work would include studying other forms of states to find more optimal information transfer.
There has been work on lossy interferometry using states of other forms, where these states are more robust to the noise than NOON states \cite{Maccone2008,Knysh2011}.
We expect a higher quantum Fisher information to be possible in this situation in a similar way.
The noise here has a similar $N$ dependence to the errors in lossy channels.

\chapter[Communication with accelerated agents]{Communication between inertial and causally disconnected, accelerated agents}
\label{ch:alicerobantirob}

\begin{figure}[thp]
\begin{center}
  \frame{\includegraphics[scale=0.6]{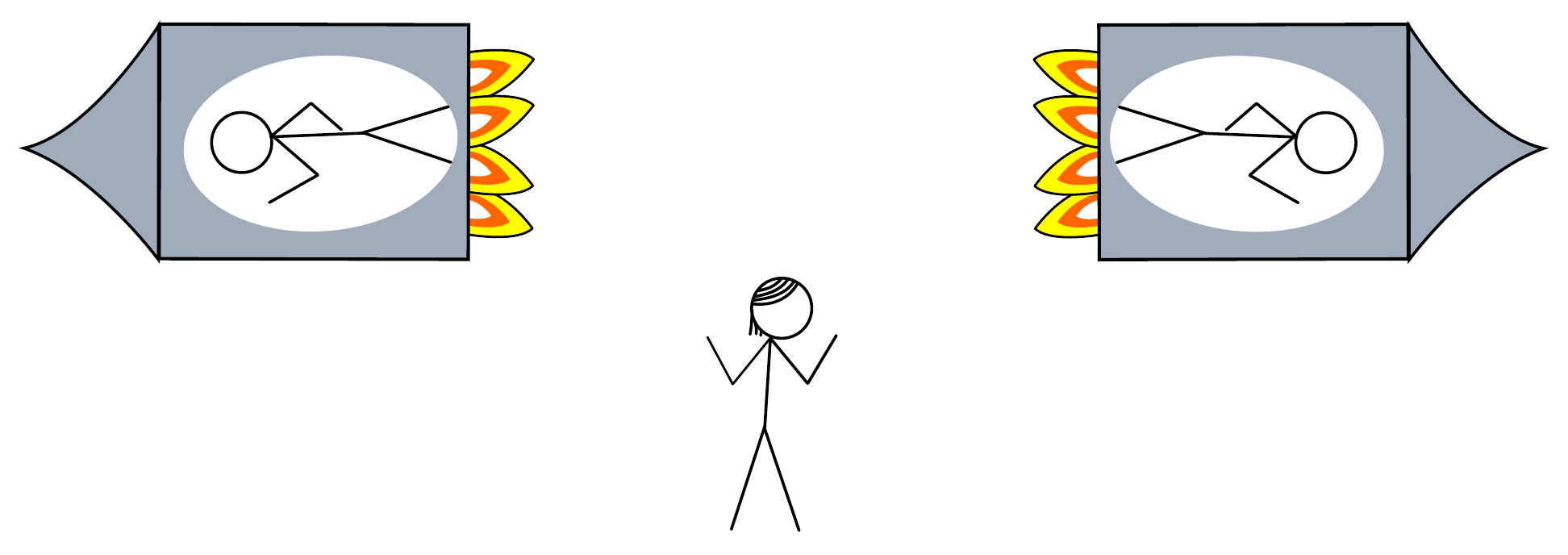}}
  \caption[Alice with counter-accelerating Rob and AntiRob]{This figure depicts Alice as an inertial observer in the centre. Rob and AntiRob are depicted accelerating away from each other. \label{fig:arr:alicerobantirob}}
\end{center}
\end{figure}

We study a different situation in this chapter.
\Cref{ch:blackholes,ch:measure} looked at Alice, an inertial observer, sending information to Rob only, who was undergoing constant acceleration or hovering above a black hole.
We approximated the field in Rob's frame using the Rindler modes.
In this chapter, we take the Rindler setup, and study the communication between two pairs of observers: An inertial observer, Alice ($A$) and two constantly accelerated complementary observers Rob ($R$) and AntiRob ($\AR$).
The complementary observers move with opposite accelerations in the two causally disconnected regions of the Rindler spacetime, the same setting as in \cite{Bruschi2010}.
This setting is depicted in \cref{fig:arr:alicerobantirob}.
This chapter is based on a paper prepared in collaboration with Eduardo Mart\'in-Mart\'inez and Miguel Montero \cite{Martin-Martinez2012a}.

\section{Introduction}
\label{sec:arr:introduction}

In \cref{ch:blackholes,ch:measure} we used the the Single Wedge Mapping (SWM, discussed in \cref{ch:SMA}) as we believed it to be optimal for the communication we studied between Alice and Rob.
Here, we discover that for classical communication at high accelerations this is not the case: we must go beyond the SWM to reach optimal communication.
The channels we study are the bipartite Alice-Rob and Alice-AntiRob channels.
It is not necessary to study any Rob-AntiRob channel because they are causally disconnected; no communication is possible.
We must also go beyond the SWM to deal with communication between Alice and AntiRob.

We discuss that for families of states built from Unruh modes, Alice and different non-inertial observers may not be able to generate entanglement using the state merging protocols (see \cref{subsec:inf:statemerging}) with these quantum channels.
In other words, when preparing the field state, Alice will have to choose with whom she wants to generate entanglement, since some of the states that she can prepare will not allow entanglement generation with some of the non-inertial observers.
We will show the monogamy property for the Unruh mode-based entanglement generation mentioned above when we go beyond the SWM.

\section{Setting}
\label{sec:arr:setting}

We will analyse here the information flow for both the single rail and dual rail encoding methods.
The single rail uses a single field mode, representing a logical zero with the vacuum and a logical one with a single excitation.
The dual rail uses two field modes with an excitation in one or the other to represent logical zero and logical one.
These encoding methods have been discussed in \cref{sec:inf:encoding}.

We study here restricted channel capacities (see \cref{subsec:inf:quantumchannelcapacity}) because we confine ourselves to either the single rail or dual rail encoding methods.
To study these restricted capacities, we must optimize over all parameters and encodings controlled by either Alice or Rob, obtaining optimal achievable rates by these encodings methods.
Alice has the freedom to choose which field mode to excite to send the best message to Rob.
We will not consider the possibility of Alice or Rob using arbitrary elements of the Fock space, as this would mean an infinite number of optimization parameters; we will stick to the simpler case of single field modes excited just once.
Rob is still allowed to measure within the full Fock basis of his mode, as this allows him to extract the information required.

\section{Field state}
\label{sec:arr:fieldstate}

Alice creates the appropriate excitations, once she knows which observer she would like to communicate with, and what their acceleration is.
As discussed in \cref{subsec:inf:quantuminfomeasures}, the classical message is a probabilistic mixture of logical zero and logical one states, and the quantum message is a particular qubit where the state merging protocol is then used to generate entanglement.
These excitations will be given by, \cref{eq:arr:classicallogical} for classical communication and \myeqref{eq:arr:quantumlogical} for quantum communication, respectively,
\begin{subequations}
  \label{eq:arr:logicalmessage}
\begin{align}
  \label{eq:arr:classicallogical}
  \sigma_A(\mbox{general bit}) =& \modsq{\alpha} \Ket[A]{\zero}\Bra{\zero} + \modsq{\beta} \Ket[A]{\one}\Bra{\one},\\
  \label{eq:arr:quantumlogical}
  \rho_A(\mbox{general qubit}) =& \left(\alpha \Ket[A]{\zero} + \beta \Ket[A]{\one}\right) \otimes \left(\alpha \Bra[A]{\zero} + \beta \Bra[A]{\one}\right).
\end{align}
\end{subequations}
Again, we use the parameter values $\modsq{\alpha}=\modsq{\beta}=\frac{1}{2}$ for the majority of this chapter.
This is optimal for dual rail which is verified numerically here, and provides representative behaviour for single rail communication.
We discuss the effect on the optimisation if we allow $\alpha$ and $\beta$ to be free variables again in \cref{sec:arr:optimisationalpha}.
We then perform the transformation into the accelerated frame as specified in \cref{sec:fields:transformation}.
AntiRob's frame is not traced out because we need to calculate information quantities in the Alice-AntiRob bipartition.

I present here the full state of the field for the four communication methods.
For classical communication, the field is given by,
\begin{subequations}
\begin{align}
  \label{eq:arr:fieldclassical}
  \sigma_{AR\AR}=\modsq{\alpha} \Ketbra{\Psi|\Psi}\thiscommandisahacktofixtheketbra + \modsq{\beta} \Ketbra{\Phi|\Phi}\thiscommandisahacktofixtheketbra,
\end{align}
and for quantum communication, by,
\begin{align}
  \label{eq:arr:fieldquantum}
  \rho_{AR\AR}= \left(\alpha \Ket{\Psi} + \beta \Ket{\Phi} \right) \otimes \left(\alpha^* \Bra{\Psi} + \beta^* \Bra{\Phi} \right)
\end{align}
where, the $\ket{\Psi}$ and $\ket{\Phi}$ depend on the encoding methods.
In the single rail, using the basis $\Ket{A,R,\AR}$ these are given by,
\begin{align}
  \label{eq:arr:fieldsingle}
  \ket{\Psi}^{(s)}=& \sum_n \frac{\tanh^n r}{\cosh r} \Ket{0,n,n},\\
  \ket{\Phi}^{(s)}=& \sum_n \frac{\tanh^n r}{\cosh^2 r} \sqrt{n+1} \sum_{p=0}^1 q_\text{L}^{1-p} q_\text{R}^p \Ket{1,n+p,n+1-p}.
\end{align}
In the dual rail, using the basis $\Ket{A,R_\zero,R_\one,\AR_\zero,\AR_\one}$ these are given by,
\begin{align}
  \label{eq:arr:fielddual}
  \ket{\Psi}^{(d)}=& \sum_{n,m} \frac{\tanh^{n+m} r}{\cosh^2 r} \sqrt{n+1}\sqrt{m+1} \sum_{p=0}^1 q_\text{L}^{1-p} q_\text{R}^p \Ket{0,n+p,m,n+1-p,m},\\
  \ket{\Phi}^{(d)}=& \sum_{n,m} \frac{\tanh^{n+m} r}{\cosh^2 r} \sqrt{n+1}\sqrt{m+1} \sum_{p=0}^1 q_\text{L}^{1-p} q_\text{R}^p \Ket{1,n,m+p,n,m+1-p}.
\end{align}
\end{subequations}

The dual rail has a clear symmetry between the zero and one modes because the form of $\ket{\Psi}$ and $\ket{\Phi}$ are the same.
There is symmetry under the exchange of $\qr$ with $q_\text{L}$ and Rob with AntiRob, so without loss of generality we need only study Rob as a receiver.
This allows us to trace out AntiRob simplifying the density matrix.

Beyond the Single Wedge Mapping, ($q_\text{R}<1$) the relevant density matrices are no longer block diagonal and they cannot be diagonalized into a closed form, as was shown by Bruschi et al. in \cite{Bruschi2010}.
This forced us to calculate the matrix representations in the Fock basis of the full field state in Matlab \cite{Matlab} with truncation at a matrix size of $k\times k$.
The partial traces and diagonalisation required were performed numerically, and the sums calculated for the entropic quantities.
This was done for many different values of $k$ until it was established that we had $k$ large enough such that the error introduced was sufficiently small.

\section{Classical communication}
\label{sec:classical}
\begin{figure}[t!]
  \begin{center}
      \includegraphics[width=272pt]{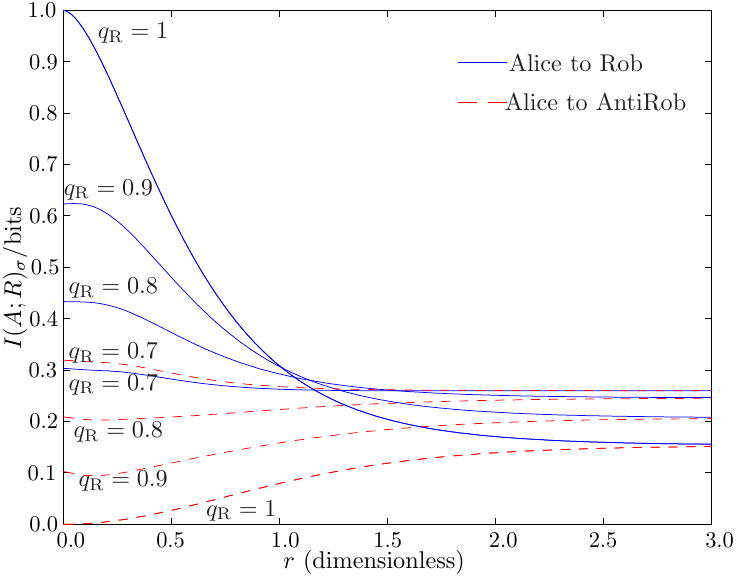}
  \caption[Single rail Holevo information]{Holevo information plotted as a function of $r$ of the single rail method of classical communication for various values of $\qr$.}
  \label{fig:singleMI}
  \end{center}
\end{figure}
\begin{figure}[b!]
  \begin{center}
      \includegraphics[width=272pt]{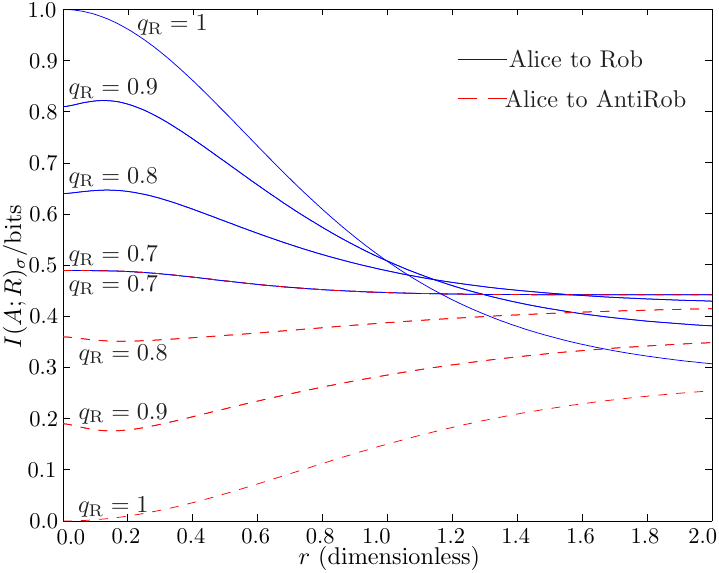}
  \caption[Dual rail Holevo information]{Holevo information plotted as a function of $r$ of the dual rail method of classical communication for various values of $\qr$.}
  \label{fig:dualMI}
  \end{center}
\end{figure}
The restricted classical channel capacity is given by the Holevo information maximized over controllable parameters.
This has been calculated and plotted for both the single rail encoding in \cref{fig:singleMI} and the dual rail encoding in \cref{fig:dualMI}.
We see from the plots that for large acceleration the maximum Holevo information is not achieved for $\qr=1$ (Single Wedge Mapping).
This shows that the assumptions made in the previous chapter are not correct for larger accelerations.
While it is true that SWM maximizes classical communication for low accelerations, for large accelerations we need to go beyond the SWM to achieve the restricted capacity.

Classical communication tends to a finite value for large accelerations (equivalent to large $r$), which is larger for dual rail communication.
No matter what the value of $\qr$ we find that as acceleration increases it is possible for Alice to communicate with both Rob and AntiRob.
At large accelerations Alice has equal communication channel capacity with both Rob and AntiRob, which is optimized at $\qr=\frac{1}{\sqrt2}$.

Note from \cref{fig:singleMI,fig:dualMI} that, at infinite acceleration, the Holevo information of the Alice-Rob and Alice-AntiRob channels both converge to the same value.
This is easily understood by looking at \cref{eq:fields:unruh-general-annihilation} and the definition of $\hat{a}_{\text{U}_\text{L}}$ and $\hat{a}_{\text{U}_\text{R}}$ in \cref{eq:fields:UrandUl}, and noting that at infinite acceleration both $\sinh\ro$ and $\cosh\ro$ tend to $e^{\ro}/2$, and therefore in this limit \cref{eq:fields:unruh-general-annihilation} may be written as,
\begin{align}
\hat{a}_\text{U}\approx \frac{e^{\ro}}{2}\left[\qr \hat{a}_{\text{R}_\text{I}}-\qr \hat{a}_{\text{R}_\text{IV}}^{\dagger}+q_\text{L} \hat{a}_{\text{R}_\text{IV}}-q_\text{L}\hat{a}_{\text{R}_\text{I}}^{\dagger}\right].
\end{align}
This expression is invariant under the replacements $\text{I}\leftrightarrow\text{IV}$ and $q_\text{L}\leftrightarrow q_\text{R}$, which takes Rob to AntiRob and vice-versa, and therefore in the infinite acceleration limit the Holevo information of both bipartitions must be the same.
In this limit, there is maximal squeezing between Rob and AntiRob's modes.
For Alice's message to be received well by either or both Rob and AntiRob then she must have it evenly spread between the two.
This ensures that both regions of spacetime have the message present, so when they mix, there is still the message.
If Alice only put the message in one region, then with maximal squeezing it would get mixed with the region where there was no message and get lost.
This also explains, why at large accelerations the parameter values $\qr=q_\text{L}=\frac{1}{\sqrt2}$ are optimal.

\section{Optimising over \textmd{$\qr$} and $\alpha$}
\label{sec:arr:optimisationalpha}

\begin{figure}[t!]
  \begin{center}
      \includegraphics[width=267pt]{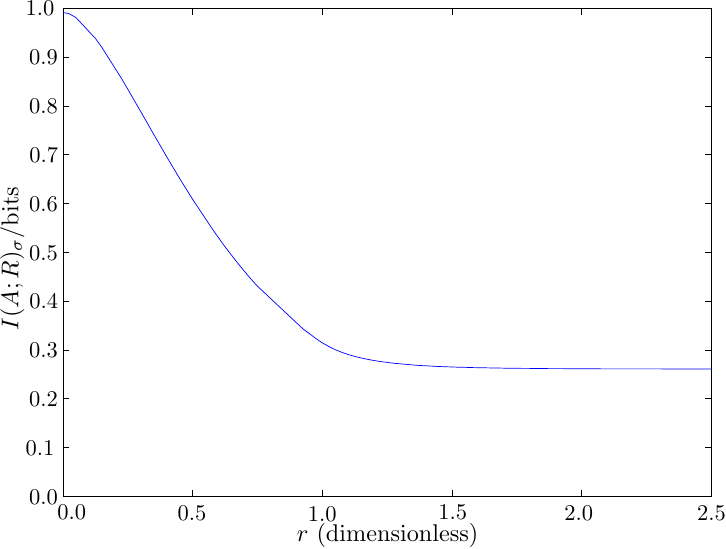}
  \caption[Single rail optimal Holevo information]{Optimal Holevo information as a function of the acceleration in the single rail case.}  \label{fig:opt}
  \end{center}
\end{figure}

\begin{figure}[b!]
  \begin{center}
      \includegraphics[width=267pt]{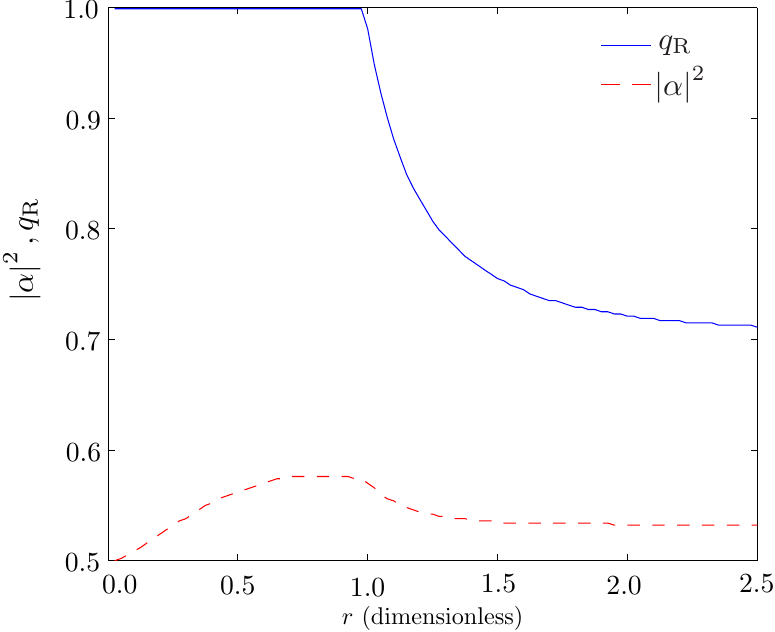}
  \caption[Single rail optimal parameter values]{$\modsq{\alpha}$ and $\qr$ optimal parameters for the results plotted in figure \ref{fig:opt}.}  \label{fig:opt2}
  \end{center}
\end{figure}

The optimal value of $\qr$ is $1$ for low acceleration, and $\qr=\frac{1}{\sqrt2}$ for high acceleration.
It is interesting to study the behaviour of the optimal parameter values between these two regimes.
We take the single rail encoding as the example because the computation time of numerically optimising over different $\qr$ in the dual rail encoding is prohibitive.
As mentioned in \cref{ch:blackholes}, there is an asymmetry in the single rail encoding.
We also noted the large variation of the quantum Fisher information as the parameter changed the state between the logical basis states in the single rail encoding in \cref{subsec:meas:singlerailamplitudemaths}.
It is interesting to study the behaviour of the optimum value for $\alpha$ in the single rail encoding.
We perform a two dimensional optimisation here over these two parameters.

We have plotted the optimal Holevo information which is the restricted channel capacity for the classical single rail case in \cref{fig:opt}.
\Cref{fig:opt2} shows the optimal values of $(\modsq{\alpha},\qr)$ used for each point on \cref{fig:opt}.
We see that the Single Wedge Mapping is optimal up to $\ro\approx1$.
Note the interesting non-monotonic behaviour of the parameter $\modsq{\alpha}$ which starts at approximately the same time $\qr=1$ becomes non-optimal.
This parameter is always close to $\frac{1}{2}$, but slightly biased to higher values.
This is due to the asymmetry in the noise when using the single rail method.

\begin{figure}[thp]
  \begin{center}
      \includegraphics[width=272pt]{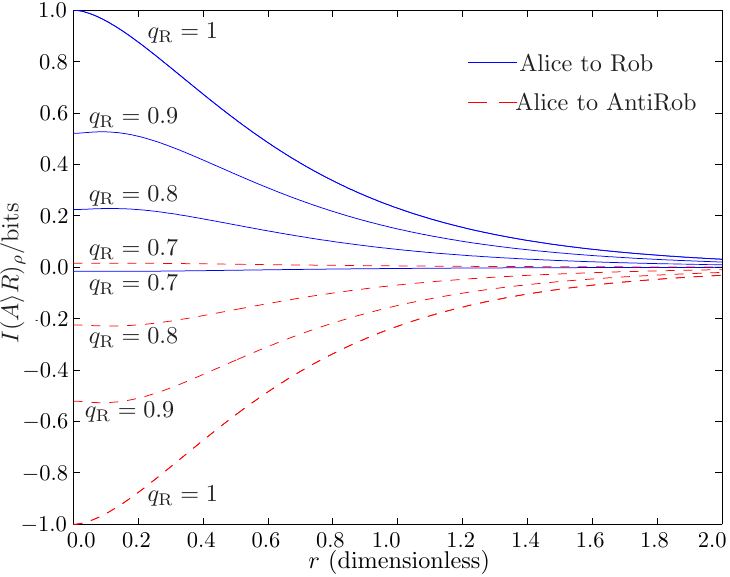}
  \caption[Single rail coherent information]{Coherent information of the single rail method of quantum entanglement generation.}  \label{fig:singleNCE}
  \end{center}
\end{figure}

\section{Quantum communication and entanglement generation}
\label{sec:arr:quantum}

\begin{figure}[bhtp]
  \begin{center}
      \includegraphics[width=272pt]{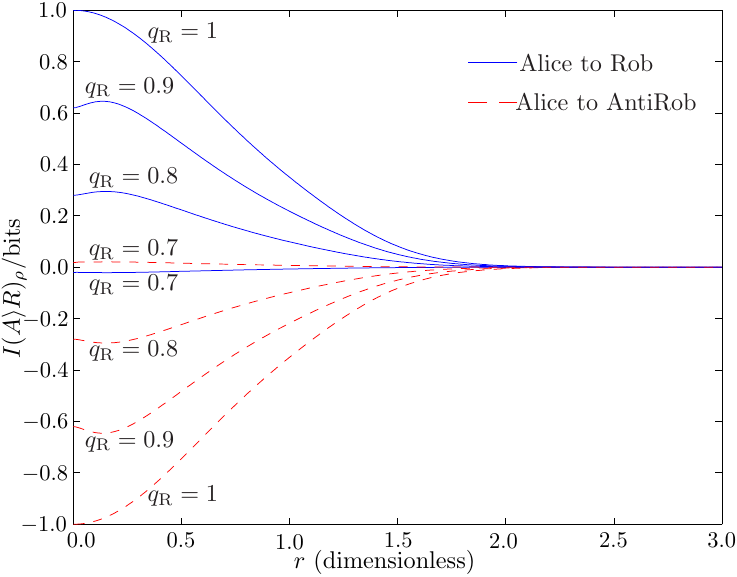}
  \caption[Dual rail Holevo information]{Coherent information of the dual rail method of quantum entanglement generation.}  \label{fig:dualNCE}
  \end{center}
\end{figure}

The quantum coherent information is given by the negative of the conditional entropy.
The results of this were plotted for both the single rail in \cref{fig:singleNCE} and the dual rail in \cref{fig:dualNCE}.

The plots show that when the coherent information is positive for one bipartition, it is negative for the other.
We call this the monogamy property, Alice must choose in advance with whom she wants to generate entanglement for quantum communication.
For large acceleration, the coherent information tends to zero, for both single rail and dual rail methods.
The dual rail methods always perform slightly better than the single rail methods.

We find that beyond the Single Wedge Mapping the decrease in information transfer is non-monotonic.
However, in most cases this is not optimal, as Alice is able to choose her modes, and therefore has control over the value of $\qr$.
This means that, after the maximization, the restricted capacities are monotonically decreasing with the acceleration parameter.
The SWM is sufficient to fully describe optimal entanglement generation for all accelerations.
In \cref{ch:blackholes}, the interesting entanglement generation results used the SWM, which we have verified here is appropriate.

\label{sec:arr:monogamy}
We find that if one of the channels is able to generate entanglement, the other cannot.
More precisely, the sum of the conditional entropies of both channels is always greater than or equal to zero.
This is a consequence of the strong subadditivity of the von Neumann entropy \cite{Lieb1973}, which implies that for any tripartite system composed of parties $A$, $B$ and $C$, the inequality,
\begin{align}
S(\rho_A)+S(\rho_C)\leq S(\rho_{AB})+S(\rho_{BC}),
\end{align}
holds.
Choosing $A=\text{AntiRob}$, $B=\text{Alice}$, $C=\text{Rob}$ and rearranging, we obtain this condition on the sum of conditional entropies,
\begin{align}
[S(\rho_{A\bar{R}})-S(\rho_{\bar{R}})]+[S(\rho_{AR})-S(\rho_R)]\geq 0.
\label{sumcond}
\end{align}
If Alice can generate quantum entanglement with Rob, she will not be able to do the same with AntiRob, and vice versa.
The results we are presenting saturate the inequality \cref{sumcond}, but the interpretation is the same: the entanglement generation ability of the state merging protocol is bound to be zero for at least one of the bipartitions.

\section{Conclusion}
\label{sec:conclusion}
We have studied the classical Holevo information and the quantum conditional information (both in the single and dual rail encodings) in the situation where an inertial observer (Alice) communicates by sending information to two counter-accelerating observers each outside of the other's acceleration horizon (Rob and AntiRob).
The dual rail encoding has consistently outperformed the single rail for both classical and quantum channel capacities in this setting.
The dual rail is symmetrical between the logical basis states.
The noise generated in each mode is independent but equal in magnitude and therefore the resulting state is symmetrical.
The single rail uses only one mode, and therefore there are differences between the logical basis states.
They acquire potentially different levels of noise.
We saw that the optimal values of the parameters deviate from symmetry.

We found that the quantum channels between the inertial observer and each of the accelerated observers are mutually exclusive.
Alice must choose with which accelerated observer, Rob or AntiRob, she wants to generate entanglement.
This is not simply a demonstration of the no-cloning theorem because the monotony holds for separate messages on separate occasions, if Alice uses the same setup.
The setup used to communicate with Rob cannot ever be used to communicate with AntiRob, the values of $\qr$ and $q_\text{L}$ need to change.
If we wanted to design technology to communicate quantum information between inertial and accelerated observers, we would have to design some way of changing the mode functions to which it couples.
A fixed mode function coupling would only ever allow quantum communication with observers that have certain specific characteristics.
We may need some sort of `tuning' mechanism that allows Alice to switch her communication channel from Rob to AntiRob.
Classical communication is not affected by this limitation.
We find again, that classical information and correlations are more robust, and allow Alice to communicate with both Rob and AntiRob using the same channel.

We showed that an Unruh mode with $\qr=1$ is always optimal to send quantum information to Rob and $\qr=0$ is optimal for communication with AntiRob.
For larger acceleration, and therefore larger $r$, we find that $\qr=1$ is no longer optimal for sending classical information to Rob, in both the single and dual rail encodings.
This is important for other research in the field because the single wedge approximation (of $\qr=1$) is very commonly used.
Researchers must understand the limitations of using this approximation, and in what situations it no longer provides an optimal channel.
The main conclusions do not change however, from extending beyond the approximation.
The channel still remains finite in the infinite acceleration limit.
We have computed the optimal $\modsq{\alpha}$ and $\qr$ for the single rail encoding, showing that the corresponding Holevo information is a monotonically decreasing function of the acceleration.
This monotonical decrease is important because it would be very surprising to have the channel increase again for increased noise levels.

The encodings of single and dual rail have restricted the states used in the channel.
The communication may be improved if we were to use a different computational basis, perhaps not using the number basis at all.
We could use some states that are common in metrology, such as the NOON state, or other multi-excitation states.

We could also study the situation using a broadcast channel formalism.
Instead of treating it as separate communication attempts, using the same system setup, we could investigate simultaneous communication between Alice, Rob and AntiRob.
Causality would prohibit Rob and AntiRob from communicating at all, but Alice could simultaneously send classical and quantum information to both of them.

\chapter{Conclusion}
\label{ch:conclusion}

Both quantum communication and metrology will be very relevant to technology in the coming years.
Quantum computing is getting closer, and with it the need for quantum communication.
Metrology has always been relevant to perform the most precise measurements possible.
Using quantum effects in metrology can increase the precision beyond what is possible classically.
When these technologies are used in situations where they are subject to acceleration or a gravitational field it is important to understand how their behaviour will change.
Studying what encodings and techniques perform the task most successfully will drive the design of this technology, particularly when it is intended to be used under similar conditions.

In this thesis we have studied relativistic quantum communication.
We started by introducing the important areas of relativistic field theory and quantum information, then reviewed the field of relativistic quantum information.
The focus of the thesis is communication between an inertial and an accelerated observer.
We have looked at both single and dual rail encodings using bosonic field modes in all situations.
The main points that have been studied are classical and quantum communication using a quantum channel between inertial Alice and Rob both accelerated and hovering above a black hole, the situation where Alice communicates with both accelerated Rob and his causally disconnected (counter-accelerated) partner AntiRob, and the ability of Rob to recover information about a continuous parameter has also been studied in the accelerated case.
Time dilation is ignored in the entire thesis because it only affects the rate of communication, and of all channels equally.

Throughout the communication studies we found that the classical information was always much more robust to the noise generated by the acceleration.
This is demonstrated by the fact that the classical channel capacity remains finite all the way up to the horizon, where the quantum channel capacity drops to zero.
The quantum communication is further restricted in the case where Alice uses the same setup to communicate with both Rob and AntiRob.
This restriction is that Alice can tune her communication device to either have a non-zero quantum capacity with Rob or AntiRob, but not both.
For parameter estimation, we find a similar situation to classical communication, in that the amount of extractable information tends towards a finite value as the acceleration increases.
Using the entanglement generating communication protocol with the parameter encoded onto the amplitudes, this limiting value is highly dependent on the particular encoding (single or dual rail) and the value of the parameter itself.

For the majority of the thesis, we use the single wedge mapping because it often provides the optimal information transfer.
Quantum communication remains optimal in the single wedge mapping in all situations that we studied.
However, we found for classical communication with an accelerated observer that it is no longer optimal.
The main conclusions do not change, only the limiting value of the classical channel capacity.
This should not affect the metrology chapter because the information transfer is fully quantum, where we continued to find the single wedge mapping optimal.
It is important to understand the limitations of such simplifications, particularly when they are heavily used in the literature.

The dual rail encoding provided greater information transfer in almost all situations studied.
Estimating the phase of a NOON state after communication through the quantum channel had better precision when using the single rail.
This was unexpected but perhaps not too surprising.
Their are some fundamental differences to the two techniques that are worth noting.
In the situations where the dual rail was preferable, the communication was based on creating entanglement between Alice and Rob.
It also stored the information in the amplitudes of the logical basis.
The single rail was preferable when the communication did not create any entanglement, when it simply transferred the state to Rob.
In this case we also put the parameter containing the information in the phase of the state.
Further investigation would be required to determine the exact cause of this discrepancy.
It is important for further research and the design of quantum communication technology to know which of these performs better in any given situation.
If we could ascertain the fundamental cause of this preference, we would be able to find the optimal encoding for other situations.
We may even get ideas for other encodings that could outperform both the single rail and dual rail.

In all situations we find that the information task has a monotonically decreasing ability with noise.
There is no case where we suddenly have more noise in the system, and more information.
This is exactly as expected.

We find that the channel we use behaves like an amplifying channel.
The number of noise excitations that are created depend on the number of excitations already present in the mode.
This was only noticeable when we studied NOON states with different numbers of excitations.
For large $N$, the extractable information dropped to zero, because the channel became flooded with noise.
This combined with the noiseless situation of a higher $N$ providing better precision meant that there was an optimal $N$ for each noise level.
This may reflect back to the communication channel capacities, if we define the logical basis to be the two parts of a NOON state.
We found this situation very similar to that of using NOON states in lossy interferometers that have been studied by other people.
They found that replacing the NOON states with other forms of states was optimal.
It would be interesting to see if those other states also improved on NOON states in this situation.

The field of relativistic quantum information theory is still quite new.
Entanglement seems to have been the favourite topic so far, studied under many conditions.
This work extends the field into quantum communication and metrology by studying the effect of an accelerated observer on the standard protocols.
It provides many pointers for areas that may prove interesting to study.
It would always be interesting to reflect the situations in which entanglement has been studied, and look at the full communication protocols.
The metrology is also important because it tells us how well we can measure quantities in the situation studied.
This all contributes to a greater understanding of how quantum information behaves under the influence of curved spacetimes.

\appendix

\chapter{The Single Wedge Mapping}
\label{ch:SMA}

I call this choice of parameters the Single Wedge Mapping (SWM).
This is because it reflects the fact that Alice creates Unruh mode excitations that map directly to only one of the Rindler wedges.
Previously in the literature, this has been called the Single Mode Approximation.
Even when we go beyond the single wedge mapping, we are still using single Rindler modes (of a single frequency) so this nomenclature is misleading.
I propose to change the name and call it the single wedge mapping instead.

To set up the transformation between an inertial packet and the accelerated observers, we must first start with an annihilation operator of a Minkowski packet $\hat{a}_\text{U}$.
We then transform it to the annihilation operators of Unruh modes, $\hat{a}_{\text{U}_\text{L}}$ and $\hat{a}_{\text{U}_\text{R}}$ for the left and right regions respectively, using the most general transformation,
\begin{equation}
  \label{eq:SMA:unruh-modes}
  \hat{a}_{\text{U}}=q_\text{L} \hat{a}_{\text{U}_\text{L}}+\qr \hat{a}_{\text{U}_\text{R}},
\end{equation}
where $\modsq{q_\text{L}}+\modsq{\qr}=1$ and the appropriate expressions for the operators $\hat{a}_{\text{U}_\text{L}}$ and $\hat{a}_{\text{U}_\text{R}}$ for the scalar case are,
\begin{subequations}
  \label[pluralequations]{eq:SMA:bogoliubov-boson}
\begin{align}
  \hat{a}_{\text{U}_\text{R}}&=\cosh r\, \hat{a}_{\text{R}_\text{I}} - \sinh r\, \hat{a}^\dagger_{\text{R}_\text{II}},\\
  \hat{a}_{\text{U}_\text{L}}&=\cosh r\, \hat{a}_{\text{R}_\text{II}} - \sinh r\, \hat{a}^\dagger_{\text{R}_\text{I}},
\end{align}
\end{subequations}
where $\hat{a}_{\text{R}_\text{I}}$, $\hat{a}_{\text{R}_\text{II}}$ are Rindler particle operators for the scalar field in each region.

The values of $\qr$ and $q_\text{L}$ are set by the specific shape of the packet function.
The inertial observer therefore has control over it, and in certain protocols we assume that $\qr=1$.
This assumption is what I am calling the Single Wedge Mapping (SWM).
When things are discussed where $\qr\ne1$, this has been called `beyond the Single Mode Approximation' \cite{Bruschi2010}.
I will adapt the nomenclature for this situation to `beyond the Single Wedge Mapping'.

In our work on black holes, \cref{ch:blackholes,ch:measure}, we use the SWM because it broke the symmetry between Rob and AntiRob.
We thought that only mapping Alice's packet to Rob's wedge mode would optimise the information transfer between Alice and Rob.
Surprisingly, this turned out not to be the case for classical communication in the regime of large accelerations.
The optimal classical communication is actually slightly higher in this regime if you move beyond the SWM.

We moved beyond the SWM in our work researching the channels between Alice, and either Rob or AntiRob, \cref{ch:alicerobantirob}.
To study any communication between Alice and AntiRob, we must move beyond the SWM to set up correlations between Alice and AntiRob.
It was in this paper where we discovered that the SWM was not sufficient for classical communication at large accelerations.

\chapter{Rindler modes}
\label{ch:rindler-modes}

Solving the Klein-Gordon equation in Rindler coordinates gives separate solutions for each region of the Rindler spacetime.
The modes given here are single frequency modes as observed by the Rindler observer travelling on the path $\xi=\frac1a$.

The modes from region I are written using the Rindler equivalent of the tortoise coordinate $\xi_* = \ln|\xi|$, as \cite{Takagi1986},
\begin{subequations}
  \label{eq:rindler:rindlermodesregioni}
\begin{align}
  u_{k,\text{R}_\text{I}}^{(\omega)}(\eta,\xi,x,y)=&\theta(\xi) \frac1{2 \pi \sqrt{2 \omega}} \exp\left(- i \omega \eta + i k_x x + i k_y y\right) h_k^{(\omega)}(\xi),\\
  h_k^{(\omega)}(\xi)=& \sqrt{\frac{2}{\pi}} A_k^{(\omega)} \left(\frac{\mu_k}{2}\right)^{i \omega} \Gamma(i\omega)^{-1} K_{i\omega}(\mu_k \xi),\\
  \mu_k=&\sqrt{k_x^2 + k_y^2},
\end{align}
\end{subequations}
where $\theta(\xi)$ is the heaviside step function, $\Gamma(x)$ is the gamma function, $\omega$ is a positive parameter that we will treat as a frequency, $K$ is a modified Bessel function of the second kind, and $A_k^{(\omega)}$ is a constant phase factor.

The modes from region IV are only different in the signs of certain parts.
\begin{subequations}
  \label{eq:rindler:rindlermodesregioniv}
\begin{align}
  u_{k,\text{R}_\text{IV}}^{(\omega)}(\eta,\xi,x,y)=&\theta(- \xi) \frac1{2 \pi \sqrt{2 \omega}} \exp\left(i \omega \eta + i k_x x + i k_y y\right) h_k^{(\omega)}(\xi),\\
  h_k^{(\omega)}(\xi)=& \sqrt{\frac{2}{\pi}} A_k^{(\omega)} \left(\frac{\mu_k}{2}\right)^{i \omega} \Gamma(i\omega)^{-1} K_{i\omega}(\mu_k \xi),\\
  \mu_k=&\sqrt{k_x^2 + k_y^2}.
\end{align}
\end{subequations}

\chapter{Analytical solutions for the Fisher information}
\label{ch:measurefullmaths}

The Fisher information involves some fairly lengthy mathematical expressions due to the infinite sums.
We give the full expressions here for completeness.

\section{Required mathematical functions}
Before we write the full expression as found by Mathematica \cite{Mathematica} we need to define some of the functions used.
These functions turn up because they are defined in terms of series sums.

The generalized hypergeometric function is defined as,
\begin{align}
  _p F_q \left[
    \begin{array}{c}
      a_1,a_2,\cdots,a_p \\
      b_1,b_2,\cdots,b_q
    \end{array};x
  \right]= \sum_{k=0}^\infty \frac{(a_1)_k (a_2)_k \cdots (a_p)_k}{(b_1)_k(b_2)_k\cdots(b_q)_k} \frac{x^k}{k!},
\end{align}
where $(a)_k$ is the Pochhammer symbol defined as,
\begin{align}
  (a)_k \equiv \frac{\Gamma(a+k)}{\Gamma(a)}= a (a+1) \cdots (a+k-1).
\end{align}
The arrays of parameters $a$ and $b$ can be any size.

The Lerch transcendent is defined as,
\begin{align}
  \Phi(z,s,a) \equiv \sum_{k=0}^\infty \frac{z^k}{(a+k)^s},
\end{align}
where $|z|<1$ and $a \notin \{\mathbb{Z}^{-},0\}$.

\section{Single rail encoding}
The single rail calculations has an analytical solution that can be presented here for the Fisher information of Rob performing the relative amplitude measurement by himself.
This does not matter about classical or quantum encoding because Rob's state is the same once Alice is traced out.
The Fisher information is,
\begin{align}
  \mathscr{F}(\theta) = \,\,\,\,\,\,&\!\!\!\!\!\!\frac{4 \cos^2\theta}{\csch^2(r) \sin^2\theta + \cos^2\theta}
  \Bigg[\sech^4(r) \sin^2\theta \\
  \times &\Bigg(\csch^2(r)\, _3F_2\left(
  \!\begin{array}{c}
    2, 2, \cot^2\theta \sinh^2(r) +1 \\
    1, \cot^2\theta \sinh^2(r) +2
  \end{array}\!\!;\tanh^2(r)\right) \nonumber \\
  &- 2 \, _2F_1 \left(
  \!\begin{array}{c}
    2,\cot^2\theta \sinh^2(r)+1 \\
    \cot^2\theta \sinh^2(r)+2
  \end{array}\!\!;\tanh^2(r)\right)\Bigg)
   \nonumber \\
   +&\Phi\left(\tanh^2(r), 1, \cot^2\theta \sinh^2(r)\right)
   \left(\tanh^2(r) \cos^2\theta + \sech^2(r) \sin^2\theta\right)\Bigg]. \nonumber
\end{align}

\section{Dual rail encoding}
The dual rail calculations have a sum to infinity over two independent indices.
Mathematica could calculate the first sum, but not the second.
I present here the Fisher information with the first sum completed.
For the graphs I calculated the rest numerically using Matlab, and a simple convergence test.

If we trace out Alice, Rob's state for both classical and quantum information being communicated become equal.
The Fisher information in this case cannot be fully calculated analytically.
I present here the furthest the calculation can be done analytically,
\begin{align}
  \mathscr{F}(\theta) =\sum_{n=0}^\infty \,\,\,\,\,&\!\!\!\!\!\frac{-\sech^4(r)\tanh^{2n}(r)}{\cos^2\theta + n \sin^2\theta}\\
    \times& \Bigg[-\,_3F_2\left(
  \!\begin{array}{c}
    2, 2,1+n\tan^2\theta \\
    1, 2+n\tan^2\theta
  \end{array}\!\!;\tanh^2(r)\right) \sech^2(r)\sin^2(2\theta) \nonumber \\
  &+ 2n \Bigg(\!\! -2  \,_2F_1\left(
  \!\begin{array}{c}
    1, n\tan^2\theta \\
    1+n\tan^2\theta
  \end{array}\!\!;\tanh^2(r)\right) \cos^2\theta \csch^2(r) \left(\cos^2\theta + n \sin^2\theta \right)\nonumber \\
  \!\!\!\!\!\!&\,\,\,\,\,\,+  \,_2F_1\left(
  \!\begin{array}{c}
    2, 1+n\tan^2\theta \\
    2+n\tan^2\theta
  \end{array}\!\!;\tanh^2(r)\right) \sech^2(r)\sin^2(2\theta)\Bigg)\Bigg]. \nonumber
\end{align}
The other infinite sum had to be calculated numerically in Matlab \cite{Matlab}.

\backmatter

\sloppy 
\printbibliography

\end{document}